\documentclass[aps,prd,twocolumn,superscriptaddress,preprintnumbers,floatfix,nofootinbib,balancelastpage]{revtex4-2}

\pdfoutput=1
\usepackage{hyperref}
\usepackage{amsmath,bbm}
\usepackage{amssymb}
\usepackage{graphicx}
\usepackage{slashed}
\usepackage{dsfont}
\usepackage{empheq}
\usepackage[usenames,dvipsnames]{xcolor}

\newcommand{\refcite}[1]{Ref.~\cite{#1}}
\newcommand{\refscite}[1]{Refs.~\cite{#1}}
\newcommand{\eq}[1]{Eq.~\eqref{eq:#1}}
\newcommand{\eqs}[2]{Eqs.~\eqref{eq:#1} and \eqref{eq:#2}}
\newcommand{\eqss}[3]{Eqs.~\eqref{eq:#1}, \eqref{eq:#2} and \eqref{eq:#3}}
\renewcommand{\sec}[1]{Sec.~\ref{sec:#1}}

\newcommand{\fig}[1]{Fig.~\ref{fig:#1}}

\newcommand{\tab}[1]{Table~\ref{tab:#1}}

\newcommand{\nn}{\nonumber}

\newcommand{\ord}[1]{\mathcal{O}(#1)}
\newcommand{\Ord}[1]{\mathcal{O}\bigl(#1\bigr)}
\newcommand{\ORd}[1]{\mathcal{O}\Bigl(#1\Bigr)}

\newcommand{\mae}[3]{\langle#1\lvert#2\rvert#3\rangle}
\newcommand{\Mae}[3]{\big\langle#1\big\lvert#2\big\rvert#3\big\rangle}

\newcommand{\ket}[1]{\lvert#1\rangle}
\newcommand{\Ket}[1]{\big\lvert#1\big\rangle}

\newcommand{\Bra}[1]{\big\langle#1\big\rvert}

\newcommand{\braket}[2]{\langle#1\vert#2\rangle}

\newcommand{\tr}{\operatorname{tr}}
\newcommand{\Tr}{\operatorname{Tr}}

\newcommand{\Sl}[1]{\slashed{#1}}

\newcommand{\df}{\mathrm{d}}
\newcommand{\id}{\mathbf{1}}
\newcommand{\img}{\mathrm{i}}

\DeclareMathOperator*{\SumInt}{%
\mathchoice%
  {\ooalign{$\displaystyle\sum$\cr\hidewidth$\displaystyle\int$\hidewidth\cr}}
  {\ooalign{\raisebox{.14\height}{\scalebox{.7}{$\textstyle\sum$}}\cr\hidewidth$\textstyle\int$\hidewidth\cr}}
  {\ooalign{\raisebox{.2\height}{\scalebox{.6}{$\scriptstyle\sum$}}\cr$\scriptstyle\int$\cr}}
  {\ooalign{\raisebox{.2\height}{\scalebox{.6}{$\scriptstyle\sum$}}\cr$\scriptstyle\int$\cr}}
}

\newcommand{\eps}{\epsilon}

\newcommand{\bn}{{\bar n}}
\newcommand{\bg}{{\bar{g}}}

\newcommand{\cA}{\mathcal{A}}

\newcommand{\cH}{\mathcal{H}}

\newcommand{\cL}{\mathcal{L}}
\newcommand{\cM}{\mathcal{M}}
\newcommand{\cO}{\mathcal{O}}

\newcommand{\cQ}{\mathcal{Q}}

\newcommand{\lqcd}{\Lambda_\mathrm{QCD}}
\newcommand{\as}{\alpha_{s}}

\newcommand{\bperp}{b_{\perp}}
\newcommand{\kperp}{k_{\perp}}

\newcommand{\FH}{F^{\beta\beta'}_{H, h_{H}h_{H}'}}

\allowdisplaybreaks[2]

\providecommand{\sectionPaper}[1]{\section{#1}}
\providecommand{\headingAcknowledgments}{}
\providecommand{\app}[1]{Appendix~\ref{app:#1}}

\begin{document}

\title{How Many Structure Functions Characterize \\
Heavy-Quark Fragmentation in the Transverse Plane?}

\author{Rebecca von Kuk}%
\email{rebecca.vonkuk@unibe.ch}%
\affiliation{Albert Einstein Center for Fundamental Physics, Institut f\"ur Theoretische Physik,\\
	Universit\"at Bern, Sidlerstrasse 5, CH-3012 Bern, Switzerland}%

\author{Kyle Lee}%
\email{kyle@anl.gov}%
\affiliation{High Energy Physics Division, Argonne National Laboratory, Lemont, IL 60439, USA}%

\author{Johannes K. L. Michel}%
\email{j.k.l.michel@uva.nl}%
\affiliation{Institute for Theoretical Physics Amsterdam and Delta Institute for Theoretical Physics, University of Amsterdam, Science Park 904, 1098 XH Amsterdam, The Netherlands}%
\affiliation{Nikhef, Theory Group, Science Park 105, 1098 XG, Amsterdam, The Netherlands}%

\author{Zhiquan Sun}%
\email{zqsun@berkeley.edu}%
\affiliation{Leinweber Institute for Theoretical Physics,\,University of California,\,Berkeley,\,CA\,94720}
\affiliation{Theory Group,\,Physics Division,\,Lawrence Berkeley National Laboratory,\,Berkeley,\,CA\,94720}

\date{November 30, 2026}

\begin{abstract}
We extend the spin density matrix formalism developed in \refcite{vonKuk:2023jfd}
to study the transverse-momentum-dependent (TMD) fragmentation functions (FFs)
of polarized heavy hadrons.
Heavy-quark spin symmetry allows us to express 
all leading-power TMD FFs of a heavy-hadron spin multiplet, with the light constituents carrying total angular momentum $s_\ell$,
in terms of the light spin-density matrix $\rho_{\ell}$.
Constraining $\rho_{\ell}$ by hermiticity, parity, and rotational covariance, 
we find that only $\lceil (2s_\ell + 1)^2/2 \rceil$ independent structure functions remain, 
yielding relations between the TMD FFs of heavy hadrons within the same multiplet.
The remaining dynamical information offers powerful new probes
of the quantum state prepared by hadronization in the presence of the heavy quark, i.e., a static color source.
The formalism we develop to deal with arbitrary $s_\ell$ is of general interest in spin physics.
Our analysis opens the door to a variety of exciting, in-depth experimental studies of hadronization at existing $B$ factories
and the future Electron-Ion Collider.

\end{abstract}

\maketitle

\tableofcontents

\makeatletter
\let\toc@pre\relax
\let\toc@post\relax
\makeatother

\sectionPaper{Introduction}
\label{sec:intro}

Understanding the dynamics of hadronization, 
the nonperturbative process in which quarks and gluons form color-confined hadrons,
is one of the greatest challenges of Quantum Chromodynamics (QCD).
The fragmentation of a heavy (charm or bottom) quark into a heavy hadron
is particularly instructive to study,
because the mass $m = m_c, m_b$ of the heavy quark provides a perturbative scale
far above the confinement scale $\lqcd$,
making the heavy quark effectively a static color source
coupling to the light degrees of freedom.
For collinear fragmentation, which describes the fraction of the quark's longitudinal momentum carried by the hadron,
the heavy-quark mass has long been exploited
to separate the perturbative and nonperturbative dynamics~\cite{Mele:1990cw, Jaffe:1993ie, Neubert:2007je, Fickinger:2016rfd, Bonino:2023icn}
and, through heavy-quark spin symmetry~\cite{Isgur:1989vq, Isgur:1990yhj, Isgur:1991wq, Neubert:1993mb, Manohar:2000dt},
to relate the production rates and polarizations of heavy hadrons
within the same spin symmetry multiplet~\cite{Falk:1993rf, Galanti:2015pqa, Masouminia:2023zhb}.

A rigorous field-theoretic framework for studying fragmentation
is provided by transverse momentum-dependent (TMD) factorization,
which allows for a detailed investigation
of the three-dimensional dynamics of hadronization.
TMD factorization separates the full hadronic cross section into 
the perturbatively calculable partonic hard scattering process and
nonperturbative functions that describe either
the structure of initial-state hadrons or their formation in the final state.
The three-dimensional nonperturbative information of quarks fragmenting into final-state hadrons
is captured by TMD fragmentation functions (FFs),
which in particular also encode the correlations
between the transverse momentum of the hadron, its polarization,
and the parent quark's polarization.
The number of such TMD FFs grows rapidly with the hadron spin $s_H$:
at leading power, $2(2s_H+1)^2$ of them are needed for each quark flavor,
i.e., $2$, $8$, $18$, and $32$ for $s_H = 0, \tfrac{1}{2}, 1, \tfrac{3}{2}$~\cite{Mulders:1995dh, Boer:1997mf, Bacchetta:2000jk, Chen:2016moq, Zhao:2022lbw},
and this rich structure has attracted great interest in recent years~\cite{Metz:2016swz, Anselmino:2020vlp, Boussarie:2023izj, Chen:2023kqw, Kumano:2024fpr, Accardi:2012qut, AbdulKhalek:2021gbh}. 
See~\refcite{Boussarie:2023izj} for a review.
However, while unpolarized TMDs are by now extracted from global fits at high perturbative accuracy~\cite{Bacchetta:2022awv, Moos:2023yfa, Bacchetta:2024qre}
and the Collins function is increasingly constrained by global analyses~\cite{Anselmino:2013vqa, Kang:2015msa, Cammarota:2020qcw, Gamberg:2022kdb},
phenomenological studies and experimental constraints on fragmentation into polarized hadrons have so far concentrated on a small subset of the allowed spin-dependent functions~\cite{Bunce:1976yb, Anselmino:2000vs, Boer:2010ya, Belle:2018ttu, Anselmino:2019cqd, DAlesio:2020wjq, Callos:2020qtu, Li:2020oto, Chen:2021hdn, Gamberg:2021iat, DAlesio:2022brl, DAlesio:2023ozw, STAR:2025jwc, Kang:2020xyq, Kang:2021ffh, Kang:2021kpt, Chen:2021zrr, Gao:2024bfp, Yang:2024kjn, ALEPH:1996oew, OPAL:1997oem, OPAL:1997vmw, Chen:2020pty}. Heavy quarks offer a unique opportunity to bring this proliferation of structures under theoretical control.

Recently, the study of TMD FFs for heavy quarks fragmenting
to heavy hadrons was initiated
by some of us in \refcite{vonKuk:2023jfd} (see also \refcite{Dai:2023rvd}).
Here, effective field theories (EFTs),
and boosted Heavy Quark Effective Theory (bHQET)~\cite{Fleming:2007qr, Fleming:2007xt} in particular,
are used to analyze the rich hierarchies of scales involved in the problem.
In \refcite{vonKuk:2023jfd} we showed that in the regime $\lqcd \lesssim k_{T} \ll m$,
the two TMD FFs (unpolarized and Collins) that characterize
the production of unpolarized heavy hadrons
from unpolarized and transversely polarized quarks, respectively,
are entirely determined by the underlying spin density matrix
of the light hadron constituents
thanks to heavy-quark spin symmetry.
This further allowed us to derive new spin symmetry relations and sum rules
between the unpolarized and Collins TMD FFs of heavy hadrons
within the same spin symmetry multiplet,
e.g.\ the pseudoscalar $B$ and vector $B^*$,
halving the number of independent nonperturbative functions.%
\footnote{
The number can be brought down further
using the exact heavy quark flavor symmetry $c \leftrightarrow b$
of HQET at leading order in $1/m$,
as well as the approximate isospin
and -- even more approximate -- flavor $\mathrm{SU}(3)$
symmetry of the light constituents.
Like the relations from charge conjugation,
these are easy to implement,
so we will simply consider a generic $Q = c,b$ in this paper,
and mostly suppress the flavor of the light constituents.
}
It is natural to ask, as we do in the title,
whether a similar reduction applies
to the much larger set of TMD FFs
describing the production of polarized hadrons off (un)polarized quarks,
and whether doing so provides access
to additional entries in the light spin density matrix.

\begin{figure}[b]
\includegraphics[width=0.45\textwidth]{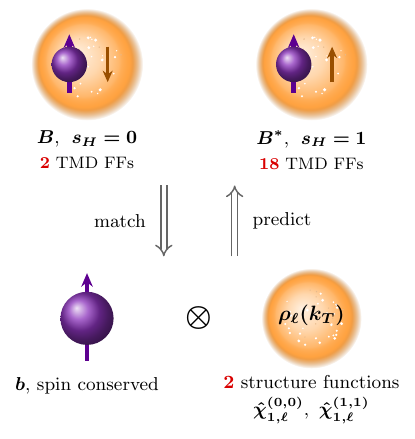}
\caption{
\label{fig:spin_multiplet_example}
$b$-quark spin symmetry for the $s_\ell = \tfrac12$ multiplet.
The conserved heavy-quark spin couples to the light degrees of freedom, 
described by the light spin-density matrix $\rho_\ell(k_T)$, 
to form the pseudoscalar $B$ and the vector $B^*$.
Since both hadrons share the same $\rho_\ell$, 
all of their 20 leading-power TMD FFs are predicted by just two light structure functions (see \sec{light_spin_density_matrix}).
}
\end{figure}

In this work, we thus extend the formalism of \refscite{vonKuk:2023jfd, vonKuk:2024uxe, vonKuk:2025kbv}
to study heavy-quark TMD FFs for \emph{polarized} heavy hadrons.
We perform a comprehensive symmetry analysis
of the light constituent dynamics,
allowing us to reduce all polarized TMD FFs for heavy hadrons
to a minimal set of universal nonperturbative bHQET structure functions.
As summarized in \tab{number_structure_functions}, the resulting reduction is drastic:
all $8$ leading-power TMD FFs of $\Lambda_Q$ $(s_H = \tfrac{1}{2})$ baryons are fixed by a single function,
the combined $2 + 18 = 20$ TMD FFs of the $B$ ($s_H = 0$) and $B^*$ ($s_H = 1$) mesons by only two,
and the combined $8 + 32 = 40$ TMD FFs of the $\Sigma_b$ ($s_H = \tfrac{1}{2}$) and $\Sigma_b^*$ ($s_H = \tfrac{3}{2}$) baryons by only five.
For example, all $18$ TMD FFs of the vector meson $B^*$
are predicted in terms of the unpolarized and Collins TMD FFs of the pseudoscalar $B$.
We show a cartoon illustration of this example in \fig{spin_multiplet_example}.
Since these relations hold point by point in transverse momentum for $k_T \ll m$,
up to corrections suppressed by powers of $1/m$,
heavy-quark fragmentation provides a unique, theoretically controlled testing ground
for the spin structure of TMD fragmentation.

Polarized heavy-hadron TMD FFs are also within experimental reach. Longitudinal and transverse momentum distributions of $B$ mesons inside jets have already been measured~\cite{ATLAS:2021agf, LHCb:2026mbl}, alongside longitudinal distributions of charm mesons and baryons~\cite{ALICE:2019cbr, ALICE:2022mur, ALICE:2023jgm} and radial profiles of $D^0$ mesons~\cite{CMS:2019jis}. Heavy-baryon polarization can be reconstructed from decay products~\cite{Falk:1993rf, ALEPH:1995aqx, OPAL:1998wmk, DELPHI:1999hkl, LHCb:2013hzx, ATLAS:2014swk, CMS:2018wjk, LHCb:2020iux, LHCb:2022sck, LHCb:2023crj}, while heavy-vector-meson spin alignment (i.e. tensor polarization) can be measured through decay angular distributions~\cite{OPAL:1997nwj, ALICE:2022byg}. Proposals have also been developed to use heavy baryons as polarimeters in flavor-tagged jets~\cite{Galanti:2015pqa, Kats:2023zxb, Afik:2025grr}, whose identification has improved substantially in recent years~\cite{ATLAS:2025dkv}. At the $B$ factories, collinear fragmentation functions of charm hadrons have already been measured~\cite{Belle:2005mtx}, while measurements of spin-dependent TMD fragmentation have focused on light hadrons so far~\cite{Belle:2005dmx, BaBar:2013jdt, Belle:2018ttu}. Belle~II, with its larger dataset and improved vertex reconstruction~\cite{Accardi:2022oog,Belle-II:2010dht}, should enable transverse-momentum measurements for charm hadrons, including the transverse polarization of $\Lambda_c$ baryons and the spin alignment of $D^*$ mesons. At the future Electron-Ion Collider, polarized electron and proton beams~\cite{AbdulKhalek:2021gbh} will produce heavy quarks in all polarization states with simultaneous sensitivity to heavy quark TMD PDFs~\cite{vonKuk:2023jfd}, providing access to spin-dependent TMD fragmentation and all the nontrivial tests of the spin-symmetry relations derived here.

The rest of the paper is structured as follows:
In \sec{light_spin_density_matrix}, we decompose the light spin-density matrix into a minimal set of scalar structure functions, in particular through the symmetry analysis in \sec{symmetry_analysis}, which counts the independent structure functions for any light spin $s_\ell$ and answers the question posed in the title.
In \sec{light_dm_to_tmdffs}, we relate these structure functions to the standard leading-power TMD FFs of polarized heavy hadrons, in particular giving explicit results for hadron spins $s_H \leq 1$ in \sec{tmdff_lightchi_relations}.
In \sec{examples}, we discuss the resulting heavy-quark spin symmetry relations between TMD FFs within the same multiplet and their physical interpretation.
We conclude in \sec{conclusions}.
Conventions and the complete dictionaries between Dirac projections, hadron analyzers, and TMD FFs are collected in the appendices.

\sectionPaper{Light spin-density matrix}
\label{sec:light_spin_density_matrix}

\begin{figure}[b]
\includegraphics[width=0.4\textwidth]{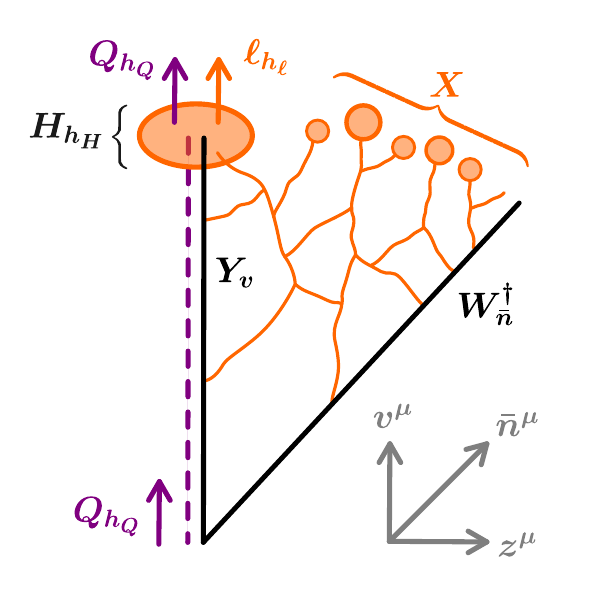}
\caption{
\label{fig:rho_ell_spacetime}
Minkowski diagram of fragmentation dynamics encoded
in the light spin-density matrix $\rho_\ell$.
The transverse direction, which points out of the plane,
and the operators in the conjugate amplitude
are suppressed for clarity.
}
\end{figure}

\subsection{Formalism}
\label{sec:formalism}

We consider the fragmentation $Q \to H X$ of a heavy quark $Q = c, b$,
where the heavy hadron contains $Q$ and additional light constituents $\ell$
with total angular momentum $s_\ell$ and flavor $f_\ell$,
as illustrated in \fig{rho_ell_spacetime}.
We are interested in the transverse recoil of the heavy hadron
against the other fragmentation products $X$ radiated into the final state,
which is equivalent to measuring the total transverse momentum $p_{X\perp}^\mu$
in the so-called ``hadron frame''~\cite{Collins:2011zzd}
where the hadron is at rest in the transverse plane, $P_{H\perp}^\mu = 0$.
By momentum conservation, we then have $p_{X\perp}^\mu = k_\perp^\mu$
with $k_\perp^\mu$ the transverse momentum injected into the quark field in the initial state,
which in turn is Fourier conjugate to a transverse spatial separation $b_\perp^\mu$
between quark fields in position space.
We also define the magnitudes $k_T \equiv \sqrt{-k_\perp^2}$
and $b_T \equiv \sqrt{-b_\perp^2}$.
In this section, we will mainly use
the position-space formulation of the relevant fragmentation correlators
since in this case, total fragmentation probabilities
(i.e., total integrals over $k_\perp^\mu$, as well as higher moments)
can easily be obtained from the limit $b_\perp^\mu \to 0$.
Parametrically, in this paper we are interested in the regime
where the recoil is small compared to the quark mass,
$1/b_T \sim k_T \ll m$,
and generically of the order of the QCD confinement scale,
$1/b_T \sim k_T \gtrsim \lqcd$,
and thus typically a nonperturbative effect
of the genuine hadronization dynamics.

Heavy-quark spin symmetry is the statement that up to corrections of $\ord{1/m}$,
the spin $\Sigma_Q^\mu$ of the heavy quark is a conserved quantity
since it does not couple to the light dynamics.%
\footnote{
In this paper, as in \refcite{vonKuk:2023jfd}, we use $\Sigma^\mu$
for the rest-frame spin operators
to distinguish them from Bloch vectors $S^\mu$,
and also write them in relativistically covariant form with $v \cdot \Sigma = 0$
for ease of connecting them with relativistic TMD FFs.
}
In our case, this means that the heavy quark remains in the spin state
prepared by the hard scattering at a high scale $\mu \sim E_\mathrm{hard} \gg m$.
Note that it is crucial here
that by our parametric assumption $k_T \ll m$,
radiation into the final state at the scale $\mu \sim m$ is power suppressed,
while virtual corrections at this scale are diagonal in spin
and can be absorbed into a scalar Wilson coefficient $C_m = 1 + \ord{\as}$~\cite{Hoang:2015vua, Hoang:2019fze, vonKuk:2023jfd}.
Thus the heavy-quark spin is also conserved
by the dynamics around the scale $\mu \sim m$.
We refer to \refcite{vonKuk:2024uxe} for the case where $1/b_T \sim k_T \sim m$
and perturbative radiation into the final state is allowed (and calculable).

The spin state of the heavy hadron in the final state lives in the product
$\cH_H = \cH_Q \otimes \cH_\ell$
of the heavy-quark and light spin Hilbert spaces,
on which the heavy-quark spin $\Sigma_Q^\mu$ and the dynamically generated
total angular momentum $\Sigma_\ell^\mu$ of the light constituents act,
with $\Sigma_Q^2 = - \tfrac{3}{4} \, \id_Q$
and $\Sigma_\ell^2 = -s_\ell (s_\ell + 1) \, \id_\ell$.
The total spin $\Sigma_H^\mu$ of the heavy hadron
is then simply given by their sum,
\begin{align} \label{eq:angular_momentum_addition}
\Sigma_H^\mu = \Sigma_Q^\mu \otimes \id_\ell + \id_Q \otimes \Sigma_\ell^\mu
\,.\end{align}
Written in position space and decomposing it in terms
of the usual eigenvectors $\ket{h_\ell}$
of the spin operator $\Sigma_{\ell,e} \equiv \vec{\Sigma} \cdot \vec{e} = - e \cdot \Sigma_\ell$ projected onto a fixed unit direction $e^2 = -1$,
$-s_\ell \leq h_\ell \leq s_\ell$,
the definition of the light spin-density matrix is~\cite{vonKuk:2023jfd}:%
\footnote{We note that as in \refcite{vonKuk:2023jfd},
and as commonly done in the literature on polarized fragmentation functions,
we keep implicit that the states $X$ in the amplitude and conjugate amplitude
are identical only up to a difference in total helicity
$h_X' = - h_H' \neq h_X = - h_H$ for offdiagonal terms,
as required by angular momentum conservation
since the source term is a spin scalar.
In \eq{def_rho_ell} we have also suppressed
the dependence on the rapidity and dimensional regulators
as well as soft subtraction terms,
none of which affect the spin structure~\cite{vonKuk:2024uxe}.
(For this reason we can also ignore the possible presence
of evanescent spin structures in $d \neq 4$ spacetime dimensions.)
}
\begin{align} \label{eq:def_rho_ell}
\rho_{\ell, h_{\ell}'h_{\ell}}
&\equiv
\frac{1}{N_{c}} \Tr
\SumInt_{X}
\Mae{0}{W_{\bn}^\dagger(b_\perp) \,
   Y_v(b_\perp) \Ket{s_\ell, h_\ell, f_\ell ; X}
\nn \\
& \quad \times
\Bra{s_\ell, h_\ell', f_\ell ; X} Y^\dagger_v(0) \,
W_{\bn}(0)}{0}
\,.\end{align}
Note the indexing indicates that the LHS is the matrix element $\mae{h_{\ell}'}{\rho_{\ell}}{h_{\ell}}$,
which reverses the index ordering on the RHS.
Here $N_c$ is the number of colors, $\Tr$ indicates a trace over color indices,
and $Y_v$ is a timelike Wilson line in the fundamental representation
that encodes the static interaction with the heavy quark
travelling along the direction
\begin{align} \label{eq:def_v}
v^\mu \equiv P_H^\mu/M_H
\,.\end{align}
Finally, $W_{\bn}$ denotes a Wilson line along a lightlike direction $\bn$
and captures the interactions of the fragmenting system
with the remainder of the hard scattering process,
which appears to the boosted fragmenting quark
as a single lightlike static source with antifundamental color charge.
The interactions with $W_{\bn}$, which are required by gauge invariance,
distinguish a fragmentation axis
\begin{align} \label{eq:def_z}
z^\mu = v^\mu - \frac{\bn^\mu}{\bn \cdot v}
\end{align}
in the heavy-quark rest frame, $v \cdot z = 0$.
The physics encoded in $\rho_\ell$ are illustrated
at the amplitude level in \fig{rho_ell_spacetime}.
We summarize the results in \refcite{vonKuk:2023jfd} relevant for the derivation of \eq{def_rho_ell} in \app{corr_spin_density}.

\subsection{Symmetry analysis}
\label{sec:symmetry_analysis}

The light spin density matrix in \eq{def_rho_ell}
depends on the three orthogonal four-vectors
\begin{align} \label{eq:def_x}
v^\mu
\,, \quad
z^\mu
\,, \quad
b_\perp^\mu \equiv b_T \, x^\mu
\,,\end{align}
where $v^2 = -z^2 = -x^2 = 1$.
The first two vectors define the transverse plane,
\begin{align} \label{eq:def_g_perp_eps_perp_v_z}
g_\perp^{\mu\nu} = g^{\mu\nu} - v^\mu v^\nu + z^\mu z^\nu
\,, \qquad
\eps_\perp^{\mu\nu} = \eps^{\mu\nu\rho\sigma} v_\rho z_\sigma
\,,\end{align}
and all three together define a fourth (unit) direction $y^\mu$,
\begin{align} \label{eq:def_y}
y^\mu =  \eps_\perp^{\nu\mu} x_\nu
\,.\end{align}
We will also occasionally make use of the Galilean metric in the heavy-hadron rest frame,
\begin{align}
\bg^{\mu\nu} \equiv g^{\mu\nu} - v^\mu v^\nu
\,.\end{align}
The triad $\hat x, \hat y, \hat z$ is right-handed.
Since $v \cdot \Sigma_\ell = 0$ for all spin operators
that could serve as building blocks for $\rho_\ell$, it is sufficient
to focus on the dependence on $z^\mu$ and $b_\perp^\mu$,
which amounts to carrying out the analysis in the heavy-quark rest frame.
There are three applicable symmetries constraining
the structure of the density matrix $\rho_\ell$.
First, from its definition in \eq{def_rho_ell} it is easy to verify
that it is hermitian up to a sign change of $b_\perp$,
\begin{align} \label{eq:hermiticity}
\rho_\ell^\dagger(z, b_\perp)
= \rho_\ell(z, -b_\perp)
\,,\end{align}
where we used momentum conservation
to translate the field operators
in the conjugate Wilson-line correlator by $-b_\perp$
and recover the form of the original definition.
Second, QCD is invariant under parity, and thus
\begin{align} \label{eq:parity}
\rho_\ell(z, b_\perp)
= \rho_\ell(-z, -b_\perp)
\,.\end{align}
Here we are thinking of an active parity transformation
that acts on the four-vectors $z^\mu$ and $b_\perp^\mu$
describing the correlator
with respect to a fixed set of external reference coordinates.
Note that under this transformation, the states $\ket{h_\ell}$
in the definition in \eq{def_rho_ell}
map onto themselves, $\ket{+ h_\ell} \mapsto \ket{+ h_\ell}$,
because they are labelled by eigenvalues
of a component of an axial vector operator
and because the magnetization axis $e$
is chosen by the observer and thus held fixed
when transforming the system.
We note that time reversal is not a good symmetry
of fragmentation correlators due to the presence of the out states~\cite{Mulders:1995dh}.
Similarly, charge conjugation relates density matrices
for different final states $\ell$,
but provides no additional constraint on the density matrix at hand,
making \eqs{hermiticity}{parity} the only relevant discrete symmetries.
As a third, final constraint, the density matrix has to be covariant
under rest-frame rotations $R_\alpha$ generated
by the light spin operator $\Sigma^\mu_\ell$, i.e.,
\begin{align} \label{eq:covariance_under_rotations}
U_\alpha \, \rho_\ell(z, b_\perp) \, U_\alpha^\dagger
= \rho_\ell \bigl( R_\alpha(z), R_\alpha(b_\perp) \bigr)
\,,\end{align}
where $\alpha^\mu$ with $\alpha \cdot v = 0$ is the rotation axis and angle,
$R_\alpha$ denotes the (active) rotation, and
\begin{align} \label{eq:rotation_spin_rep}
U_\alpha = \exp \bigl( - \img \alpha \cdot \Sigma_\ell \Bigr)
\,.\end{align}
Our goal in this subsection is to decompose $\rho_\ell$
in terms of a basis of spin operators $\{ \cO_\ell \}$ on $\cH_\ell$
that are built out of covariant combinations
of $\Sigma_\ell^\mu$, $x^\mu$, $y^\mu$, and $z^\mu$,
and are individually consistent with the discrete symmetries.
It then follows that their coefficients are scalar structure functions
that can only depend on the magnitude $b_T$,
so counting the operators available for a given $s_\ell$
will answer the question in the title.

To start, consider the following set
of tensor spin operators of rank $N \leq 2 s_\ell$
\begin{align} \label{eq:def_cQ}
\cQ_\ell^{\mu_1 \cdots \mu_N}
\equiv \Sigma_\ell^{\{ \mu_1} \cdots \Sigma_\ell^{\mu_N \}}
\,,\end{align}
where the superscript curly braces denote total symmetrization
including a normalization factor of $1/N!$ as usual,
e.g.\ $\cQ_\ell^{\mu_1 \mu_2} = \tfrac{1}{2!} \bigl( \Sigma_\ell^{\mu_1} \Sigma_\ell^{\mu_2} + \Sigma_\ell^{\mu_2} \Sigma_\ell^{\mu_1} \bigr)$.
Here we have used that the antisymmetric parts of tensor operators at rank $N$
can be reduced to operators of rank $N - 1$ using the Lie algebra.
Operators that implement the covariance under rotation
are then easily obtained by contracting $\cQ_\ell^{\mu_1 \dots \mu_N}$
with strings of $k$ instances of $x^\mu$, $n - k$ instances of $y^\mu$,
and $N - n$ instances of $z^\mu$,
\begin{align} \label{eq:cQ_contracted_with_general_strings}
\cQ_\ell^{\mu_1 \cdots \mu_N} \,
x_{\mu_1} \cdots x_{\mu_{k}} \,
y_{\mu_{k + 1}} \cdots y_{\mu_{n}} \,
z_{\mu_{n + 1}} \cdots z_{\mu_N}
\,.\end{align}
These operators, however, are not yet linearly independent.
The conventional approach at this point would
be to construct irreducible spherical tensor operators~\cite{Sakurai:2011zz},
which however are not particularly useful in our case
because of the explicit breaking of rotational symmetry by $z^\mu$ and $b_\perp^\mu$.
Instead, as a first attempt at constructing the desired basis,
we may simply eliminate all operators involving repeated contractions
with $y^\mu$ using $(y \cdot \Sigma_\ell)^2 = s_\ell(s_\ell + 1) \id_\ell - (x \cdot \Sigma_\ell)^2 - (z \cdot \Sigma_\ell)^2$.
This leaves only the strings with $n - k = 0$ or $1$,
\begin{align} \label{eq:def_tilde_T}
\tilde{T}_{\mu_1 \dots \mu_N}^{(N,n,+)}(x, z)
&\equiv
x_{\mu_1} \cdots x_{\mu_{n}} \,
z_{\mu_{n + 1}} \cdots z_{\mu_N}
\,,\\[0.4em]
\tilde{T}_{\mu_1 \dots \mu_N}^{(N,n,-)}(x, z)
&\equiv
x_{\mu_1} \cdots x_{\mu_{n - 1}} x^\nu \,
\eps_{\perp\,\nu\mu_n}
z_{\mu_{n + 1}} \cdots z_{\mu_N}
\nn
\,.\end{align}
At each rank $N \leq 2s_\ell$ this leaves $2N + 1$ strings.
Since $\sum_{N=0}^{2s_\ell} (2N+1) = (2s_\ell+1)^2$,
their contractions with $\cQ_\ell^{\mu_1 \cdots \mu_N}$
span the full space of operators on $\cH_\ell \simeq \mathbb{C}^{2s_\ell+1}$.

We next apply the constraint from parity, \eq{parity}.
For the operators with $n - k = 0$ (no $y^\mu$),
this simply restricts $N$ to be even.
On the other hand, taking $x^\mu \mapsto - x^\mu$ and $z^\mu \mapsto -z^\mu$
leaves $y^\mu$ invariant, so for $n - k = 1$ (single insertion of $y^\mu$),
$N$ must be odd.
Including the identity,
we are thus left with a basis of hermitian operators
\begin{align} \label{eq:def_tilde_cO}
\tilde{\cO}_\ell^{(0,0)}
&\equiv \id_\ell
\,, \\
\tilde{\cO}_\ell^{(N,n)}
&\equiv \cQ_\ell^{\mu_1 \cdots \mu_N}
\begin{cases}
\tilde{T}_{\mu_1 \cdots \mu_N}^{(N,n,+)}
\,, \quad
N \in 2 \mathbb{N}
\,, \quad
0 \leq n \leq N
\,, \\[0.4em]
\tilde{T}_{\mu_1 \cdots \mu_N}^{(N,n,-)}
\,, \quad
N \notin 2 \mathbb{N}
\,, \quad
1 \leq n \leq N
\,,\end{cases}
\nn \end{align}
where the $+$ ($-$) case should be picked for $N$ even (odd).
These operators span the space of physically allowed
TMD light spin density matrices,
\begin{align} \label{eq:result_rho_ell_tilde_cO}
\rho_\ell(v, z, b_\perp)
= \sum_{N,n} \tilde{\chi}_{1,\ell}^{(N,n)}(b_T) \,
(-\img)^n \tilde{\cO}_\ell^{(N,n)}\Bigl(v, z, \frac{b_\perp}{b_T}\Bigr)
\,.\end{align}
Here we included a factor of $(-\img)^n$ in \eq{result_rho_ell_tilde_cO}
to ensure that the coefficients $\tilde{\chi}_{1,\ell}^{(N,n)}$
are real functions of the magnitude $b_T$
according to the constraint from hermiticity, \eq{hermiticity}.
The tildes over the expressions in \eqs{def_tilde_cO}{result_rho_ell_tilde_cO} indicate
that these are not yet our final basis, which we develop in \sec{orthogonality_and_final_operator_basis}.

However, already at this point we are ready to count out the number
of linearly independent, physically allowed operators
---\,and thus scalar structure functions\,---
present for any given value of $s_\ell$.
The sum in \eq{result_rho_ell_tilde_cO} runs over all values $N \leq 2 s_\ell$ and $n \leq N$,
subject to the constraint that only $n \geq 1$ contributes for odd $N$.
At each $N$, we thus have $2\lfloor N/2\rfloor + 1$ new operators ($N+1$ for even $N$ and $N$ for
odd $N$), resulting in a total of
\begin{align} \label{eq:answer_to_question_in_title}
\Bigl\lceil \frac{(2s_\ell + 1)^2}{2} \Bigr\rceil
= 1
\,, \,
2
\,, \,
5
\,, \,
8 \dots
\,, \,
\end{align}
TMD structure functions describing
the fragmentation of heavy quarks
to heavy hadrons with light spin $
s_\ell = 0
\,, \,
\tfrac{1}{2}
\,, \,
1
\,, \,
\tfrac{3}{2}
\,, \,
\dots$
at leading order in $1/m$.
This number of independent TMD structure functions in bHQET is compared
to the number of leading-power TMD FFs for various relevant
hadron multiplets in \tab{number_structure_functions}.%
\begin{table}
\caption{\label{tab:number_structure_functions}%
The number of independent structure functions characterizing
bottom quark fragmentation to various heavy hadrons
at leading power in $1/m$, compared to the respective number
of leading-power TMD FFs in the full theory for each spin symmetry multiplet.%
The hadron spins $s_H$ and the corresponding numbers $2(2s_H+1)^2$ of LP TMD FFs
are listed in the same order as the hadrons in each multiplet.
Equivalent results hold for charm quarks.
We do not separately count the different structure functions
for hadrons of identical spin, but different light flavor content.}
\vspace{0.5em}
\begin{ruledtabular}
\begin{tabular}{ccccc}
Hadron multiplet & $s_H$ & $s_\ell$ &LP TMD FFs & $\chi_{1,\ell}^{(N,n)}$
\\
\colrule
$\Lambda_b$ & $1/2$ & $0$ & $8$ & $1$
\\
$B, B^*$ & $0, 1$ & $1/2$ & $2 + 18$ & $2$
\\
$\Sigma_b, \Sigma_b^*$ & $1/2, 3/2$ & $1$ & $8 + 32$ & $5$
\\
$B_1, B_2^*$ & $1, 2$ & $3/2$ & $18 + 50$ & $8$
\end{tabular}
\end{ruledtabular}
\end{table}
For the lightest $s_H = 0$ pseudoscalar mesons with $s_\ell = 1/2$,
the above result precisely coincides with the two structure functions
relevant for general unpolarized hadrons (but possibly polarized quarks)
that we identified in \refcite{vonKuk:2023jfd}.
For $s_H > 0$ and $s_\ell > 1/2$, the latter two form a subset
of the structure functions that are accessible
when general hadron and quark polarization are considered.

\begin{figure}
\noindent\hspace*{-3mm}\includegraphics[width=0.5\textwidth]{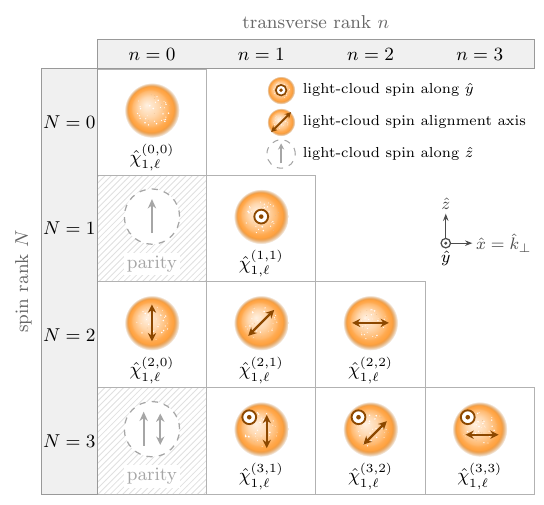}
\caption{
\label{fig:sell_triangle}
The independent light structure functions $\hat\chi^{(N,n)}_{1,\ell}$, 
organized by the spin rank $N$ of the light-spin operator and its transverse rank $n$, 
i.e., the number of its indices carried by the transverse direction $\hat x = \hat k_\perp$ (or $\hat b_\perp$).
Each cloud depicts the corresponding spin structure of the light degrees of freedom 
in the plane spanned by $\hat z$ and $\hat x$, with $\hat y$ pointing out of the page.
$\odot$ denotes light-cloud spin along $\hat y$, 
and double-headed arrows denote an alignment axis:
along which the light spin is more (or less) likely to align, rather than perpendicular to it.
Odd-rank structures always carry exactly one $\hat y$ on top of the corresponding even-rank structure, 
cf.\ \eq{transverse_tensor_minus}, 
while the purely longitudinal odd-rank structures along $\hat z$ are forbidden by parity.
A heavy-hadron multiplet with light spin $s_\ell$ involves all rows $N \le 2s_\ell$, 
giving $\lceil (2s_\ell+1)^2/2 \rceil = 1, 2, 5, 8$ structure functions for $s_\ell = 0, \tfrac12, 1, \tfrac32$, 
cf.\ \tab{number_structure_functions}.
The symbols for $N = 3$ are schematic and indicate the index structure rather than the full octupole.}
\end{figure}

\subsection{Orthogonality and final operator basis}
\label{sec:orthogonality_and_final_operator_basis}

The basis we developed above is still somewhat cumbersome
to work with, for two related reasons.
First, they are not orthogonal to each other,
making it less transparent to obtain the $\chi_{1,\ell}^{(N,n)}$
as suitable expectation values measured on the density matrix.
Second, the Fourier transform of the string $x \cdots x$
with respect to $x^\mu = b_\perp/b_T$ is a hassle to compute.
To remedy both of these, we use \emph{harmonic tensors}
(also known as Cartesian tensors)
that transform according to irreducible representations of the relevant rotation groups
and are defined by requiring that they are ``trace-free'',
i.e., that all their partial pairwise traces vanish:
For a given fully symmetric tensor $T^{\mu_1 \dots \mu_k}$ of rank $k$
and a given non-degenerate metric $\eta^{\mu\nu}$ in $\eta^{\mu\nu} \eta_{\nu\mu} = d$ dimensions,
the harmonic tensor $T^{\langle\mu_1 \dots \mu_k\rangle}$, indicated by angular brackets around the indices,
is defined as the unique tensor
that is linear in $T^{\mu_1 \dots \mu_k}$ with unit coefficient and satisfies
\begin{align}
\label{eq:traceless_tensor}
\eta_{\mu_i \mu_j} \, T^{\langle\mu_1 \dots \mu_k\rangle} = 0
\quad \text{for all} \quad
1 \leq i, j \leq k
\,.\end{align}
Its general closed-form expression can be obtained by induction~\cite{Toth:2021cpx},
where one iteratively subtracts, e.g.,
\begin{align}
T^{\langle \mu\nu \rangle}
= T^{\mu\nu} - \frac{1}{d} \eta_{\rho \sigma} T^{\rho\sigma} \, \eta^{\mu\nu}
\,.\end{align}

We first deal with rotations in the transverse plane,
$d = 2$ and $\eta^{\mu\nu} = g_\perp^{\mu\nu}$,
where we require the trace-free counterpart
\begin{align}
\label{eq:transverse_tensor_plus}
T^{(n,+)}_{\perp,\mu_1 \dots \mu_n}(x) \equiv x_{\langle \mu_1} \cdots x_{\mu_n \rangle}
\,,\end{align}
of the string $x \cdots x$ in \eq{def_tilde_T}.
Note that $T_\perp^{\mu_1 \cdots \mu_n}$ also depends on $v^\mu$ and $z^\mu$ through $g_\perp^{\mu\nu}$,
which we leave implicit.
We also leave implicit that for tensors with purely perpendicular indices, like $T_\perp$,
the trace-free property indicated by the angle brackets will always hold with respect to $g_\perp^{\mu\nu}$.
In addition, to set up our operators at odd rank $N$,
we need the transverse tensor obtained by performing the same operation
on the string $x \cdots x \, y$,
\begin{align}
\label{eq:transverse_tensor_minus}
T^{(n,-)}_{\perp,\mu_1 \dots \mu_n}(x)
&\equiv - x_{\langle \{ \mu_1} \cdots x_{\mu_{n - 1}} y_{\mu_n \}\rangle}
\,.\end{align}
These tensors transform according to the highest irreducible representation of $\mathrm{SO}(2)$
available at the respective rank $n$.

To encode longitudinal polarization of the light spin state,
we next have to append a string $z \cdots z$ as in \eq{def_tilde_T},
but again need to make sure that partial traces are subtracted,
in this case with respect to the Galilean metric $\bg^{\mu\nu}$ in $d = 3$.
The full analogue of \eq{def_tilde_T} is then given by
\begin{align} \label{eq:def_T}
T_{\mu_1 \dots \mu_N}^{(N,n,\pm)}(x, z)
&\equiv
T^{(n,\pm)}_{\perp, \langle\{ \mu_1 \cdots \mu_n}(x) \,
z_{\mu_{n + 1}} \cdots z_{\mu_N \}\rangle }
\,.\end{align}
We note that we could have chosen to directly subtract
traces with respect to $\bg^{\mu\nu}$ on the full string $x \cdots x \, (y) \, z \cdots z$,
which would reflect the fact
that $x^\mu$ and $z^\mu$ enter the density matrix in nearly identical ways
(i.e., as four vectors it must have covariant dependence on).
However, we find that the above, sequential construction
makes the eventual Fourier transform with respect to $x^\mu = b_\perp^\mu/b_T$
more transparent
and more closely aligns with common conventions for transverse (tensor) polarization,
where one typically uses a basis containing entries such as $\Sigma_x^2 - \Sigma_y^2$.

Contracting the above tensors with the rank-$N$ spin operators,
our final operator basis reads
\begin{align} \label{eq:def_cO}
\cO_\ell^{(0,0)}
&\equiv \id_\ell
\,, \nn \\
\cO_\ell^{(N,n)}(x, z)
&= \cQ_\ell^{\mu_1 \cdots \mu_N}
T_{\mu_1 \cdots \mu_N}^{(N,n,\pm)}(x, z)
\,,\end{align}
where as in \eq{def_tilde_cO},
$+$ ($-$) should be picked for $N$ even (odd),
and the operators are defined for $0 \leq n \leq N$ ($1 \leq n \leq N$) for $N$ even (odd).
These have the same good properties as the $\tilde{\cO}_\ell^{(N,n)}$
in \eq{def_tilde_cO} with regards to covariance, hermiticity,
and parity,
and there are exactly as many of them as before,
leading again to \tab{number_structure_functions}.
In addition, however,
they are now orthogonal with respect to the operator trace,
\begin{align} \label{eq:orthogonality}
\Tr_\ell \Bigl[ \cO_\ell^{(N', n')} \cO_\ell^{(N, n)} \Bigr]
= \delta_{NN'} \delta_{nn'} \, C_{\ell}^{(N,n)}
\,,\end{align}
where $C_\ell^{(N,n)}$ is a normalization factor that depends on $N$, $n$, and $s_\ell$.
We refer to \app{rhoell} for details
and for closed-form expressions suitable for numerical implementation.

In terms of this operator basis,
we arrive at the following, final result
for the decomposed light-spin density matrix in position space,
\begin{align} \label{eq:result_rho_ell_cO}
\rho_\ell(v, z, b_\perp)
= \sum_{N,n} \frac{\chi_{1,\ell}^{(N,n)}(b_T)}{C_\ell^{(N,n)}} \,
(-\img)^n \cO_\ell^{(N,n)}\Bigl(v, z, \frac{b_\perp}{b_T}\Bigr)
\,,\end{align}
where we have included a factor of $1/C^{(N,n)}$ for convenience
and the factor of $(-\img)^n$ ensures hermiticity, as in \eq{result_rho_ell_tilde_cO}.
This in particular implies
that $\Tr_\ell \bigl[ \rho_\ell \bigr] = \chi_{1,\ell}^{(0,0)}$
and, more generally,
\begin{align}
\chi_{1,\ell}^{(N,n)}
= (+\img)^n \Tr_\ell \bigl[
\cO_{\ell}^{(N,n)} \, \rho_\ell \bigr]
\,.\end{align}

To be explicit, the density matrices for the lowest few $s_\ell$ read
\begin{align}
\label{eq:rho_ell_explicit_pos_space}
s_\ell = 0 \, : ~
&\rho_\ell
= \chi_{1,\ell}^{(0,0)}
\,, \\
s_\ell = \tfrac{1}{2} \, : ~
&\rho_\ell
= \frac{\id}{2} \, \chi_{1,\ell}^{(0,0)}
+ 2\img \Sigma_{y} \, \chi_{1,\ell}^{(1,1)}
\,, \nn \\
s_\ell = 1 \, : ~
&\rho_\ell
= \frac{\id}{3} \, \chi_{1,\ell}^{(0,0)}
+ \tfrac{\img}{2} \Sigma_{y} \, \chi_{1,\ell}^{(1,1)}
+ \bigl( \tfrac{3}{2} \Sigma_{z}^2 - \id \bigr) \chi_{1,\ell}^{(2,0)}
\nn \\[0.2em] & \qquad
- \img \bigl( \Sigma_x \Sigma_z + \Sigma_z \Sigma_x \bigr) \chi_{1,\ell}^{(2,1)}
- \bigl( \Sigma_x^2 - \Sigma_y^2 \bigr) \chi_{1,\ell}^{(2,2)}
\nn \,,\end{align}
where we have suppressed all arguments
and the subscript $\ell$ on the operators
on the right-hand side for brevity.
Note that the structure functions for different $s_\ell$ are distinct
and depend on $s_\ell$ (as well as the light flavor $f_\ell$)
through the label $\ell$.

\subsection{Fourier transform}

The operator basis above has a further advantage
when computing the Fourier transform of the spin-density matrix,
\begin{align}
\hat{\rho}_\ell(v,z,k_\perp)
= \int \! \frac{\df^2 b_\perp}{(2\pi)^2} \,
e^{+\img k_\perp \cdot b_\perp} \, \rho_\ell(v, z, b_\perp)
\,,\end{align}
as required for evaluating momentum-space TMD FFs.
(We denote the momentum-space objects
by an additional hat.)
This is made straightforward by the fact that for any scalar test function $f(b_T)$,
\begin{align} \label{eq:fourier_transform_T}
&\int \! \frac{\df^2 b_\perp}{(2\pi)^2} \, e^{+ \img k_\perp \cdot b_\perp} \,
T_\perp^{(n,\pm) \, \mu_1 \cdots \mu_n}\Bigl( \frac{b_\perp}{b_T} \Bigr) \, f(b_T)
\\
&= (-\img)^n \, T_\perp^{(n, \pm) \, \mu_1 \cdots \mu_n}\Bigl( \frac{k_\perp}{k_T} \Bigr)
\int_0^\infty \! \frac{\df b_T \, b_T}{2\pi} \, J_n(b_T k_T) \, f(b_T)
\nn \,,\end{align}
i.e., the operator basis is self-similar under Fourier transform.
To easily see that \eq{fourier_transform_T} holds, note that it must be possible
in general to write the result of the integral as a sum of dimensionless tensors that only depend on $k_\perp/k_T$
times scalar coefficient functions of $k_T$ involving $f(b_T)$, but no vectorial information.
However, the fact that both sides of the equation must be trace free
implies that the unique term on the right-hand side is $T_\perp^{(n, \pm) \, \mu_1 \cdots \mu_n}(k_\perp/k_T)$.
Its scalar coefficient function is then easily found to be the $n$\textsuperscript{th} Hankel transform of $f$
by contracting both sides of the equation
with $T_\perp^{(n, \pm) \, \mu_1 \cdots \mu_n}(k_\perp/k_T)$ itself
and taking successive derivatives of the $n = 0$ base case with respect to $b_\perp$ to pull down the required powers of $k_\perp$.

Thanks to \eq{fourier_transform_T}, we can immediately write down
the momentum-space spin-density matrix as
\begin{align}
\hat{\rho}_\ell(v, z, k_\perp)
= \sum_{N,n} \frac{\hat\chi_{1,\ell}^{(N,n)}(k_T)}{C_\ell^{(N,n)}} \,
\cO_\ell^{(N,n)}\Bigl(v, z, \frac{k_\perp}{k_T}\Bigr)
\,,\end{align}
where
\begin{align}
\label{eq:FT_chi}
\hat \chi_{1,\ell}^{(N,n)}(k_T)
&\equiv (-1)^{n} \!\! \int_0^\infty \! \frac{\df b_T \, b_T}{2\pi}
\chi_{1,\ell}^{(N,n)}(b_T) \, J_n(b_T k_T)
\nn \\
&= \Tr_\ell \bigl[
\cO_{\ell}^{(N,n)} \, \hat\rho_\ell \bigr]
\,.\end{align}
For the lowest few spin multiplets we have
the following explicit expressions in momentum space:
\begin{align}
\label{eq:rho_ell_explicit_mom_space}
s_\ell = 0 \, : ~
&\hat\rho_\ell
= \hat \chi_{1,\ell}^{(0,0)}
\,, \\
s_\ell = \tfrac{1}{2} \, : ~
&\hat\rho_\ell
= \frac{\id}{2} \, \hat \chi_{1,\ell}^{(0,0)}
- 2\Sigma_{y} \, \hat \chi_{1,\ell}^{(1,1)}
\,, \nn \\
s_\ell = 1 \, : ~
&\hat\rho_\ell
= \frac{\id}{3} \, \hat \chi_{1,\ell}^{(0,0)}
- \tfrac{1}{2} \Sigma_{y} \, \hat \chi_{1,\ell}^{(1,1)}
+ \bigl( \tfrac{3}{2} \Sigma_{z}^2 - \id \bigr) \hat\chi_{1,\ell}^{(2,0)}
\nn \\[0.2em] & \qquad
+ \bigl( \Sigma_x \Sigma_z + \Sigma_z \Sigma_x \bigr) \hat \chi_{1,\ell}^{(2,1)}
+ \bigl( \Sigma_x^2 - \Sigma_y^2 \bigr) \hat \chi_{1,\ell}^{(2,2)}
\nn \,,\end{align}
where arguments $v$, $z$, $k_\perp$, or $k_T$ are understood.

\subsection{Inclusive heavy-quark fragmentation}
\label{sec:inclusive_frag}

The limit $b_\perp \to 0$ of the light spin-density matrix is of particular interest, 
since it connects our analysis to the well-studied case of inclusive heavy-quark fragmentation \cite{Mele:1990cw, Jaffe:1993ie, Neubert:2007je, Fickinger:2016rfd, Bonino:2023icn, Falk:1993rf}.
For the bare correlator in \eq{def_rho_ell}, 
$\rho_\ell(b_\perp = 0)$ reduces to the inclusive light-constituent matrix element 
that defines the collinear heavy-quark fragmentation coefficients~\cite{Falk:1993rf}.
With no transverse direction left, all structures with $n \geq 1$ drop out, 
and by parity only the even-rank, purely longitudinal structures survive,
\begin{equation} 
\label{eq:rho_ell_inclusive}
\rho_\ell(b_\perp = 0) = \int \! \df^2 k_\perp \, \hat{\rho}_\ell(k_\perp) = \sum_{N \,\text{even}} \frac{\chi^{(N,0)}_{1,\ell}(0)}{C^{(N,0)}_\ell}\, O^{(N,0)}_\ell
\,,\end{equation}
so that the inclusive light spin-density matrix is diagonal in $h_\ell$ along the fragmentation axis, 
depends only on $|h_\ell|$, and in particular carries no vector polarization.
The number of independent structure functions thus reduces from $\lceil (2s_\ell+1)^2/2 \rceil$ to $\lfloor s_\ell \rfloor + 1$,
illustrating how much additional information becomes accessible once the transverse momentum is resolved.

Normalized to the total fragmentation probability, 
the surviving nonperturbative constants $\chi_{1,\ell}^{(N,0)}(b_T = 0)$
are in one-to-one correspondence with the Falk--Peskin parameters $w_{s_\ell}$~\cite{Falk:1993rf},
which give the probability that the light constituents carry maximal helicity $|h_\ell| = s_\ell$ along the fragmentation axis.
From \eq{rho_ell_explicit_pos_space} and \eqs{rho_ell_32}{rho_ell_32_terms} we find
\begin{align}
\label{eq:falk_peskin}
w_1 &= \frac{2}{3} + \frac{\chi^{(2,0)}_{1,\ell}(0)}{\chi^{(0,0)}_{1,\ell}(0)} \,, \qquad
w_{3/2} &= \frac{1}{2}\left[1 + \frac{\chi^{(2,0)}_{1,\ell}(0)}{\chi^{(0,0)}_{1,\ell}(0)}\right] ,
\end{align}
where the isotropic values $w_1 = 2/3$ and $w_{3/2} = 1/2$ correspond to vanishing alignment, $\chi^{(2,0)}_{1,\ell}(0) = 0$.
These relations hold for the transverse-momentum-integrated light spin-density matrix at bare level and at leading power in $1/m$.
In terms of the normalized structure functions in momentum space $w^{(N,n)}(k_T) \equiv \hat\chi^{(N,n)}_{1,\ell}(k_T)/\hat\chi^{(0,0)}_{1,\ell}(k_T)$ used in Ref.~\cite{vonKuk:2025kbv}, \eq{rho_ell_explicit_pos_space} reproduces the transverse-momentum-dependent generalization of the Falk--Peskin parametrization introduced there, with $w^{(2,0)} = 0$ corresponding to $w_1 = 2/3$ after integrating over $k_T$.

We stress that the diagonal structure of the inclusive hadron polarization state produced by an unpolarized parton follows from symmetry alone, independent of the heavy-quark expansion.
For an unpolarized parton, the fragmentation axis $\hat z$ is the only available direction, 
so rotational invariance about $\hat z$ implies $[\rho_H, J_z] = 0$, 
and parity combined with a rotation by $\pi$ about a transverse axis implies $\rho_{h_H h_H} = \rho_{-h_H,-h_H}$.
The inclusive polarization state of any hadron, light or heavy, 
is thus characterized by $\lfloor s_H \rfloor + 1$ structure functions,
e.g., $D_1$ and $D_{1LL}$ for a vector meson, as long as the parent parton is unpolarized.
Longitudinal parton polarization then breaks the $h_H \to -h_H$ symmetry,
and transverse polarization or a measured $k_\perp$ generate off-diagonal structures.
Heavy-quark spin symmetry adds that the inclusive parameters of all members of a multiplet 
are fixed by the same $\chi^{(N,0)}_{1,\ell}(0)$, 
implying, e.g., $D_{1LL}^{H^*} = 0$ (see \sec{example_sellhalf} for details), which symmetry alone does not enforce.

It is straightforward to extend the above analysis to subleading terms in $\lqcd$
if we \emph{assume} that $\rho_\ell(b_\perp)$ at bare level is not just continuous as $b_\perp \to 0$,
as is apparent from its matrix-element definition,
but infinitely differentiable at the origin.
In this case, for $(\partial/\partial b_\perp^\mu)^n \rho_\ell(b_\perp)$ to be continuous,
the nonperturbative onset of each structure function at the bare level
must precisely be governed by its transverse polarization rank $n$,
\begin{align} \label{eq:small_b_ope_bare}
\chi^{(N,n)}_{1,\ell}(b_T) = \Ord{ b_T^{\,n} \lqcd^n}
\,.\end{align}
Beyond the bare correlator, the renormalized structure functions have no finite limit as $b_T \to 0$, since radiation off $Y_v$ and $W_{\bar n}$ generates logarithms of $b_T$, 
which are captured by nontrivial Wilson coefficients $C_{1}^{(N,n,k)}$,
\begin{align}
\label{eq:small_b_ope}
\chi^{(N,n)}_{1,\ell}(b_T, \mu, \nu) &= \sum_k C_1^{(N,n,k)}(b_T, \mu, \nu) \; b_T^{\,n}\; \lambda^{(N,n,k)}_{\ell}(\mu)
\nn \\
&\quad \times \Big[1 + \mathcal{O}\big(b_T\Lambda_\mathrm{QCD}\big)\Big] \,.
\,,\end{align}
Here the $\lambda^{(N,n,k)}_\ell \sim \lqcd^n$ are bHQET matrix elements
that for $n > 0$ involve additional soft field insertions on top of the Wilson line structure present at $b_\perp = 0$;
we refer to \refcite{vonKuk:2023jfd} for a worked example where the heavy-quark Collins TMD FF was found to be governed by a subleading bHQET matrix element with $n = 1$.
The $C_1^{(N,n,k)}$ account for the overall rapidity and virtuality evolution of the bHQET TMD matrix elements~\cite{vonKuk:2024uxe},
while there will in general be RGE evolution and operator mixing between the scales $\mu \sim 1/b_T$ and $\mu \sim \lqcd$.
Note that we base our expectation for the small-$b_T$ scaling of the $\chi_{1,\ell}^{(N,n)}$
on the bare (or classical, or tree-level) result in \eq{small_b_ope_bare},
which should at most be modified by anomalous powers of $\lqcd \; b_T$ through the RGE.
In particular, we expect that exactly one tree-level term with $k = 0$ and $C_1^{(N,n,0)} = 1 + \ord{\as}$
is present at each power, while other subleading operators with $k > 0$ have Wilson coefficients suppressed by $\as(1/b_T)$.
The unique leading-power Wilson coefficient $C_1(b_T, \mu, \nu) = C_1^{(N,0,0)}$ was calculated to NLO in \refcite{vonKuk:2024uxe}, is independent of $N$,
and simply multiplies the respective combination of Falk-Peskin parameters that do not run.
We anticipate that there will be a large set of relations between the $\lambda^{(N,n,k)}_{\ell}(\mu)$, as well,
since the bHQET operator basis is finite at each order in $\lqcd$,
but leave a detailed analysis of the OPE to future work.

As an application of \eq{small_b_ope}, we consider again the momentum-space density matrix.
Since polynomial terms in $b_\perp$ do not contribute at $k_T > 0$, 
the large-$k_T$ tails stem from the logarithms in $C_1^{(N,n,k)}$, 
giving $\hat\chi^{(N,n)}_{1,\ell} \sim \alpha_s \Lambda_\mathrm{QCD}^n/k_T^{n+2}$ 
and thus the model-independent scaling $w^{(N,n)} = \hat\chi^{(N,n)}_{1,\ell}/\hat\chi^{(0,0)}_{1,\ell} \sim (\Lambda_\mathrm{QCD}/k_T)^n$ for $k_T \gg \Lambda_\mathrm{QCD}$,
which we already made use of in \refcite{vonKuk:2025kbv}.

\sectionPaper{From the light spin-density matrix to polarized TMD FFs}
\label{sec:light_dm_to_tmdffs}

In this section, we connect the light structure functions of \sec{light_spin_density_matrix} to the standard leading-power TMD FFs, which describe the production of a heavy hadron in a given spin state from a quark with given polarization.
Experimentally, the hadron spin state is inferred from the angular distribution of the hadron's decay products, so the decay acts as an analyzer of the hadron spin.
In many cases of interest, this decay is itself subject to heavy-quark spin symmetry.
In the strong decay $\Sigma_Q^* \to \Lambda_Q \pi$, for example, the heavy quark is a spectator at leading order in $1/m$, and the total angular momentum $s_\ell = 1$ of the light constituents is simply transferred into the pion~\cite{Isgur:1991wq, Falk:1993rf, Cho:1992gg, Yan:1992gz}.
In such cases, the decay distribution directly probes the light spin-density matrix $\rho_\ell$, and one does not need the TMD FFs as an intermediate step; see \refcite{vonKuk:2025kbv} for an example.
Nevertheless, here we give complete results in the conventional basis of TMD FFs and the operator basis for the light spin density matrix presented in \sec{light_spin_density_matrix}, to make contact with the existing literature and to demonstrate the power of the formalism.
To this end, we first describe the polarization state of the heavy hadron and the corresponding analyzer in \sec{analyzer}, and then derive the relations between the light structure functions and the TMD FFs in \sec{tmdff_lightchi_relations}.

\subsection{Polarized hadron state and analyzer}
\label{sec:analyzer}

\begin{table*}
\begin{tabular}{@{}llllll@{}}
\hline
$\Gamma$ & $\mathcal{O}_{Q}^{[\Gamma]}$ & $\cO_{H}=\id$ & $\cO_{H}=2J^x = \sigma^{x}$ & $\cO_H=2J^y = \sigma^{y}$ & $\cO_H=2J^z = \sigma^{z}$\\
\hline
$\Sl{\bn}$ & $\id_Q$ & $D_1$ & $0$ & $- iM_Hb_TD_{1T}^{\perp(1)}$ & $0$\\
$\Sl{\bn}\gamma_5$ & $2\Sigma_{Q,z}$ & $0$ & $+iM_Hb_TG_{1T}^{\perp(1)}$ & $0$ & $G_1$\\
$x_\mu\,i\sigma^{\mu\nu}\bar n_\nu\gamma_5$ & $-2\Sigma_{Q,x}$ & $0$
& $-H_1+\tfrac14(M_Hb_T)^2H_{1T}^{\perp(2)}$ & $0$
& $-iM_Hb_TH_{1L}^{\perp(1)}$\\
$y_\mu\,i\sigma^{\mu\nu}\bar n_\nu\gamma_5$ & $-2\Sigma_{Q,y}$
& $+iM_Hb_TH_1^{\perp(1)}$ & $0$
& $-H_1-\tfrac14(M_Hb_T)^2H_{1T}^{\perp(2)}$ & $0$\\
\hline
\end{tabular}
\caption
{Dirac projections and the corresponding quark operators
$\mathcal{O}^{[\Gamma]}_{Q}$ (rows) versus hadron analyzers $\mathcal{O}_H$ (columns) for spin-1/2 hadron.
The TMD FFs are written in position space here.}  
\label{tab:operators_spin1half}
\end{table*}

Similar to the density matrix of the light degree of freedom, 
the polarization state of a heavy hadron can be described by a density matrix
$\rho_{H, h_{H}h_{H}'}$, whose structure depends on the spin $s_{H}$ of the hadron.
For example,
\begin{align}
\rho_{H}^{s_{H}=0} &= \id_{1\times1}
\,,\nn \\
\rho_{H}^{s_{H}=\frac12} &= \frac 12 \Bigl( 
\id_{2\times2} + \sigma^{a} S^{a} \Bigr)
\,, \end{align}
where $\sigma^{a}$ are the Pauli matrices, repeated indices are summed, and 
$S^{a} = (S^x, S^y, S^z) = (S_{\perp}^{x}, S_{\perp}^{y}, S_{L})$ is the spin (Bloch) vector of the hadron, with $|\vec S| \le 1$.
Higher-spin hadrons may have tensor components as well \cite{Bacchetta:2000jk}:
\begin{align}
\rho_{H}^{s_{H}=1} &= \frac 13 \Bigl( 
\id_{3\times3} + \frac 32 J^{a} S^{a}
+ 3 Q^{ab} T^{ab}
\Bigr)
\,,\end{align}
where $J^{a}$ are the standard spin matrices for a spin-1 system, 
$Q^{ab} = \frac{1}{2} \{J^{a}, J^{b}\} - \frac{2}{3} \delta^{ab}$, 
and $T^{ab}$ is a symmetric traceless rank-two spin tensor 
with parameters characterizing both longitudinal and transverse spin,
\begin{align}
\label{eq:spin1_hadron_params}
T^{ab} = \frac{1}{2}
\begin{pmatrix}
  -S_{LL} + S_{TT}^{xx} & S_{TT}^{xy}           & S_{LT}^{x} \\
  S_{TT}^{xy}           & -S_{LL} - S_{TT}^{xx} & S_{LT}^{y} \\
  S_{LT}^{x}            & S_{LT}^{y}            & 2S_{LL}
\end{pmatrix}
\,,\end{align}
so that $S^{a} = \Tr[\rho_{H} J^{a}]$ and $T^{ab} = \Tr[\rho_{H} Q^{ab}]$.
Our convention differs from the one in \refcite{Bacchetta:2000jk} only by
$S_{LL} = \frac{2}{3} S_{LL}^\text{\refcite{Bacchetta:2000jk}}$.

Generally, we can write
\begin{align}
\label{eq:hadron_density_general}
\rho_{H}^{s_{H}} = \frac{1}{d} \Big( \id_{d\times d} + \sum_{\alpha} \cO_{H}^{(\alpha)}c_{\alpha} \Big)
\,,\end{align}
where $d=2s_{H} +1$.
The overall factor $1/d$ is fixed by the unit trace $\Tr[\rho_{H}] = 1$ of a density matrix.
The $\cO_{H}^{(\alpha)}$ are Hermitian, traceless, mutually orthogonal operators that span the spin space of the hadron;
we take them to be symmetric traceless products of the $J^{a}$,
exactly like the operators $\cO_{\ell}^{(N,n)}$.
$c_{\alpha}$ are the generalized polarization parameters.
Only the products $\cO_{H}^{(\alpha)}c_{\alpha}$ appear, so the relative normalization of the two
has to be fixed by convention.
One can pick normalization factors $n_{\alpha}$ for $\cO_{H}^{(\alpha)}$ 
through $\Tr[\cO_{H}^{(\alpha)} \cO_{H}^{(\beta)}] = n_{\alpha}\delta^{\alpha\beta}$,
or pick the normalization of the $c_{\alpha}$,
e.g.\ at rank 1 we usually identify the $c_{\alpha}$'s with the components of the Bloch vector $\vec S$, with $|\vec S| = 1$ for a pure state.
These two normalizations are related by
\begin{align}
c_{\alpha} = \frac{d}{n_{\alpha}} \Tr[ \rho_{H} \cO_{H}^{(\alpha)}] \,.
\end{align}

In the fragmentation process the hadron spin enters in two distinct roles, which should be distinguished.
The TMD FF correlator with open hadron helicity indices, $\Delta_{h_{H}h_{H}'}$, built from the amplitude
$\mae{H, h_{H}'; X}{\bar \psi} {0}$ and its conjugate $\mae {0}{\psi} {H, h_{H}; X}$, and summed over $X$, 
cf. \eq{def_tmd_ff_correlator},
is (up to normalization) the spin density matrix of the hadron produced by the fragmenting quark, i.e.\ a state.
The polarization parameters $c_{\alpha}$ appearing in the parametrization of the TMD FFs, however, do not describe this state.
They specify the hadron spin state selected by the \emph{measurement}, which enters through an analyzer $\cA_{H}(c)$ on the hadron Hilbert space $\cH_{H}$ traced against the produced state,
\begin{align}
\label{eq:hadron_analyzer_trace}
\Delta^{[\Gamma]}(c) = \sum_{h_{H}, h_{H}'} \big[\cA_{H}(c)\big]_{h_{H}'h_{H}}\,\Delta^{[\Gamma]}_{h_{H}h_{H}'}
= \Tr_{H}\big[\cA_{H}(c)\,\Delta^{[\Gamma]}\big]
\,.\end{align} 
For example, selecting a hadron in the pure spin state $|\chi\rangle$ corresponds to $\cA_{H}\propto|\chi\rangle\langle\chi|$,
whose first index is contracted with the conjugate amplitude, as for any expectation value $\langle\Psi|\cA|\Psi\rangle$.
Whereas a density matrix must be Hermitian, positive semi-definite, and of unit trace,
an analyzer only needs to be Hermitian, so its normalization is a matter of convention.
We choose the analyzer to be \eq{hadron_density_general} without the factor $1/d$,
\begin{align}
\label{eq:hadron_analyzer_convention}
\cA_{H}(c) = d\,\rho_{H}(c) = \id_{d\times d} + \sum_{\alpha} \cO_{H}^{(\alpha)}c_{\alpha}
\,,\end{align}
which is allowed only because $\big(\cO_{H}^{(\alpha)}\big)^{\dagger} = \cO_{H}^{(\alpha)}$.
The unpolarized analyzer is then the $d$-dimensional identity, so that unpolarized TMD FFs such as $D_1$ are summed,
rather than averaged, over the spin states of the hadron.
Inserting $\cA_{H}(c)$ into \eq{hadron_analyzer_trace} gives
\begin{align}
\label{eq:hadron_analyzer}
\Delta^{[\Gamma]}(c) = \Tr_{H}\big[\Delta^{[\Gamma]}\big] + \sum_{\alpha} c_{\alpha}\,\Tr_{H}\big[\cO_{H}^{(\alpha)}\Delta^{[\Gamma]}\big]
\,,\end{align}
so a single analyzer $\cO_{H}^{(\alpha)}$ extracts a polarized structure through $\partial\Delta^{[\Gamma]}/\partial c_{\alpha}$.
On the full space $\cH_{Q}\otimes\cH_{\ell}$, 
each such analyzer is represented by its projection $\cM_{H}$ onto the full Hilbert space $\cH$, 
as defined in \eq{projected_measurement}.

\subsection{Relating light structure functions to TMD FFs}
\label{sec:tmdff_lightchi_relations}

Recall that the full Hilbert space is $\cH = \cH_{Q} \otimes \cH_{\ell} = \cH_{s_{H} = s_{\ell} + \frac{1}{2}} \otimes \cH_{s_{H} = s_{\ell} - \frac{1}{2}}$.
We have just discussed that specifying an operator $\cO_{H}^{(\alpha)}$, which live in the Hilbert space $\cH_{s_{H}}$, 
is equivalent to specifying the spin measurement on the hadron.
Similarly, we can also specify the spin information of the heavy quark with operators defined on the $\cH_{Q}$ subspace. 
The heavy quark structure is conventionally imposed by Dirac projectors 
$\Gamma = \{ \Sl{\bn}, \Sl{\bn} \gamma^{5}, \img\sigma^{\mu \nu} \bn_{\nu}\gamma^{5}\}$,
which translates to operators on $\cH_{Q}$ 
\begin{align} \label{eq:def_OQ_Gamma}
\cO^{[\Gamma]}_{Q} = \frac{1}{\bn \cdot v} \tr_\text{Dirac} \Big[ \Big(\id_{Q} + 2 \gamma_{5} \Sl{\Sigma}_{Q} \Big) \frac{1+\Sl{v}}{2} \frac{\Gamma}{2} \Big]
\,,\end{align}
where $\Sigma^{\mu}_{Q} = \frac{1}{2} (0, \sigma_{x}, \sigma_{y}, \sigma_{z})$ in the rest frame.
The heavy quark and light degrees of freedom part of the correlator is indeed a state
given by $\cO_{Q}^{[\Gamma]} \otimes \rho_{\ell}$ (with specified quark spin given by $\Gamma$)
living in the space of $\cH = \cH_{Q} \otimes \cH_{\ell}$.

Therefore, performing a measurement on the TMD FF correlator $\Delta$ in the limit of HQET
relates the standard leading-power TMD FFs to the structure functions of the light density matrix $\rho_{\ell}$
(cf. \eq{FH_hq_CG_rhoell}):
\begin{align}
\label{eq:master_formula}
\Tr_{H} \Big[ \cO_{H} \Delta^{[\Gamma]}\Big] = \frac{\bn \cdot v}{2} \Tr \Big[ \Big( \cO_{Q}^{[\Gamma]} \otimes \rho_{\ell} \Big) 
\cM_{H} \Big] 
\,,\end{align}
where $\Delta^{[\Gamma]} \equiv \tr [\frac{\Gamma}{2} \Delta]$,
the trace on the RHS is over the whole Hilbert space, 
and $\cM_{H}$ is the measurement hadronic analyzer projected onto the whole Hilbert space
\begin{align}
&\cM_{H} = P_{s_{H}}^{\dagger} \cO_{H} P_{s_{H}}
\,, \\
&P_{s_{H}} = P_{s_{\ell} \pm \frac{1}{2}} = \frac{\pm 2 \vec \Sigma_{Q} \cdot \vec \Sigma_{\ell} + \frac{1}{2} (2s_{\ell} + 1 \pm 1)}{2s_{\ell} + 1}
\label{eq:projected_measurement}
\,.\end{align}
Note that because $\cO_{H}$ are generically built out of the spin matrices $J^{a}$ and $[J^{a}, P_{s_{H}}]=0$,
we have $P_{s_{H}}^{\dagger} \cO_{H} P_{s_{H}} = \cO_{H} P_{s_{H}}$.
We note that the LHS of \eq{master_formula} is $\partial \Delta^{[\Gamma]} / \partial c_{\alpha}$, cf. \eq{hadron_analyzer},
where $c_{\alpha}$ are the spin parameter coefficients.

For a spin-1/2 hadron, 
we can decompose the full TMD FF correlator $\Delta(b_{T})$ in position-space
into 8 leading-power TMD FFs grouped by the Dirac structure $\Gamma= \{ \Sl{\bn}, \Sl{\bn} \gamma^{5}, \img\sigma^{\mu \nu} \bn_{\nu}\gamma^{5}\}$ of the heavy quark%
\footnote{We follow the parametrization in the TMD handbook \cite{Boussarie:2023izj},
but since our definition of $\eps_{\perp}$ contracts the auxiliary direction rather than the hadron direction first,
we include a minus sign for the terms with $\epsilon_{\perp}$ relative to the equations in the TMD handbook.
We also find opposite sign of the term proportional to $G_{1T}^{\perp(1)}$, 
which we believe is a typo in Eq.~(2.135) of~\refcite{Boussarie:2023izj}.}%
:
\begin{align} \label{eq:spin_half_tmdffs}
\Delta^{[\Sl{\bn}]} &= D_{1} +  \img M_{H} \eps_{\perp}^{\rho\sigma} b_{\perp\rho}S_{\perp\sigma} D_{1T}^{\perp(1)}
\,, \nn \\
\Delta^{[\Sl{\bn}\gamma_{5}]} &= S_{L} G_{1} - \img {M_{H}} \bperp\cdot S_{\perp} G_{1T}^{\perp(1)} 
\,, \nn \\
\Delta^{[\img\sigma^{\mu\nu}\bn_{\nu}\gamma_{5}]} 
&= S_{\perp}^{\mu} H_{1} + \img {M_{H}} S_{L}\bperp^{\mu} H_{1L}^{\perp(1)}
+ \img M_{H} \eps_{\perp}^{\mu\nu} b_{\perp\nu} H_{1}^{\perp(1)} 
\nn \\
&\quad + \frac12 {M_{H}^{2}} \Big[b_\perp^\mu\,b_\perp\!\cdot\!S_\perp
-\tfrac12 b_\perp^2\,S_\perp^\mu\Big]  H_{1T}^{\perp (2)}
\,, \end{align}
where specifying $\cO_{H}$ further on each Dirac structure will single out one TMD FF (except $H_{1}$ and $H_{1T}^{\perp(2)}$ mix).
\tab{operators_spin1half} shows the TMD FFs for any given combination of $\cO_{H}$ and $\cO_{Q}^{[\Gamma]}$.
Its entries follow from the parametrization in \eq{spin_half_tmdffs} alone,
by evaluating the left-hand side of \eq{master_formula} for each pair $(\Gamma, \cO_{H})$,
while the column $\cO_{Q}^{[\Gamma]}$ is \eq{def_OQ_Gamma} evaluated for each $\Gamma$.
\tab{operators_spin1half} thus serves as a dictionary between pairs $(\Gamma, \cO_{H})$ and individual TMD FFs,
and \eq{master_formula} is the bridge from this dictionary to $\rho_\ell$:
evaluating its right-hand side with \eq{result_rho_ell_cO} expresses each TMD FF
in terms of the light structure functions $\chi_{1,\ell}^{(N,n)}$.

For a spin-1 hadron, with the parameters defined in \eq{spin1_hadron_params},
we have the 18 leading TMD FFs,
which are given in momentum space in \eq{spin1_hadorn_TMDFFs},
following the convention of \refcite{Bacchetta:2000jk}
except the $S_{LL} = \frac{2}{3} S_{LL}^\text{\refcite{Bacchetta:2000jk}}$ difference mentioned above.
The corresponding table relating TMD FFs for to $\cO_{H}$ and $\cO_{Q}^{[\Gamma]}$ is given in \tab{operators_spin1} in \app{ff_structures}.

Explicitly, \eq{master_formula} allows us to express each standard TMD FF in terms of the light structure functions $\chi^{(N,n)}_{1,\ell}$.
We present these relations for all heavy hadrons with spin $s_H = 0, \frac{1}{2}, 1$, which requires light spins up to $s_\ell = \frac{3}{2}$:
the lowest-lying pseudoscalar mesons $H$ with $s_\ell = \frac{1}{2}$ in \tab{tmd_ffs_ps_mesons},
the $\Lambda_Q$ baryons with $s_\ell = 0$ in \tab{tmd_ffs_lambda_baryons},
the $\Sigma_Q$ baryons with $s_\ell = 1$ in \tab{tmd_ffs_sigma_baryons},
the vector mesons $H^*$ with $s_\ell = \frac{1}{2}$ in \tab{tmd_ffs_vector_mesons},
and the axial-vector mesons $H_1$ with $s_\ell = \frac{3}{2}$ in \tab{tmd_ffs_axial_vector_mesons}.
Each table gives the results in both momentum and position space, where an overall factor of $\delta(1 - z_H)\, C_m$ is understood, with $C_m$ the spin-independent matching coefficient at $\mu \sim m$ from \sec{formalism}.
The two differ by constant factors because the Hankel transforms conventionally used for TMD FFs differ from the one we defined for the light structure functions.
For the TMD FFs, we follow the TMD handbook~\cite{Boussarie:2023izj} convention and use
\begin{align}
\Big(\frac{k_{T}}{ M}\Big)^{n} F(k_{T}) &= \frac{1}{n!} \!\int_{0}^{\infty}\! \frac{\df b_{T} \, b_{T} }{2\pi} J_{n} (b_{T} k_{T}) (b_{T} M)^{n} F^{(n)} (b_{T})
\,. \end{align}
Combined with the Hankel transform for the light structure functions \eq{FT_chi},
we have
\begin{align}
\Big(\frac{k_{T}}{M}\Big)^{n} F(k_{T}) &= a\, \hat \chi_{1,\ell}^{(N,n)}
\nn \\ \Leftrightarrow
(b_{T} M)^{n} F^{(n)} (b_{T}) &= (-1)^{n} n! \, a\, \chi_{1,\ell}^{(N,n)}
\,.\end{align}
In the tables, this appears as a relative sign flip for structures of transverse rank $n = 1$, a factor of $2$ for $n = 2$, and a factor of $-6$ for the $n = 3$ structure $H^{\perp\perp}_{1TT}$ of the axial-vector meson in \tab{tmd_ffs_axial_vector_mesons}.

\begin{table}
\caption{\label{tab:tmd_ffs_ps_mesons}%
Results for TMD FFs to lowest-lying pseudoscalar mesons $Q \to H$, $s_{\ell} = \frac{1}{2}, s_{H} = 0$.
}
\begin{ruledtabular}
\begin{tabular}{lc|lc}
\multicolumn{2}{c|}{momentum space} & \multicolumn{2}{c}{position space}\\
\hline
\makebox[2cm][l]{$D_{1}$} & \makebox[2.2cm]{$\tfrac{1}{4} \hat\chi_{1,\ell}^{(0,0)}$}
 & \makebox[2cm][l]{$D_{1}$} & \makebox[2.2cm]{$\tfrac{1}{4} \chi_{1,\ell}^{(0,0)}$}
\\
$\tfrac{k_T}{M_H} H_{1}^{\perp}$ & $-\tfrac{1}{2} \hat\chi_{1,\ell}^{(1,1)}$
 & $b_T M_H H_{1}^{\perp(1)}$ & $\tfrac{1}{2} \chi_{1,\ell}^{(1,1)}$
\end{tabular}
\end{ruledtabular}
\end{table}

\begin{table}
\caption{\label{tab:tmd_ffs_lambda_baryons}%
Results for TMD FFs to $\Lambda_Q$ baryons,
$s_{\ell} = 0, s_{H} = \frac{1}{2}$.
}
\begin{ruledtabular}
\begin{tabular}{lc|lc}
\multicolumn{2}{c|}{momentum space} & \multicolumn{2}{c}{position space}\\
\hline
\makebox[2cm][l]{$D_{1}$} & \makebox[2.2cm]{$\hat\chi_{1,\ell}^{(0,0)}$}
 & \makebox[2cm][l]{$D_{1}$} & \makebox[2.2cm]{$\chi_{1,\ell}^{(0,0)}$}
\\
$G_{1}$ & $\hat\chi_{1,\ell}^{(0,0)}$
 & $G_{1}$ & $\chi_{1,\ell}^{(0,0)}$
 \\
$H_{1}$ & $\hat\chi_{1,\ell}^{(0,0)}$
 & $H_{1}$ & $\chi_{1,\ell}^{(0,0)}$
 \\
$\tfrac{k_T}{M_H} D_{1T}^{\perp}$ & $0$
 & $b_T M_H D_{1T}^{\perp(1)}$ & $0$
\\
$\tfrac{k_T}{M_H} G_{1T}^{\perp}$ & $0$
 & $b_T M_H G_{1T}^{\perp(1)}$ & $0$
\\
$\tfrac{k_T}{M_H} H_{1L}^{\perp}$ & $0$
 & $b_T M_H H_{1L}^{\perp(1)}$ & $0$
\\
$\tfrac{k_T}{M_H} H_{1}^{\perp}$ & $0$
 & $b_T M_H H_{1}^{\perp(1)}$ & $0$
\\
$(\tfrac{k_T}{M_H})^2 H_{1T}^{\perp}$ & $0$
 & $(b_T M_H)^2 H_{1T}^{\perp (2)}$ & $0$
\end{tabular}
\end{ruledtabular}
\end{table}

\begin{table}
\caption{\label{tab:tmd_ffs_sigma_baryons}%
Results for TMD FFs to $\Sigma_Q$ baryons, $s_{\ell} = 1, s_{H} = \frac{1}{2}$.
}
\begin{ruledtabular}
\begin{tabular}{lc|lc}
\multicolumn{2}{c|}{momentum space} & \multicolumn{2}{c}{position space}\\
\hline
\makebox[1.2cm][l]{$D_{1}$} & \makebox[2cm]{$\tfrac{1}{3} \hat\chi_{1,\ell}^{(0,0)}$}
 & \makebox[1.2cm][l]{$D_{1}$} & \makebox[2cm]{$\tfrac{1}{3} \chi_{1,\ell}^{(0,0)}$}
\\
$G_{1}$ & $-\tfrac{1}{9} \hat\chi_{1,\ell}^{(0,0)} \!\!-\!\! \tfrac{2}{3} \hat\chi_{1,\ell}^{(2,0)}$
 & $G_{1}$ & $-\tfrac{1}{9} \chi_{1,\ell}^{(0,0)} \!\!-\!\! \tfrac{2}{3} \chi_{1,\ell}^{(2,0)}$
\\
$H_{1}$ & $-\tfrac{1}{9} \hat\chi_{1,\ell}^{(0,0)} \!\!+\!\! \tfrac{1}{3} \hat\chi_{1,\ell}^{(2,0)}$
 & $H_{1}$ & $-\tfrac{1}{9} \chi_{1,\ell}^{(0,0)} \!\!+\!\! \tfrac{1}{3} \chi_{1,\ell}^{(2,0)}$
\\
$\tfrac{k_T}{M_H} D_{1T}^{\perp}$ & $\tfrac{1}{3} \hat\chi_{1,\ell}^{(1,1)}$
 & $b_T M_H D_{1T}^{\perp(1)}$ & $-\tfrac{1}{3} \chi_{1,\ell}^{(1,1)}$
\\
$\tfrac{k_T}{M_H} G_{1T}^{\perp}$ & $-\tfrac{2}{3} \hat\chi_{1,\ell}^{(2,1)}$
 & $b_T M_H G_{1T}^{\perp(1)}$ & $\tfrac{2}{3} \chi_{1,\ell}^{(2,1)}$
\\
$\tfrac{k_T}{M_H} H_{1L}^{\perp}$ & $-\tfrac{2}{3} \hat\chi_{1,\ell}^{(2,1)}$
 & $b_T M_H H_{1L}^{\perp(1)}$ & $\tfrac{2}{3} \chi_{1,\ell}^{(2,1)}$
\\
$\tfrac{k_T}{M_H} H_{1}^{\perp}$ & $-\tfrac{1}{3} \hat\chi_{1,\ell}^{(1,1)}$
 & $b_T M_H H_{1}^{\perp(1)}$ & $\tfrac{1}{3} \chi_{1,\ell}^{(1,1)}$
\\
$(\tfrac{k_T}{M_H})^2 H_{1T}^{\perp}$ & $-\tfrac{4}{3} \hat\chi_{1,\ell}^{(2,2)}$
 & $( b_T M_H\!)^2 \! H_{1T}^{\perp (2)}$ & $-\tfrac{8}{3} \chi_{1,\ell}^{(2,2)}$
\end{tabular}
\end{ruledtabular}
\end{table}

\begin{table}[h]
\caption{\label{tab:tmd_ffs_vector_mesons}%
Results for TMD FFs to vector mesons $Q \to H^{*}$, $s_{\ell} = \frac{1}{2}, s_{H} = 1$,
$J^{P} = 1^{-}$.
}
\begin{ruledtabular}
\begin{tabular}{lc|lc}
\multicolumn{2}{c|}{momentum space} & \multicolumn{2}{c}{position space}\\
\hline
\makebox[2cm][l]{$D_1$} & \makebox[2cm]{$\frac{3}{4}\hat\chi_{1,\ell}^{(0,0)}$}
 & \makebox[2cm][l]{$D_1$} & \makebox[2cm]{$\frac{3}{4}\chi_{1,\ell}^{(0,0)}$}\\
$D_{1LL}$ & $0$ & $D_{1LL}$ & $0$\\
$G_{1}$ & $\frac{3}{4}\hat\chi_{1,\ell}^{(0,0)}$ & $G_{1}$ & $\frac{3}{4}\chi_{1,\ell}^{(0,0)}$\\
$H_{1}$ & $\frac{3}{4}\hat\chi_{1,\ell}^{(0,0)}$ & $H_{1}$ & $\frac{3}{4}\chi_{1,\ell}^{(0,0)}$\\
$H_{1LT}$ & $0$ & $H_{1LT}$ & $0$\\
$\frac{k_T}{M_H}D_{1T}^\perp$ & $\frac{3}{2}\hat\chi_{1,\ell}^{(1,1)}$
 & $b_TM_HD_{1T}^{\perp(1)}$ & $-\frac{3}{2}\chi_{1,\ell}^{(1,1)}$\\
$\frac{k_T}{M_H}D_{1LT}^\perp$ & $0$ & $b_TM_HD_{1LT}^{\perp(1)}$ & $0$\\
$\frac{k_T}{M_H}G_{1T}^\perp$ & $0$ & $b_TM_HG_{1T}^{\perp(1)}$ & $0$\\
$\frac{k_T}{M_H}G_{1LT}^\perp$ & $-\frac{3}{2}\hat\chi_{1,\ell}^{(1,1)}$
 & $b_TM_HG_{1LT}^{\perp(1)}$ & $\frac{3}{2}\chi_{1,\ell}^{(1,1)}$\\
$\frac{k_T}{M_H}H_1^\perp$ & $\frac{1}{2}\hat\chi_{1,\ell}^{(1,1)}$
 & $b_TM_HH_1^{\perp(1)}$ & $-\frac{1}{2}\chi_{1,\ell}^{(1,1)}$\\
$\frac{k_T}{M_H}H_{1L}^\perp$ & $0$ & $b_TM_HH_{1L}^{\perp(1)}$ & $0$\\
$\frac{k_T}{M_H}H_{1LL}^\perp$ & $\frac{3}{2}\hat\chi_{1,\ell}^{(1,1)}$
 & $b_TM_HH_{1LL}^{\perp(1)}$ & $-\frac{3}{2}\chi_{1,\ell}^{(1,1)}$\\
$\frac{k_T}{M_H}H_{1TT}^\perp$ & $-\frac{3}{2}\hat\chi_{1,\ell}^{(1,1)}$
 & $b_TM_HH_{1TT}^{\perp(1)}$ & $\frac{3}{2}\chi_{1,\ell}^{(1,1)}$\\
$(\frac{k_T}{M_H})^2D_{1TT}^\perp$ & $0$ & $(b_TM_H)^2D_{1TT}^{\perp(2)}$ & $0$\\
$(\frac{k_T}{M_H})^2G_{1TT}^\perp$ & $0$ & $(b_TM_H)^2G_{1TT}^{\perp(2)}$ & $0$\\
$(\frac{k_T}{M_H})^2H_{1T}^\perp$ & $0$ & $(b_TM_H)^2H_{1T}^{\perp(2)}$ & $0$\\
$(\frac{k_T}{M_H})^2H_{1LT}^\perp$ & $0$ & $(b_TM_H)^2H_{1LT}^{\perp(2)}$ & $0$\\
$(\frac{k_T}{M_H})^3H_{1TT}^{\perp\perp}$ & $0$ & $(b_TM_H)^3H_{1TT}^{\perp\perp(3)}$ & $0$\\
\end{tabular}
\end{ruledtabular}
\end{table}

\begin{table*}
\caption{\label{tab:tmd_ffs_axial_vector_mesons}%
Results for TMD FFs to axial vector mesons $H_{1}$, $s_{\ell} = \frac{3}{2}, s_{H} = 1$, $J^{P} = 1^{+}$.
}
\begin{ruledtabular}
\begin{tabular}{lc|lc}
\multicolumn{2}{c|}{momentum space} & \multicolumn{2}{c}{position space}\\
\hline
\makebox[4cm][l]{$D_1$} & \makebox[4cm]{$\frac{3}{8}\hat\chi_{1,\ell}^{(0,0)}$}
 & \makebox[4cm][l]{$D_1$} & \makebox[4cm]{$\frac{3}{8}\chi_{1,\ell}^{(0,0)}$}\\
$D_{1LL}$ & $\frac{9}{16}\hat\chi_{1,\ell}^{(2,0)}$
 & $D_{1LL}$ & $\frac{9}{16}\chi_{1,\ell}^{(2,0)}$\\
$G_{1}$ & $-\frac{3}{16}\hat\chi_{1,\ell}^{(0,0)}-\frac{3}{8}\hat\chi_{1,\ell}^{(2,0)}$
 & $G_{1}$ & $-\frac{3}{16}\chi_{1,\ell}^{(0,0)}-\frac{3}{8}\chi_{1,\ell}^{(2,0)}$\\
$H_{1}$ & $-\frac{3}{16}\hat\chi_{1,\ell}^{(0,0)}+\frac{3}{16}\hat\chi_{1,\ell}^{(2,0)}$
 & $H_{1}$ & $-\frac{3}{16}\chi_{1,\ell}^{(0,0)}+\frac{3}{16}\chi_{1,\ell}^{(2,0)}$\\
$H_{1LT}$ & $0$ & $H_{1LT}$ & $0$\\
$\frac{k_T}{M_H}D_{1T}^\perp$ & $\frac{3}{8}\hat\chi_{1,\ell}^{(1,1)}$
 & $b_TM_HD_{1T}^{\perp(1)}$ & $-\frac{3}{8}\chi_{1,\ell}^{(1,1)}$\\
$\frac{k_T}{M_H}D_{1LT}^\perp$ & $\frac{3}{8}\hat\chi_{1,\ell}^{(2,1)}$
 & $b_TM_HD_{1LT}^{\perp(1)}$ & $-\frac{3}{8}\chi_{1,\ell}^{(2,1)}$\\
$\frac{k_T}{M_H}G_{1T}^\perp$ & $-\frac{3}{8}\hat\chi_{1,\ell}^{(2,1)}$
 & $b_TM_HG_{1T}^{\perp(1)}$ & $\frac{3}{8}\chi_{1,\ell}^{(2,1)}$\\
$\frac{k_T}{M_H}G_{1LT}^\perp$ & $\frac{3}{40}\hat\chi_{1,\ell}^{(1,1)}+\frac{3}{4}\hat\chi_{1,\ell}^{(3,1)}$
 & $b_TM_HG_{1LT}^{\perp(1)}$ & $-\frac{3}{40}\chi_{1,\ell}^{(1,1)}-\frac{3}{4}\chi_{1,\ell}^{(3,1)}$\\
$\frac{k_T}{M_H}H_1^\perp$ & $-\frac{1}{4}\hat\chi_{1,\ell}^{(1,1)}$
 & $b_TM_HH_1^{\perp(1)}$ & $\frac{1}{4}\chi_{1,\ell}^{(1,1)}$\\
$\frac{k_T}{M_H}H_{1L}^\perp$ & $-\frac{3}{8}\hat\chi_{1,\ell}^{(2,1)}$
 & $b_TM_HH_{1L}^{\perp(1)}$ & $\frac{3}{8}\chi_{1,\ell}^{(2,1)}$\\
$\frac{k_T}{M_H}H_{1LL}^\perp$ & $-\frac{3}{40}\hat\chi_{1,\ell}^{(1,1)}+\frac{9}{8}\hat\chi_{1,\ell}^{(3,1)}$
 & $b_TM_HH_{1LL}^{\perp(1)}$ & $\frac{3}{40}\chi_{1,\ell}^{(1,1)}-\frac{9}{8}\chi_{1,\ell}^{(3,1)}$\\
$\frac{k_T}{M_H}H_{1TT}^\perp$ & $\frac{3}{40}\hat\chi_{1,\ell}^{(1,1)}-\frac{3}{16}\hat\chi_{1,\ell}^{(3,1)}$
 & $b_TM_HH_{1TT}^{\perp(1)}$ & $-\frac{3}{40}\chi_{1,\ell}^{(1,1)}+\frac{3}{16}\chi_{1,\ell}^{(3,1)}$\\
$(\frac{k_T}{M_H})^2D_{1TT}^\perp$ & $\frac{3}{8}\hat\chi_{1,\ell}^{(2,2)}$
 & $(b_TM_H)^2D_{1TT}^{\perp(2)}$ & $\frac{3}{4}\chi_{1,\ell}^{(2,2)}$\\
$(\frac{k_T}{M_H})^2G_{1TT}^\perp$ & $-\frac{3}{4}\hat\chi_{1,\ell}^{(3,2)}$
 & $(b_TM_H)^2G_{1TT}^{\perp(2)}$ & $-\frac{3}{2}\chi_{1,\ell}^{(3,2)}$\\
$(\frac{k_T}{M_H})^2H_{1T}^\perp$ & $-\frac{3}{4}\hat\chi_{1,\ell}^{(2,2)}$
 & $(b_TM_H)^2H_{1T}^{\perp(2)}$ & $-\frac{3}{2}\chi_{1,\ell}^{(2,2)}$\\
$(\frac{k_T}{M_H})^2H_{1LT}^\perp$ & $-\frac{3}{2}\hat\chi_{1,\ell}^{(3,2)}$
 & $(b_TM_H)^2H_{1LT}^{\perp(2)}$ & $-3\chi_{1,\ell}^{(3,2)}$\\
$(\frac{k_T}{M_H})^3H_{1TT}^{\perp\perp}$ & $-\frac{3}{2}\hat\chi_{1,\ell}^{(3,3)}$
 & $(b_TM_H)^3H_{1TT}^{\perp\perp(3)}$ & $9\chi_{1,\ell}^{(3,3)}$\\
\end{tabular}
\end{ruledtabular}
\end{table*}

\sectionPaper{Examples of spin-symmetry relations}
\label{sec:examples}

Heavy hadrons in the same heavy-quark spin multiplet, i.e.\ with the same light-spin $s_{\ell}$,
share the same light-spin density matrix $\rho_{\ell}$ and hence the same light structure functions
$\chi_{1,\ell}^{(N,n)}$; they differ only in how the heavy-quark spin is coupled to $s_{\ell}$.
Since a multiplet contains many more TMD FFs than structure functions (see \tab{number_structure_functions}),
heavy-quark spin symmetry implies a set of relations among the TMD FFs of its members,
which follow directly from \eq{master_formula}.
Each choice of Dirac projection $\Gamma$, analyzer $\cO_{H}$ and multiplet member $s_{H}$
selects one entry $F_{\Gamma,\cO_{H}} \equiv \Tr_{H}\big[\cO_{H}\Delta^{[\Gamma]}\big]$,
which by the parametrization of the correlator is a known linear combination of TMD FFs,
i.e. the entries of \tab{operators_spin1half} and alike.
On the right-hand side of \eq{master_formula}, the partial trace over $\cH_{Q}$ can be performed
before anything is specified about the light degrees of freedom,
\begin{align}
\label{eq:Qtrace}
L_{\Gamma,\cO_{H}} \equiv \frac{\bn \cdot v}{2}\, \Tr_{Q} \Big[ \big( \cO_{Q}^{[\Gamma]} \otimes \id_{\ell} \big)\, \cO_{H} P_{s_{H}} \Big] \,,
\end{align}
which leaves an operator on $\cH_{\ell}$ alone. 
The remaining trace can be left formal,
\begin{align}
F_{\Gamma,\cO_{H}} = \Tr_{\ell} \big[ \rho_{\ell}\, L_{\Gamma,\cO_{H}} \big] \equiv \big\langle L_{\Gamma,\cO_{H}} \big\rangle_{\ell} \,,
\end{align}
so every entry, for every member of the multiplet, is an expectation value in the same $\rho_{\ell}$.
Any linear dependence among the light operators $L_{\Gamma,\cO_{H}}$ 
therefore translates to relations among the corresponding entries,
which may be for TMD FFs of the same hadron or of different hadrons within the same multiplet.
The simplest example is for $s_{\ell} = 0$, where the only multiplet member is the $\Lambda$ baryon with $s_{H} = 1/2$,
and the hadron spin is the heavy-quark spin $J^{a} = \Sigma_{Q}^{a}$.
From Pauli matrix identities we arrive at
$D_{1} = G_{1} = H_1 = \langle \id \rangle_{\ell} = \chi_{1,\ell}^{(0,0)}$,
whereas the rest of the TMD FFs vanish.
Physically, for a spin-less light cloud, measuring the hadron spin along any axis is the same as measuring the heavy-quark spin,
and the vanishing functions all require nontrivial light spin.

For $s_{\ell}>0$ we have more dependency relations.
Since $\cH_{\ell}$ only supports operators of rank at most $2s_{\ell}$, and parity restricts which
components have nonvanishing expectation values (see detailed analysis in \sec{symmetry_analysis}), 
these dependencies follow from a few angular-momentum identities for the partial traces in \eq{Qtrace}.

We give a few of illustrative examples that can be directly verified from the tables in \sec{tmdff_lightchi_relations}.
The mathematics used here is standard angular-momentum algebra,
so we focus our discussion on physical interpretations of the relations.

\subsection{$s_{\ell} = 1/2$ multiplet}
\label{sec:example_sellhalf}

\begin{figure*}[t]
\includegraphics[width=0.9\textwidth]{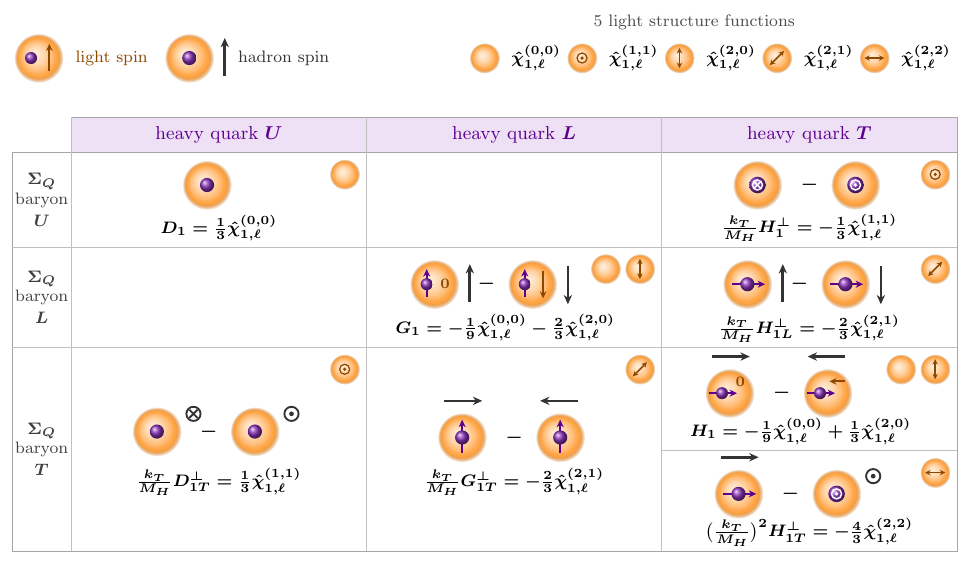}
\caption{
\label{fig:Sigma_tmdff_table}
Leading-power TMD FFs of the $\Sigma_Q$ baryon ($s_H = \tfrac12$, $s_\ell = 1$) in momentum space, expressed through the five light structure functions $\hat\chi^{(N,n)}_{1,\ell}$ (cf.\ \tab{tmd_ffs_sigma_baryons}).
Columns label the polarization $(U, L, T)$ of the fragmenting heavy quark and rows that of the hadron.
Each entry depicts the TMD FF as the difference of two spin configurations, 
with the heavy-quark spin (purple) and the light-cloud spin (orange; ``0'' denoting $m_\ell = 0$) inside the hadron 
and the hadron spin outside.
The small clouds in the corner of each entry indicate which $\hat\chi^{(N,n)}_{1,\ell}$ it depends on, using the symbols in \fig{sell_triangle}.
Entries sharing a light structure function are related by heavy-quark spin symmetry, $D_{1T}^\perp = -H_1^\perp$, $G_{1T}^\perp = H_{1L}^\perp$, and $D_1 + G_1 + 2H_1 = 0$ (see \sec{example_sell1}).
An overall factor of $\delta(1 - z_H)\,C_m$ is understood.
}
\end{figure*}

The $s_{\ell} = 1/2$ multiplet consists of the pseudoscalar meson $H$ (\tab{tmd_ffs_ps_mesons}) and the
vector meson $H^*$ (\tab{tmd_ffs_vector_mesons}). The $2 + 18 = 20$ TMD FFs are fixed by the two
structure functions $\chi_{1,\ell}^{(0,0)}$ and $\chi_{1,\ell}^{(1,1)}$, giving 18 relations.
On $\cH_{\ell}$ the only operators are $\id$ and $\vec\sigma_{\ell}$, and parity leaves only
$\langle \id \rangle_{\ell} = \hat\chi_{1,\ell}^{(0,0)}$ and $\langle \sigma_{\ell,y} \rangle_{\ell} = -2\hat\chi_{1,\ell}^{(1,1)}$
nonvanishing. The projectors onto the hadrons are
\begin{align}
\text{singlet:}\quad P_{0} &= \frac{1}{4}\Big(1 - \vec\sigma_{Q}\cdot\vec\sigma_{\ell}\Big) \,, 
\\
\text{triplet:}\quad P_{1} &= \frac{1}{4}\Big(3 + \vec\sigma_{Q}\cdot\vec\sigma_{\ell}\Big) \,,
\end{align}
with hadron spin $\vec J = \frac12(\vec\sigma_{Q} + \vec\sigma_{\ell})$ (cf. \eq{angular_momentum_addition}). 
We organize the relations by the hadron analyzer.
In each case the partial traces in \eq{Qtrace} follow from $\Tr_{Q}[\sigma_{Q}^{a}\sigma_{Q}^{b}] = 2\delta^{ab}$
and rotational covariance.

\paragraph{Unpolarized analyzer $\cO_{H} = \id$.}
The partial traces can be easily evaluated with
\begin{align}
\Tr_{Q} P_{0} = \tfrac12 \,, \quad \Tr_{Q} P_{1} = \tfrac32 \,, \quad
\Tr_{Q}\big[\sigma_{Q}^{a} P_{0,1}\big] = \mp\tfrac12\,\sigma_{\ell}^{a} \,.
\end{align}
The $\slashed{\bn}$ projection gives $D_1^{H^*} = 3D_1^{H}$, the ratio of the multiplicities.
The $y_\mu\,i\sigma^{\mu\nu}\bar n_\nu\gamma_5$ projection gives $H_1^{\perp H^*} = -H_1^{\perp H}$. 
$P_0 + P_1 = \id$ also gives
\begin{align}
D_1^{H} + D_1^{H^*} = \langle \id \rangle_{\ell} \,, \qquad H_1^{\perp H} + H_1^{\perp H^*} = 0 \,.
\end{align}
These relations are already commented on in \refcite{vonKuk:2023jfd}.

\paragraph{Vector analyzer $\cO_{H} = J^{a}$.}
The triplet states are symmetric under exchanging heavy quark and light cloud spin, 
which manifests as identities
\begin{align}
\Tr_{Q}\big[J^{a} P_{1}\big] = \sigma_{\ell}^{a} \,, \qquad
\Tr_{Q}\big[\sigma_{Q}^{a} J^{b} P_{1}\big] = \delta^{ab} \,.
\nn
\end{align}
The second identity contains no $\epsilon^{abc}\sigma_{\ell}^{c}$ term, so a polarized heavy quark transfers
its spin to $H^*$ isotropically, and an unpolarized one produces $H^*$ polarization only through the
light-cloud polarization $\langle\sigma_{\ell,y}\rangle_{\ell}$. 
In terms of TMD FFs,
\begin{gather}
G_{1}^{H^*} = H_{1}^{H^*} = D_1^{H^*} \,, \qquad D_{1T}^{\perp H^*} = -3 H_1^{\perp H} \,, 
\\
G_{1T}^{\perp H^*} = H_{1L}^{\perp H^*} = H_{1T}^{\perp H^*} = 0 \,.
\end{gather}
The first equality can be seen by first noting that $D_{1LL}^{H^{*}} = 0$, a fact from analyzing the tensor polarizations in the next paragraph. 
With $D_{1LL}^{H^{*}} = 0$, $G_{1}$ parametrizes the difference in parallel and antiparallel hadron helicity from a heavy quark with definite helicity. 
Since with a spin-$\frac12$ light cloud, the antiparallel hadron state is never produced, $G_{1}^{H^{*}} = D_{1}^{H^{*}}$.
The same argument but for transverse spin gives $H_{1}^{H^*} = D_{1}^{H^{*}}$.
The second equality follows from the fact that both functions are governed by the polarization of the
light cloud perpendicular to the plane spanned by the fragmentation axis and $\vec k_T$ (through by $\Sigma_{\ell,y}$).
The FFs on the second line vanish due to lack of $\hat k_{\perp}$ dependence in the $s_{\ell} = \frac12$ density matrix.

\paragraph{Tensor analyzer $\cO_{H} = Q^{ab}$.}
A single spin-$\frac12$ cannot carry rank-two polarization, 
so the $H^*$ tensor polarizations can only arise from a correlation between the two spins,
\begin{gather}
\Tr_{Q}\big[Q^{ab} P_{1}\big] = 0 \,, 
\nn \\ 
\Tr_{Q}\big[\sigma_{Q}^{c} Q^{ab} P_{1}\big] = \tfrac12\big(\delta^{ca}\sigma_{\ell}^{b} + \delta^{cb}\sigma_{\ell}^{a}\big)
- \tfrac13\delta^{ab}\sigma_{\ell}^{c} \,.
\nn
\end{gather}
An unpolarized heavy quark therefore produces no tensor polarization for the heavy hadron, 
$D_{1LL} = D_{1LT}^{\perp} = D_{1TT}^{\perp} = 0$,
whose collinear part is the known result~\cite{Falk:1993rf}.
With only $\langle\sigma_{\ell,y}\rangle_{\ell}$ nonvanishing, the remaining nonzero entries are those with
one index along $y$, giving
\begin{gather}
H_{1LL}^{\perp H^*} = -G_{1LT}^{\perp H^*} = -H_{1TT}^{\perp H^*} = D_{1T}^{\perp H^*} \,, 
\\
G_{1TT}^{\perp H^*} = H_{1LT}^{H^*} = H_{1LT}^{\perp H^*} = H_{1TT}^{\perp\perp H^*} = 0 \,.
\end{gather}
Together with the vector analyzer, every $\langle\sigma_{\ell,y}\rangle_{\ell}$-dependent function of $H^*$
is fixed by the Collins function of $H$.

\subsection{$s_{\ell} = 1$ multiplet}
\label{sec:example_sell1}

Here $\cH_{\ell}$ supports $\id$, $\vec\Sigma_{\ell}$ and the light tensor component
$Q_{\ell}^{ab} = \frac12\{\Sigma_{\ell}^{a}, \Sigma_{\ell}^{b}\} - \frac23\delta^{ab}$.
The multiplet includes the $\Sigma_{Q}$ baryon with $s_{H} = 1/2$
and the $\Sigma_{Q}^{*}$ baryon with $s_{H} = 3/2$, which we do not include here for brevity.

The 8 TMD FFs for the $s_{H} = 1/2$ hadron (\tab{tmd_ffs_sigma_baryons}) already depend on only 5 structure functions,
which leaves 3 relations. 
The projector is $P_{1/2} = \frac13\big(1 - \vec\sigma_{Q}\cdot\vec\Sigma_{\ell}\big)$, 
and $\vec J = \frac12\vec\sigma_{Q} + \vec\Sigma_{\ell}$. 
A spin-$\frac12$ hadron has only the unpolarized and
vector analyzers, and the relevant partial traces are
\begin{gather}
\Tr_{Q} P_{1/2} = \tfrac23 \,, \quad
\Tr_{Q}\big[\sigma_{Q}^{a} P_{1/2}\big] = -\tfrac23\Sigma_{\ell}^{a} \,, 
\nn \\
\Tr_{Q}\big[J^{a} P_{1/2}\big] = \tfrac13\Sigma_{\ell}^{a} \,,  \quad
\Tr_{Q}\big[\sigma_{Q}^{a} J^{b} P_{1/2}\big] = -\tfrac19\delta^{ab} - \tfrac23 Q_{\ell}^{ab} \,.
\nn
\end{gather}
The first line ties the unpolarized and vector analyzers through $\langle\Sigma_{\ell,y}\rangle_{\ell}$,
giving $D_{1T}^{\perp} = -H_1^{\perp}$. The last identity is symmetric in $a \leftrightarrow b$, so exchanging
the axes of the quark and hadron spins gives $G_{1T}^{\perp} = H_{1L}^{\perp}$.
Contracting the indices of the last identity gives
$\Tr_Q[\vec\sigma_Q\cdot\vec J\,P] = -\frac13 = -\frac12\Tr_Q P$, 
which in terms of TMD FFs gives $D_1 + G_1 + 2H_1 = 0$.
Therefore, the three relations amongst the TMD FFs are
\begin{align}
D_{1T}^{\perp} = -H_1^{\perp} \,, \quad G_{1T}^{\perp} = H_{1L}^{\perp} \,, \quad D_1 + G_1 + 2H_1 = 0 \,.
\end{align}
 
A comment on the last identity: 
Physically, since $G_1$ and $H_1$ are the longitudinal and transverse diagonal entries of the
quark--hadron spin correlation, 
and their sum is the polarization transfer from the heavy quark to the hadron averaged over all spin directions. 
Since the heavy-quark and light-cloud spins are separately
conserved, this rotational scalar is fixed by the spin quantum numbers alone, and is thus proportional to $D_1$. 

More generally, the operator
$\vec\sigma_Q\cdot\vec J$ commutes with $P_{s_H}$ and is constant,
\begin{gather}
\vec\sigma_Q\cdot\vec J\,P_{s_H} = c_H\,P_{s_H}\,, 
\nn \\
c_H \equiv s_H(s_H+1) - s_\ell(s_\ell+1) + \tfrac34 \,,
\end{gather}
so that $\Tr_Q[\vec\sigma_Q\cdot\vec J\,P_{s_H}] = c_H\,\Tr_Q P_{s_H}$.
For a hadron with vector analyzer
$\kappa_H\vec J$ this gives $G_{1} + 2 H_{1} = \kappa_H \,c_H\,D_1$, which equals $3D_1$ for
$\Lambda_Q$ and $H^*$ and $-D_1$ for $\Sigma_Q$.

\sectionPaper{Conclusions}
\label{sec:conclusions}
In this work, we extended the spin-density matrix formalism of Ref.~\cite{vonKuk:2023jfd} 
to the TMD FFs of polarized heavy hadrons, 
exploiting that heavy-quark spin symmetry reduces the fragmentation dynamics for $\Lambda_\mathrm{QCD} \lesssim k_T \ll m$
to the light spin-density matrix $\rho_\ell$.
A symmetry analysis based on hermiticity, parity, and rotational covariance 
showed that $\rho_\ell$ is characterized by only $\lceil (2s_\ell + 1)^2/2 \rceil$ independent structure functions $\chi^{(N,n)}_{1,\ell}$, 
which answers the question posed in the title.
The resulting reduction is drastic, see~\tab{number_structure_functions}.

We gave explicit expressions for all TMD FFs of heavy hadrons with spin $s_H \leq 1$ in terms of these structure functions $\chi^{(N,n)}_{1,\ell}$, 
and showed that the resulting relations and sum rules, 
both within a single hadron and across members of the same multiplet, 
are all direct manifestations of heavy-quark spin symmetry. 
The physics interpretation of our master formula \eq{master_formula} is simple: 
since the heavy-quark spin decouples from the light dynamics, 
every TMD FF is an expectation value in the same light spin-density matrix, 
with the hadron spin entering only through angular-momentum coupling.
Since these relations hold point by point in transverse momentum up to corrections suppressed by $1/m$, 
they provide clean, theoretically controlled targets for measurements at Belle~II and the future Electron-Ion Collider.%

More broadly, our results show how the perturbative gap $m \gg \lqcd$ provided by the heavy-quark mass
opens a new opportunity to use the transverse-momentum-dependent spin structure of heavy-quark fragmentation as a nontrivial probe of hadronization,
with far fewer complications than for light hadrons~\cite{Metz:2016swz, Anselmino:2020vlp, Boussarie:2023izj}.
The perturbative mass scale tames the proliferation of independent TMD FFs,
which would otherwise be challenging to extract because of their sheer number,
and yields nontrivial, model-independent predictions for genuinely nonperturbative quantities.
This reflects a more general lesson, namely that heavy quarks as a static color source offer a powerful handle on confinement and hadron formation in QCD
that is not available for light hadrons, see~\cite{Lee:2026bcq,vonKuk:2025kbv,Ko:2026ojy} for recent examples.
While collinear heavy-quark fragmentation functions have been studied in this spirit
for several decades~\cite{Mele:1990cw, Jaffe:1993ie, Falk:1993rf, Neubert:2007je, Fickinger:2016rfd, Bonino:2023icn},
our work marks the beginning of a program that takes the next step,
bringing the full spin and transverse-momentum structure of heavy-quark fragmentation under theoretical control
and thus opening a unique window into the three-dimensional dynamics of hadronization.

\vspace{1cm}
\begin{acknowledgments}
\headingAcknowledgments
We thank Dean Robinson and Bruno Scheihing for helpful comments on the manuscript.
We thank the Erwin Schrödinger Institute at the University of Vienna for its hospitality.
K.L. is supported by the U.S. Department of Energy under contract DE-AC02-06CH11357.
J.M. was supported by the D-ITP consortium, a program of NWO that is funded by the Dutch Ministry of Education, Culture and Science (OCW).
Z.S. is supported in part by Department of Energy award DESC0025293 and National Science Foundation grant PHY-2515115.

\paragraph*{AI statement}
We have used Claude (Opus 5.5) to improve the language of the manuscript, 
and to check and organize the tables presented in \sec{light_dm_to_tmdffs}.
Chats with Claude (Opus 5.5) have also helped to develop insights into generalizing the symmetry relations
discussed in \sec{examples}.
The illustrations \fig{spin_multiplet_example}, \fig{sell_triangle}, and \fig{Sigma_tmdff_table} are also drafted in TikZ by Claude (Opus 5.5).

\end{acknowledgments}

\vspace{1cm}
\appendix
\section{Conventions}
\label{app:conventions}

\subsection{Lightcone conventions}
\label{app:lightcone}
We decompose four-vectors $p^\mu$
in terms of lightlike vectors $n^\mu$, $\bn^\mu$ with $n^2 = \bn^2 = 0$
and $n \cdot \bn = 2$,
\begin{align}
p^\mu
&= \bn \cdot p \, \frac{n^\mu}{2} + n \cdot p \, \frac{\bn^\mu}{2} + p_\perp^\mu
\nn \\
&\equiv p^- \frac{n^\mu}{2} + p^+ \frac{\bn^\mu}{2} + p_\perp^\mu
\equiv (p^-, p^+, p_\perp)
\,,\end{align}
such that e.g.\ $p^2 = p^- p^+ + p_\perp^2$.
As another part of this convention, collinear momenta
near the mass shell will typically
have large components $p^- \gg p_\perp \gg p^+$.
We always take transverse vectors
with subscript $\perp$ to be Minkowskian, $p_\perp^2 \equiv p_\perp \cdot p_\perp < 0$,
and denote their magnitude by $p_T = \sqrt{- p_\perp^2}$.
Finally, we define the metric and antisymmetric tensor in transverse space as
\begin{align}
g^{\mu \nu}_\perp \equiv g^{\mu \nu} - \frac{n^\mu \bar{n}^\nu}{2} - \frac{\bar{n}^\mu {n}^\nu}{2}
\,, \qquad
\epsilon^{\mu \nu}_\perp \equiv \epsilon^{\mu \nu \rho \sigma} \frac{\bar{n}_\rho n_\sigma}{2}
\,.\end{align}
Our convention for the antisymmetric tensor is $\eps^{0123} = +1$.

In the heavy-hadron rest frame, $v^\mu = (1, 0, 0, 0)$,
\eq{def_z} gives $\bn^\mu = (\bn \cdot v)(v^\mu - z^\mu)$,
so the spatial direction of $\bn$ is $-\hat z$,
and $\hat z$ points along the spatial direction of $n$,
i.e.\ along the large momentum of the hadron.
Taking $z^\mu = (0, 0, 0, 1)$ and $x^\mu = b_\perp^\mu / b_T = (0, 1, 0, 0)$,
we have $\eps_\perp^{12} = -1$,
and \eq{def_y} gives $y^\mu = \eps_\perp^{\nu\mu} x_\nu = (0, 0, 1, 0)$,
so that $\hat x$, $\hat y$, $\hat z$ form a right-handed triad.

\subsection{bHQET conventions}
\label{app:bhqet}
The HQET velocity $v^\mu = P_H^\mu / M_H$, with $v^2 = 1$, is defined in \eq{def_v}.
The tree-level matching of the massive QCD quark field onto $h_v$ at $\mu \sim m$ reads:
\begin{align} \label{eq:bhqet_tree_matching}
\psi_Q(x) = e^{-\img m v \cdot x} h_v(x)
\Bigl[ 1 + \ORd{\frac{1}{m}} \Bigr]
\,.\end{align}
The $h_v(x)$ are implemented as Dirac spinors satisfying the projection relations
\begin{align} \label{eq:hqet_projection_identities}
P_{\pm} &\equiv \frac{1 \pm \Sl v}{2}
\,, \nn \\
P_+ h_v &= h_v
\,, \quad
P_- h_v = 0
\,, \quad
\bar{h}_v P_+ = \bar{h}_v
\,, \quad
\bar{h}_v P_- = 0
\,.\end{align}
For external states, the matching reads $\ket{H, h_H; X} = \sqrt{m} \, \ket{H_v, h_H; X}$,
and we use a nonrelativistic normalization convention for the bHQET states.

The spin degrees of freedom of the heavy-quark can be explicitly decoupled
from the light dynamics at leading power in $1/m$
by performing a field redefinition
involving static Wilson lines $Y_v(x)$~\cite{Korchemsky:1991zp, Bauer:2001yt},
\begin{align}
\label{eq:sterile_decoupling}
h_v(x) &= Y_v(x) \, h_v^{(0)}(x)
\,, \nn \\
Y_v(x) &= \bar P \Bigl[ \exp \Bigl(
  - \img g \int_0^\infty \! \df s \, v \cdot A(x + v s)
\Bigr) \Bigr]
\,.\end{align}
In this way, the heavy-quark Lagrangian becomes that of a free theory,
$\cL^{(0)} = \bar h_v^{(0)} (\img v \cdot \partial) h_v^{(0)}$,
while the $Y_v(x)$ takes the place of $h_v(x)$ in all external operators,
acting as a static source of soft gluons.
Specifically, the action of $h_v(x)$ on a product state in the decoupled theory
is given by
\begin{align}
\label{eq:decoupling_outstate}
&h_v(x) \, \ket{s_Q, h_Q; s_\ell, h_\ell, f_\ell; X}
\nn \\
&\qquad = u(v, h_Q) \, Y_{v}(x) \, \ket{s_\ell, h_\ell, f_\ell ; X}
\,,\end{align}
where $s_Q = \tfrac{1}{2}$ and $h_Q = \pm \tfrac{1}{2}$ are the spin and helicity of the heavy quark,
$u(v, h_Q) = u(m v, h_Q)/\sqrt{m}$ is an HQET spinor,
and $s_\ell$, $h_\ell$, and $f_\ell$ are the total angular momentum,
helicity, and flavor content of the light degrees of freedom inside the hadron.

\subsection{Spin-density matrix}
\label{app:corr_spin_density}

We work in QCD with $n_{\ell}$ light flavors and one additional heavy quark
with pole mass $m_{Q} \equiv m_{c}, m_{b} \gg \lqcd$.
The TMD FF correlator in position space
describing a heavy quark $Q$ fragmenting into a heavy hadron $H$
with open helicity indices $h_H$, $h_H'$ is defined as
\begin{align} \label{eq:def_tmd_ff_correlator}
&\Delta_{H/Q,\, h_H h_H'}^{\beta \beta'}(z_H, b_\perp)
= \frac{1}{2 z_H N_c} \int \! \frac{\df b^+}{4\pi} \, e^{\img b^+ (P_H^-/z_H)/2}
\nn \\[0.4em] & \quad \times
\Tr \SumInt_{X}
\Mae{0}{W^\dagger(b) \, \psi_Q^\beta(b)}{H, h_H; X}
\nn \\[0.2em] & \quad \times
\Mae{H, h_H'; X}{\bar\psi_Q^{\beta'}(0) \, W(0)}{0}
\,. \end{align}
In the above expression,
$z_H$ is the fraction of the quark's lightcone momentum retained by $H$,
$\beta, \beta'$ are the open spin indices of the quark fields,
$\Tr$ denotes a trace over fundamental color indices,
and $b \equiv (0, b^+, b_\perp)$ is the transverse separation between the quark fields.
$P_{H}^{\mu}$ is the momentum carried by the heavy hadron,
and can be decomposed in light cone momenta as
\begin{align}
P_{H}^{\mu} = P_H^- \frac{n^\mu}{2} + \frac{M_H^2}{P_H^-} \frac{\bn^\mu}{2}
\,,\end{align}
where we take $P_{H}^{-} \gg P_{H}^{+} = \frac{M_H^2}{P_H^-}$
and $P_{H,\perp}=0$ sets the ``hadron frame'' for the fragmentation process.
The Wilson line $W(x)$ is defined as an anti-path ordered exponential of gauge fields
extending to positive infinity along the lightcone direction $\bn^\mu$,
\begin{align} \label{eq:def_wilson_line}
W(x) = \bar{P} \Bigl[ \exp \Bigl(
   - \img g \int_0^\infty \! \df s \, \bn \cdot A(x + \bn s)
\Bigr) \Bigr]
\,.\end{align}

After matching onto bHQET at tree-level,
the TMD FF correlator at leading order in $\lqcd / m$ becomes
\begin{align} \label{eq:ff_bhqet}
&\Delta^{\beta\beta'}_{H/Q,\, h_H h_H'}(z_H, b_\perp)
\nn \\
&\quad = \frac{m}{ z_H} \int \! \frac{\df b^+}{4\pi} \,
e^{\img b^+ ( P_H^-/z_H - m v^- )/2}
\nn \\
&\qquad \times \Bigl[
   \FH(b_\perp)
   + \ord{\as} + \ORd{\frac{1}{m}}
\Bigr]
\,,\end{align}
where $\FH$ is the bHQET bispinor,
\begin{widetext}
\begin{align} 
\label{eq:FH_hq_CG_rhoell}
\FH (b_\perp)
&\equiv \frac{1}{2 N_c} \Tr
\SumInt_{X}
\Mae{0}{W^\dagger(b_\perp) \,
   h_v^{\beta}(b_\perp) \Ket{H_v, h_H; X} 
   \Bra{H_v, h_H'; X} \bar h^{\beta'}_v(0) \,
W(0)}{0}
\\
&=\frac12 \sum_{h_{Q}h_{Q}'}\sum_{h_{\ell}h_{\ell}'}
\frac{1+\Sl{v}}{2} \bigl( \delta_{h_Q' h_Q} + 2 \gamma_5 \Sl \Sigma_{Q,h_Q' h_Q} \bigr)
\braket{s_Q, h_Q; s_\ell, h_\ell}{s_H, h_H}
\braket{s_H, h_H'}{s_Q, h_Q'; s_\ell, h_\ell'}
\rho_{\ell,h_{\ell}'h_{\ell}} (\bperp)
\,,  \nn
\end{align}
\end{widetext}
where $\rho_{\ell,h_{\ell}'h_{\ell}}$ is the light spin-density matrix defined in \eq{def_rho_ell}, 
analyzed in \sec{light_spin_density_matrix}.

We retain the open hadron helicity indices $h_{H} \neq h_{H}'$
for polarized final-state hadrons.
Decoupling the sterile heavy quark with \eq{decoupling_outstate}
and writing out the outer product of the HQET spinors
gives the second line \eq{FH_hq_CG_rhoell}.

\section{TMD FF Structures}
\label{app:ff_structures}

\begin{table*}
\centering\small
\setlength{\tabcolsep}{5pt}
\renewcommand{\arraystretch}{1.3}
\begin{tabular}{@{}llllll@{}}
\hline
$\Gamma$ & $\mathcal{O}^{[\Gamma]}_{Q}$ & $\mathcal{O}_H=\mathbf{1}$ & $\mathcal{O}_H=\tfrac32J^x$
  & $\mathcal{O}_H=\tfrac32J^y$ & $\mathcal{O}_H=\tfrac32J^z$\\[-2pt]
 & & & {\footnotesize$(S^x)$} & {\footnotesize$(S^y)$} & {\footnotesize$(S^z)$}\\
\hline
$\slashed{\bar n}$ & $\mathbf{1}_Q$ & $D_1$ & $0$ & {$-\tfrac{k_T}{M_H}D_{1T}^{\perp}$} & $0$\\
$\slashed{\bar n}\gamma_5$ & $2\Sigma_{Q,z}$ & $0$ & $+\tfrac{k_T}{M_H}G_{1T}^{\perp}$ & $0$ & $G_{1}$\\
$x_\mu\,i\sigma^{\mu\nu}\bar n_\nu\gamma_5$ & $-2\Sigma_{Q,x}$ & $0$
  & $-H_{1}-\tfrac{k_T^2}{2M_H^2}H_{1T}^{\perp}$ & $0$ & $-\tfrac{k_T}{M_H}H_{1L}^{\perp}$\\
$y_\mu\,i\sigma^{\mu\nu}\bar n_\nu\gamma_5$ & $-2\Sigma_{Q,y}$ & {$+\tfrac{k_T}{M_H}H_1^{\perp}$} & $0$
  & $-H_{1}+\tfrac{k_T^2}{2M_H^2}H_{1T}^{\perp}$ & $0$\\
\hline
\end{tabular}

\vspace{15pt}

\setlength{\tabcolsep}{3pt}
\begin{tabular}{@{}llllll@{}}
\hline
$\Gamma$ & $\mathcal{O}_H=\tfrac92Q^{zz}$ & $\mathcal{O}_H=3Q^{xz}$ & $\mathcal{O}_H=3Q^{yz}$
  & $\mathcal{O}_H=\tfrac32(Q^{xx}-Q^{yy})$ & $\mathcal{O}_H=3Q^{xy}$\\[-2pt]
 & {\footnotesize$(S_{LL})$} & {\footnotesize$(S_{LT}^x)$} & {\footnotesize$(S_{LT}^y)$}
  & {\footnotesize$(S_{TT}^{xx})$} & {\footnotesize$(S_{TT}^{xy})$}\\
\hline
$\slashed{\bar n}$ & $D_{1LL}$ & $+\tfrac{k_T}{M_H}D_{1LT}^{\perp}$ & $0$
  & $+\tfrac{k_T^2}{M_H^2}D_{1TT}^{\perp}$ & $0$\\
$\slashed{\bar n}\gamma_5$ & $0$ & $0$ & {$+\tfrac{k_T}{M_H}G_{1LT}^{\perp}$}
  & $0$ & {$-\tfrac{k_T^2}{M_H^2}G_{1TT}^{\perp}$}\\
$x_\mu\,i\sigma^{\mu\nu}\bar n_\nu\gamma_5$ & $0$ & $0$
  & {$H_{1LT}+\tfrac{k_T^2}{2M_H^2}H_{1LT}^{\perp}$} & $0$
  & {$\begin{array}[t]{@{}l@{}}-\tfrac{k_T}{M_H}H_{1TT}^{\perp} \;+\tfrac{k_T^3}{2M_H^3}H_{1TT}^{\perp\perp}\end{array}$}\\[10pt]
$y_\mu\,i\sigma^{\mu\nu}\bar n_\nu\gamma_5$ & {$-\tfrac{k_T}{M_H}H_{1LL}^{\perp}$}
  & {$-H_{1LT}+\tfrac{k_T^2}{2M_H^2}H_{1LT}^{\perp}$} & $0$
  & {$\begin{array}[t]{@{}l@{}}+\tfrac{k_T}{M_H}H_{1TT}^{\perp}\;+\tfrac{k_T^3}{2M_H^3}H_{1TT}^{\perp\perp}\end{array}$} & $0$\\[10pt]
\hline
\end{tabular}
\caption{
Spin-1 hadron in momentum space: Dirac projections and quark operators
$\mathcal{O}^{[\Gamma]}_{Q}$ (rows) versus hadron analyzers $\mathcal{O}_H$ (columns;
the polarization parameter each one extracts is shown in parentheses), with
$\hat x$ along $\hat k_\perp$, $\hat y^\mu=\epsilon_\perp^{\nu\mu}\hat x_\nu$, and $k_T=|\mathbf k_T|$. 
Analyzers are following the sum convention in \eq{hadron_analyzer_convention}, and
$Q^{ab}=\tfrac12\{J^a,J^b\}-\tfrac23\delta^{ab}$.}
\label{tab:operators_spin1}
\end{table*}

In this appendix, we collect the decompositions of the TMD FF correlator
into TMD FFs for spin-$\frac12$ and spin-$1$ hadrons,
and derive from them the analyzer tables, \tab{operators_spin1half} and \tab{operators_spin1}.
The hadron is $n$-collinear, $\sigma^{\mu \nu} = \frac{\img}{2} [\gamma^{\mu},\gamma^{\nu}]$,
and $\eps_\perp^{\mu\nu}$ is defined in \app{lightcone}.
Since we compare to the TMD handbook~\cite{Boussarie:2023izj} throughout,
we first relate the two conventions.
The TMD handbook \cite{Boussarie:2023izj} defines $\epsilon^{\mu \nu}_T = \eps^{\mu\nu\rho\sigma} n_{a, \rho} n_{b, \sigma}$,
with normalization $n_{a} \cdot n_{b} = 1 = n \cdot \bn /2$.
The hadron in the TMD handbook is $n_{a}$-collinear, just as ours is $n$-collinear,
so the two conventions are related by $n_{a} \leftrightarrow n$ and $n_{b} \leftrightarrow \bn$ (up to normalization),
and $S_{L}$ is the same in both.
However, the two lightcone vectors enter the antisymmetric tensor in the opposite order,
so that $\epsilon_{T}^{\mu\nu} = -\eps_{\perp}^{\mu\nu}$, i.e.\ $\epsilon_{T}^{12} = +1$ while $\eps_{\perp}^{12} = -1$.
Therefore, the terms with $\eps_{\perp}$ in the expressions below have the opposite sign compared to \refcite{Boussarie:2023izj}.
Our Dirac structures are formulated with $\bn$, and those of the TMD handbook with $\gamma^{+} = n_{b} \cdot \gamma$, i.e.\ in both cases with the auxiliary direction rather than the direction of the large momentum.

We obtain the momentum-space TMD FF correlator from the TMD PDF correlator of \refcite{Ebert:2021jhy} by replacing each TMD PDF with the corresponding TMD FF, keeping the hadron along the same lightcone direction.
(\refcite{Bacchetta:2006tn} instead reverses the hadron direction between PDFs and FFs, which is the origin of the differences in notation described there.)
This gives
\begin{align}\label{eq:tmdff_decomposition}
&\Delta_{H/Q}(z_{H},\kperp) 
= \Bigl\{ D_{1} 
+ D_{1T}^{\perp} \frac{\eps_{\perp}^{\rho\sigma} k_{\perp\rho}S_{\perp\sigma}}{M_{H}} 
+ G_{1} S_{L} \gamma_{5} 
\nn \\
&\quad - G_{1T}^{\perp} \frac{\kperp\cdot S_{\perp}}{M_{H}} \gamma_{5}
+ H_{1} \gamma_{5} \Sl{S}_{\perp} 
+ H_{1L}^{\perp} S_{L} \gamma_{5} \frac{\Sl{k}_{\perp}}{M_{H}}
\nn \\
&\quad -H_{1T}^{\perp} \frac{\kperp^{2}}{M_{H}^{2}} \Big(\frac{1}{2} g_{\perp}^{\mu\nu} - \frac{k_{\perp}^{\mu} k_{\perp}^{\nu}}{k_{\perp}^{2}} \Big) S_{\perp \mu}\gamma_{\nu} \gamma_{5}
\nn \\
&\quad + \img H_{1}^{\perp} \frac{\Sl{k}_{\perp}}{M_{H}}
\Bigr\}\frac{\Sl{n}}{4} 
\,.\end{align}
Note that the TMD FFs on the right hand side are all scalar functions of $k_{T}$.
We are using the decomposition $H_{1} = H_{1T} - \frac12 \frac{k_{T}^{2}}{M_{H}^{2}} H_{1T}^{\perp}$ \cite{Bacchetta:2006tn}. 

Defining $\Delta_{H/Q}^{[\Gamma]} \equiv \tr [\frac{\Gamma}{2} \Delta_{H/Q}]$,
we have projected components:
\begin{align}
\label{eq:spin1half_hadorn_TMDFFs_mom}
\Delta_{H/Q}^{[\Sl{\bn}]} &= D_{1} +  \frac{\eps_{\perp}^{\rho\sigma} k_{\perp\rho}S_{\perp\sigma}}{M_{H}} D_{1T}^{\perp}
\nn \\
\Delta_{H/Q}^{[\Sl{\bn}\gamma_{5}]} &= S_{L} G_{1} - \frac{\kperp\cdot S_{\perp}}{M_{H}} G_{1T}^{\perp} 
\nn \\
\Delta_{H/Q}^{[\img\sigma^{\mu\nu}\bn_{\nu}\gamma_{5}]} 
&= S_{\perp}^{\mu} H_{1} + S_{L}\frac{\kperp^{\mu}}{M_{H}} H_{1L}^{\perp}
+ \frac{\eps_{\perp}^{\mu\nu} k_{\perp\nu}}{M_{H}} H_{1}^{\perp}
\nn \\
&\quad + \frac{\kperp^{2}}{M_{H}^{2}} \Big( \frac{1}{2} g_{\perp}^{\mu\nu} - \frac{k_{\perp}^{\mu} k_{\perp}^{\nu}}{k_{\perp}^{2}} \Big) S_{\perp \nu}H_{1T}^{\perp}
\,.\end{align}
The above decomposition completely agrees with Eq.~(2.134) of \refcite{Boussarie:2023izj}, 
where our $k_{\perp}$ is their $p_{T}'$.

We use the Fourier transform convention in \refcite{Boussarie:2023izj,Ebert:2021jhy},
\begin{align}
&\frac{\kperp^{\mu}}{M_{H}} D(k_{T}) \to \img M_{H} \bperp^{\mu} D^{(1)}(b_{T})
\nn \\
&\frac{k_{T}^{2}}{M_{H}^{2}} \Bigl( \frac{1}{2} g_{\perp}^{\mu\nu} + \frac{\kperp^{\mu}\kperp^{\nu}}{k_{T}^{2}} \Bigr) D(k_{T})
\nn \\
&\quad \to -\frac12 M_{H}^{2} b_{T}^{2}  \Bigl( \frac{1}{2} g_{\perp}^{\mu\nu} + \frac{\bperp^{\mu}\bperp^{\nu}}{b_{T}^{2}} \Bigr) D^{(2)}(b_{T}) 
\,,\end{align}
where $D(b_T) = \int d^2\mathbf{k}_T\, e^{i\mathbf{b}_T\cdot\mathbf{k}_T} D(k_T)$ and $D^{(n)}(b_T) = n!\,\big(-\tfrac{1}{M_H^2 b_{T}}\partial_{b_T}\big)^n D(b_T)$.

This gives the position-space TMD FF correlator,
\begin{align}\label{eq:tmdff_decomposition_bT}
&\Delta_{H/Q}(z_{H},\bperp) 
= \Bigl\{ D_{1} 
+ \img  M_{H} D_{1T}^{\perp(1)} \eps_{\perp}^{\rho\sigma} b_{\perp\rho}S_{\perp\sigma}
\nn \\
&\quad + G_{1} S_{L} \gamma_{5} 
- \img M_{H}  G_{1T}^{\perp(1)} \bperp\cdot S_{\perp} \gamma_{5}
\nn \\
&\quad + H_{1} \gamma_{5} \Sl{S}_{\perp} 
+ \img M_{H} H_{1L}^{\perp(1)} S_{L} \gamma_{5} \Sl{b}_{\perp}
- M_{H} H_{1}^{\perp (1)} \Sl{b}_{\perp}
\nn \\
&\quad + \frac12 M_{H}^{2} b_{\perp}^{2} H_{1T}^{\perp(2)} \Big( \frac{1}{2} g_{\perp}^{\mu\nu} - \frac{b_{\perp}^{\mu} b_{\perp}^{\nu}}{b_{\perp}^{2}}\Big) S_{\perp \mu}\gamma_{\nu} \gamma_{5}
\Bigr\}\frac{\Sl{n}}{4} 
\,,\end{align}
which projects to the spin-1/2 TMD FFs in \eq{spin_half_tmdffs}.

\subsection{Spin-1/2 hadron analyzer table}

The complete analyzer for spin-$1/2$ hadrons is
\begin{align}
\cA_{H}^{s_{H}=\frac12} &= 
\id_{2\times2} + \sum_{a=x,y,z} \sigma^{a} S^{a} \,.
\end{align}
We start by writing the spin vector $S^\mu=S_Lz^\mu+S_\perp^\mu$ together with the convention $e\cdot S = - S_{e}$, then:
\begin{align}
S_L &= -z\cdot S = S^z
\,, \nn \\
b_\perp\!\cdot\!S_\perp &= b_T\,x\cdot S = -b_TS^x
\,, \nn \\
\epsilon_\perp^{\rho\sigma}b_{\perp\rho}S_{\perp\sigma} &= b_T\,y\cdot S = -b_TS^y
\,,\end{align}
and, for the open-index case,
$\epsilon_\perp^{\mu\nu}b_{\perp\nu}=-b_Ty^\mu$, $b_\perp\!\cdot\!x=-b_T$,
$b_\perp\!\cdot\!y=0$, $b_\perp^2=-b_T^2$, $x\cdot S_\perp=-S^x$,
$y\cdot S_\perp=-S^y$.
The last term of \eq{spin_half_tmdffs}, proportional to $H_{1T}^{\perp(2)}$, therefore contracts to
\begin{align}\label{eq:tracelesscontr}
& \left[-\tfrac12 M_H^2 b_\perp^2 \Big(\tfrac12 g_\perp^{\mu\nu} - \frac{b_\perp^\mu b_\perp^\nu}{b_\perp^2}\Big) S_{\perp\nu}\right]x_\mu  = +\tfrac14(M_Hb_T)^2S^x
\,, \nn \\
& \left[-\tfrac12 M_H^2 b_\perp^2 \Big(\tfrac12 g_\perp^{\mu\nu} - \frac{b_\perp^\mu b_\perp^\nu}{b_\perp^2}\Big) S_{\perp\nu}\right]y_\mu  = -\tfrac14(M_Hb_T)^2S^y
\,,\end{align}
i.e.\ with opposite signs in the two transverse directions, as required by tracelessness.

To derive the spin-$1/2$ hadron analyzer table in \tab{operators_spin1half}, 
we can explicitly perform the derivatives $\partial \Delta^{[\Gamma]} / \partial c_{\alpha}$
with respect to the spin parameters $c_{\alpha} = \{S^{x}, S^{y}, S^{z}\}$,
for each Dirac structure $\Gamma$ in \eq{spin_half_tmdffs}.
This is equivalent to performing the trace on the LHS of \eq{master_formula},
see the discussion in \sec{tmdff_lightchi_relations}.
This operation straightforwardly gives the entries in \tab{operators_spin1half} in the main text.

\subsection{Spin-1 hadron analyzer table}

The spin-1 hadron TMD FFs we use are~\cite{Bacchetta:2000jk} (see also~\cite{Chen:2016moq})%
\footnote{$G_{1}$ and $H_{1}$ are commonly labeled as $G_{1L}$ and $H_{1T}$ in literature for spin-1 hadrons.
They do represent the same structure as the $G_{1}$ and $H_{1}$ for spin-1/2 hadrons as in \eq{spin1half_hadorn_TMDFFs_mom},
so we keep the spin-1/2 naming convention.}
\begin{widetext}
\begin{align}
\label{eq:spin1_hadorn_TMDFFs}
\Delta^{[\slashed{\bar n}]}_{H/Q} &= D_1
 +  \frac{\epsilon_\perp^{\rho\sigma}k_{\perp\rho}S_{\perp\sigma}}{M_H}\,D_{1T}^\perp
  + S_{LL}\,D_{1LL}
  - \frac{k_\perp\!\cdot S_{LT}}{M_H}\,D_{1LT}^\perp
  + \frac{S_{TT\,\rho\sigma}\,k_\perp^{\langle\rho\sigma\rangle}}{M_H^2}\,D_{1TT}^\perp ,
\nn\\
\Delta^{[\slashed{\bar n}\gamma_5]}_{H/Q} &= S_L\,G_{1}
  - \frac{k_\perp\!\cdot S_\perp}{M_H}\,G_{1T}^\perp
  -  \frac{\epsilon_\perp^{\rho\sigma}k_{\perp\rho}S_{LT\,\sigma}}{M_H}\,G_{1LT}^\perp
  +  \frac{\epsilon_{\perp\sigma}{}^{\rho}\,S_{TT}^{\sigma\tau}\,k_{\perp\langle\rho\tau\rangle}}{M_H^2}\,G_{1TT}^\perp ,
\nn\\
\Delta^{[i\sigma^{\mu\nu}\bar n_\nu\gamma_5]}_{H/Q} &= S_\perp^\mu\,H_{1}
  + S_L\,\frac{k_\perp^\mu}{M_H}\,H_{1L}^\perp
  - \frac{k_\perp^{\langle\mu\nu\rangle}S_{\perp\nu}}{M_H^2}\,H_{1T}^\perp
  +  \frac{\epsilon_\perp^{\mu\nu}k_{\perp\nu}}{M_H}\,H_1^\perp
  -  S_{LL}\,\frac{\epsilon_\perp^{\mu\nu}k_{\perp\nu}}{M_H}\,H_{1LL}^\perp
  -  \epsilon_\perp^{\mu\nu}S_{LT\,\nu}\,H_{1LT}
\nn \\
&\quad 
  +  \frac{\epsilon_{\perp\nu}{}^{\mu}\,k_\perp^{\langle\nu\rho\rangle}S_{LT\,\rho}}{M_H^2}\,H_{1LT}^\perp
  -  \frac{\epsilon_\perp^{\mu\nu}S_{TT\,\nu\rho}\,k_\perp^\rho}{M_H}\,H_{1TT}^\perp
  - \frac{\epsilon_{\perp\nu}{}^{\mu}\,k_\perp^{\langle\nu\rho\sigma\rangle}S_{TT\,\rho\sigma}}{M_H^3}\,H_{1TT}^{\perp\perp}
  \,.
\end{align}
\end{widetext}
where the $T^{\langle\dots\rangle}$ notation is defined in \eq{traceless_tensor}.
Repeating the procedure in the previous section, 
we arrive at the analyzer dictionary \tab{operators_spin1} for vector and tensor components,
in momentum space. 
Note that the vector component table is simply \tab{operators_spin1half} (which is for spin-1/2 hadrons) in momentum space.

\section{Spin operator basis and explicit \texorpdfstring{$\rho_\ell$}{rho\_l}}
\label{app:rhoell}

In this appendix we give additional details on the final operator basis $\cO_\ell^{(N,n)}$
developed in \sec{orthogonality_and_final_operator_basis}.
In contrast to the main text, we here work with three vectors in the rest frame of $v$ to simplify possible implementations,
where $z^\mu$, $x^\mu = b_\perp^\mu/b_T$ (or $k_\perp^\mu/k_T$), and $y^\mu$ of \eqss{def_z}{def_x}{def_y}
reduce to spatial unit vectors $\hat z$, $\hat x$, $\hat y$
forming a right-handed triad (see \app{lightcone}).
Latin indices $i, j, \ldots = 1, 2, 3$ denote Euclidean components,
repeated indices are summed,
and $\Sigma_\ell^i$ are the components of the light spin operator,
with $\Sigma_{\ell,x} = \hat x^i \Sigma_\ell^i$ etc.
We suppress the subscript $\ell$ on spin operators where unambiguous.

\subsection{Harmonic tensors}

The transverse dependence enters through the harmonic tensors
$T_\perp^{(n,\pm) \, i_1 \cdots i_n}$, defined in \eqs{transverse_tensor_plus}{transverse_tensor_minus},
For $n \geq 1$ they have the convenient closed form
\vspace{-1em}
\begin{align} \label{eq:def_T_harmonic}
T_\perp^{(n,+) \, i_1 \cdots i_n}
&= 2^{1-n} \, \mathrm{Re} \bigl[ ( x + \img  y)^{i_1} \cdots ( x + \img  y)^{i_n} \bigr]
\,, \nn \\
T_\perp^{(n,-) \, i_1 \cdots i_n}
&= - 2^{1-n} \, \mathrm{Im} \bigl[ ( x + \img  y)^{i_1} \cdots ( x + \img  y)^{i_n} \bigr]
\,,\end{align}
and $T_\perp^{(0)} = 1$.
For example, $T_\perp^{(1, +) \, i} =  x^i$, $T_\perp^{(1, -) \, i} = -  y^i$,
and
\begin{align}
T_\perp^{(2, +) \, ij} &=  x^i  x^j - \tfrac12 \bigl(  x^i  x^j +  y^i  y^j \bigr)
\,, \nn \\
T_\perp^{(2, -) \, ij}  &= - \tfrac12 \bigl(  x^i  y^j +  y^i  x^j \bigr)
\,.\end{align}
\vspace{1em}

They satisfy
$T_\perp \cdot T_\perp = \tilde T_\perp \cdot \tilde T_\perp = 2^{1-n}$
and $T_\perp \cdot \tilde T_\perp = 0$,
where the dot denotes the full contraction of all indices.
The parity-odd dual can also be defined by an azimuthal derivative,
$\tilde T_\perp = - \frac{1}{n} \partial_{\phi} T_\perp$,
where $\phi$ is the azimuthal angle of $\hat x$ about $\hat z$.
An alternative closed-form expression for $T_\perp^{(n,+)}$ is~\cite{Toth:2021cpx}
\begin{widetext}
\begin{align} \label{eq:T_perp_closed_form}
T_\perp^{(n,+)\, i_1 \cdots i_n}(x)
= x^{i_1} \cdots x^{i_n}
+ \sum_{p = 1}^{\lfloor k/2 \rfloor} \frac{k! (2k-2p -2)!!}{2^p p! (k-2p)! (2k-2)!!} \,
g_\perp^{\{ \nu_1 \nu_2} \cdots g_\perp^{i_{2p-1} i_{2p}} \,
x^{i_{2p+1}} \cdots x^{i_n \}}
\,.\end{align}
For the final, three-dimensional harmonic tensors defined in \eq{def_T},
a possible closed-form expression is~\cite{Toth:2021cpx}
\begin{align} \label{eq:T_closed_form}
T^{(N,n,\pm)\,i_1 \dots i_N}
= T_{\perp}^{(n,\pm)\,\{i_1 \cdots i_n}
\sum_{p = 0}^{p_\mathrm{max}}
\frac{(2N-2p-1)!!}{2^p p! (2N-1)!!} {N-n \choose 2p}
z_{i_{n+1}} \dots z_{i_{N-2p}} \delta_{i_{N-2p+1}i_{N-2p+1}} \cdots \delta_{i_{N-1} i_N \}}
\,.\end{align}
\end{widetext}

\subsection{Orthogonal operator basis}

The final set of operators in \eq{def_cO},
including also the parity-forbidden branches
with sign $+$ ($-$) for $N$ odd (even),
forms a complete orthogonal basis on $\cH_\ell$.
To see that the orthogonality property in \eq{orthogonality} holds,
note that for even $N + N'$,
\begin{align}
\Tr_\ell \Bigl[ \cQ^{i_1 \dots i_N} \cQ^{j_1 \dots j_{N'}} \Bigr]
\,.\end{align}
is a sum over various possible contractions of the different indices.
Contractions within one tensor vanish by their defining properties,
which also implies that the trace vanishes for $N \neq N'$
(since otherwise one tensor would have to contracted within itself).
On the other hand, for $N - n \neq N' - n'$ at least one $z^i$ is contracted with a transverse index, which implies that
\begin{align}
\Tr_\ell \Bigl[ \cO_\ell^{(N', n')} \cO_\ell^{(N, n)} \Bigr]
\sim \delta_{NN'} \delta_{nn'} \, T^{(N,n,\pm)} \cdot T^{(N,n,\pm)}
\,.\end{align}
The normalization constant is given by
\begin{align} \label{eq:C_closed_form}
C_\ell^{(N,n)}
= \frac{(2 - \delta_{n0}) \, N! \, (2s_\ell + N + 1)! \, (N+n)! \, (N-n)!}
{2^{N+2n} \, (2N+1)! \, (2s_\ell - N)! \, (2N-1)!!}
\,,\end{align}
where $(-1)!! = 1$.
Explicitly, with $s \equiv s_\ell$,
\begin{align}
\cO_\ell^{(0,0)} &= \id
\,, \qquad
\cO_\ell^{(1,1)} = - \Sigma_y
\,, \nn \\
\cO_\ell^{(2,0)} &= \Sigma_z^2 - \frac{s(s+1)}{3} \, \id
\,, \nn \\
\cO_\ell^{(2,1)} &= \tfrac12 \bigl( \Sigma_x \Sigma_z + \Sigma_z \Sigma_x \bigr)
\,, \nn \\
\cO_\ell^{(2,2)} &= \tfrac12 \bigl( \Sigma_x^2 - \Sigma_y^2 \bigr)
\,, \nn \\
\cO_\ell^{(3,1)} &= - \tfrac13 \bigl( \Sigma_y \Sigma_z^2 + \Sigma_z \Sigma_y \Sigma_z + \Sigma_z^2 \Sigma_y \bigr)
\nn \\
&\quad + \frac{3s(s+1) - 1}{15} \, \Sigma_y
\,, \nn \\
\cO_\ell^{(3,2)} &= - \tfrac16 \bigl( \Sigma_x \Sigma_y \Sigma_z + \Sigma_x \Sigma_z \Sigma_y + \Sigma_y \Sigma_x \Sigma_z
\nn \\
&\qquad + \Sigma_y \Sigma_z \Sigma_x + \Sigma_z \Sigma_x \Sigma_y + \Sigma_z \Sigma_y \Sigma_x \bigr)
\,, \nn \\
\cO_\ell^{(3,3)} &= \tfrac14 \bigl( \Sigma_y^3 - \Sigma_x^2 \Sigma_y - \Sigma_x \Sigma_y \Sigma_x - \Sigma_y \Sigma_x^2 \bigr)
\,.\end{align}
At rank four we write $\Sigma_{\{abcd\}} \equiv \Sigma_{\{a} \Sigma_b \Sigma_c \Sigma_{d\}}$
for the symmetrized product of the components along $a, b, c, d \in \{x, y, z\}$,
and $c_4 \equiv [6s(s+1) - 5]/42$. Then
\begin{align}
\cO_\ell^{(4,0)} &= \Sigma_z^4 - 6 c_4 \, \Sigma_z^2 + \frac{3 s(s+1) \bigl[ s(s+1) - 2 \bigr]}{35} \, \id
\,, \nn \\
\cO_\ell^{(4,1)} &= \Sigma_{\{xzzz\}} - \tfrac32 c_4 \bigl( \Sigma_x \Sigma_z + \Sigma_z \Sigma_x \bigr)
\,, \nn \\
\cO_\ell^{(4,2)} &= \tfrac12 \bigl( \Sigma_{\{xxzz\}} - \Sigma_{\{yyzz\}} \bigr)
- \tfrac12 c_4 \bigl( \Sigma_x^2 - \Sigma_y^2 \bigr)
\,, \nn \\
\cO_\ell^{(4,3)} &= \tfrac14 \bigl( \Sigma_{\{xxxz\}} - 3 \, \Sigma_{\{xyyz\}} \bigr)
\,, \nn \\
\cO_\ell^{(4,4)} &= \tfrac18 \bigl( \Sigma_x^4 - 6 \, \Sigma_{\{xxyy\}} + \Sigma_y^4 \bigr)
\,.\end{align}

\subsection{Explicit density matrices for $s_\ell = 3/2$ and $2$}

The density matrices for $s_\ell \leq 1$ are given in \sec{orthogonality_and_final_operator_basis}.
For $s_\ell = \frac32$, inserting the operators above into \eq{result_rho_ell_cO}
the matrix elements $\langle h_\ell' | \rho_\ell | h_\ell \rangle$,
with rows labelled by $h_\ell'$ and columns by $h_\ell$,
both ordered as $\frac32, \frac12, -\frac12, -\frac32$, read
\begin{align}
\label{eq:rho_ell_32}
\rho_\ell^{s_\ell = 3/2}
= \begin{pmatrix}
\rho_{11} & \rho_{12} & \rho_{13} & \rho_{14} \\
- \rho_{12}^* & \rho_{22} & \rho_{23} & \rho_{13}^* \\
\rho_{13}^* & - \rho_{23} & \rho_{22} & \rho_{12}^* \\
- \rho_{14} & \rho_{13} & - \rho_{12} & \rho_{11}
\end{pmatrix}
\,,\end{align}
with
\begin{align}
\label{eq:rho_ell_32_terms}
\rho_{11} &= \tfrac14 \bigl( \chi_{1,\ell}^{(0,0)} + \chi_{1,\ell}^{(2,0)} \bigr)
\,, \nn \\
\rho_{22} &= \tfrac14 \bigl( \chi_{1,\ell}^{(0,0)} - \chi_{1,\ell}^{(2,0)} \bigr)
\,, \nn \\
\rho_{12} &= \frac{\sqrt3}{10} \, \chi_{1,\ell}^{(1,1)} + \frac{\sqrt3}{6} \, \chi_{1,\ell}^{(3,1)}
- \frac{\img \sqrt3}{6} \, \chi_{1,\ell}^{(2,1)}
\,, \nn \\
\rho_{23} &= \tfrac15 \, \chi_{1,\ell}^{(1,1)} - \tfrac12 \, \chi_{1,\ell}^{(3,1)}
\,, \nn \\
\rho_{13} &= - \frac{\sqrt3}{6} \, \chi_{1,\ell}^{(2,2)} - \frac{\img \sqrt3}{3} \, \chi_{1,\ell}^{(3,2)}
\,, \nn \\
\rho_{14} &= - \tfrac23 \, \chi_{1,\ell}^{(3,3)}
\,,\end{align}
where all $\chi_{1,\ell}^{(N,n)}$ are real functions of $b_T$, and $\Tr_\ell \rho_\ell = \chi_{1,\ell}^{(0,0)}$.

For $s_\ell = 2$, with the same ordering $h_\ell, h_\ell' = 2, 1, 0, -1, -2$,
the 13 structure functions enter as
\begin{align}
&\rho_\ell^{s_\ell = 2}
\nn \\
&= \begin{pmatrix}
\rho_{11} & \rho_{12} & \rho_{13} & \rho_{14} & \rho_{15} \\
- \rho_{12}^* & \rho_{22} & \rho_{23} & \rho_{24} & \rho_{14}^* \\
\rho_{13}^* & - \rho_{23}^* & \rho_{33} & \rho_{23}^* & \rho_{13}^* \\
- \rho_{14}^* & \rho_{24} & - \rho_{23} & \rho_{22} & \rho_{12}^* \\
\rho_{15} & - \rho_{14} & \rho_{13} & - \rho_{12} & \rho_{11}
\end{pmatrix}
\,,\end{align}
with
\begin{align}
\rho_{11} &= \tfrac15 \, \chi_{1,\ell}^{(0,0)} + \tfrac17 \, \chi_{1,\ell}^{(2,0)} + \tfrac{1}{24} \, \chi_{1,\ell}^{(4,0)}
\,, \nn \\
\rho_{22} &= \tfrac15 \, \chi_{1,\ell}^{(0,0)} - \tfrac{1}{14} \, \chi_{1,\ell}^{(2,0)} - \tfrac16 \, \chi_{1,\ell}^{(4,0)}
\,, \nn \\
\rho_{33} &= \tfrac15 \, \chi_{1,\ell}^{(0,0)} - \tfrac17 \, \chi_{1,\ell}^{(2,0)} + \tfrac14 \, \chi_{1,\ell}^{(4,0)}
\,, \nn \\
\rho_{12} &= \tfrac{1}{10} \, \chi_{1,\ell}^{(1,1)} + \tfrac18 \, \chi_{1,\ell}^{(3,1)}
- \img \bigl( \tfrac17 \, \chi_{1,\ell}^{(2,1)} + \tfrac{1}{12} \, \chi_{1,\ell}^{(4,1)} \bigr)
\,, \nn \\
\rho_{23} &= \sqrt6 \, \Bigl[ \tfrac{1}{20} \, \chi_{1,\ell}^{(1,1)} - \tfrac{1}{24} \, \chi_{1,\ell}^{(3,1)}
\nn \\
&\qquad - \img \bigl( \tfrac{1}{42} \, \chi_{1,\ell}^{(2,1)} - \tfrac{1}{12} \, \chi_{1,\ell}^{(4,1)} \bigr) \Bigr]
\,, \nn \\
\rho_{13} &= - \sqrt6 \, \Bigl[ \tfrac{1}{21} \, \chi_{1,\ell}^{(2,2)} + \tfrac{1}{12} \, \chi_{1,\ell}^{(4,2)}
+ \tfrac{\img}{12} \, \chi_{1,\ell}^{(3,2)} \Bigr]
\,, \nn \\
\rho_{24} &= - \tfrac17 \, \chi_{1,\ell}^{(2,2)} + \tfrac13 \, \chi_{1,\ell}^{(4,2)}
\,, \nn \\
\rho_{14} &= - \tfrac16 \, \chi_{1,\ell}^{(3,3)} + \tfrac{\img}{3} \, \chi_{1,\ell}^{(4,3)}
\,, \nn \\
\rho_{15} &= \tfrac13 \, \chi_{1,\ell}^{(4,4)}
\,.\end{align}

These matrices are given for $\hat x$ along $b_\perp$.
If instead $b_\perp$ has azimuthal angle $\phi_b$ with respect to a fixed $\hat x$ axis,
the matrix elements acquire a phase,
$\langle h_\ell' | \rho_\ell | h_\ell \rangle \to e^{- \img (h_\ell' - h_\ell) \phi_b} \langle h_\ell' | \rho_\ell | h_\ell \rangle$.

\bibliography{refs}

\begin{thebibliography}{99}%
\makeatletter
\providecommand \@ifxundefined [1]{%
 \@ifx{#1\undefined}
}%
\providecommand \@ifnum [1]{%
 \ifnum #1\expandafter \@firstoftwo
 \else \expandafter \@secondoftwo
 \fi
}%
\providecommand \@ifx [1]{%
 \ifx #1\expandafter \@firstoftwo
 \else \expandafter \@secondoftwo
 \fi
}%
\providecommand \natexlab [1]{#1}%
\providecommand \enquote  [1]{``#1''}%
\providecommand \bibnamefont  [1]{#1}%
\providecommand \bibfnamefont [1]{#1}%
\providecommand \citenamefont [1]{#1}%
\providecommand \href@noop [0]{\@secondoftwo}%
\providecommand \href [0]{\begingroup \@sanitize@url \@href}%
\providecommand \@href[1]{\@@startlink{#1}\@@href}%
\providecommand \@@href[1]{\endgroup#1\@@endlink}%
\providecommand \@sanitize@url [0]{\catcode `\\12\catcode `\$12\catcode
  `\&12\catcode `\#12\catcode `\^12\catcode `\_12\catcode `\%12\relax}%
\providecommand \@@startlink[1]{}%
\providecommand \@@endlink[0]{}%
\providecommand \url  [0]{\begingroup\@sanitize@url \@url }%
\providecommand \@url [1]{\endgroup\@href {#1}{\urlprefix }}%
\providecommand \urlprefix  [0]{URL }%
\providecommand \Eprint [0]{\href }%
\providecommand \doibase [0]{https://doi.org/}%
\providecommand \selectlanguage [0]{\@gobble}%
\providecommand \bibinfo  [0]{\@secondoftwo}%
\providecommand \bibfield  [0]{\@secondoftwo}%
\providecommand \translation [1]{[#1]}%
\providecommand \BibitemOpen [0]{}%
\providecommand \bibitemStop [0]{}%
\providecommand \bibitemNoStop [0]{.\EOS\space}%
\providecommand \EOS [0]{\spacefactor3000\relax}%
\providecommand \BibitemShut  [1]{\csname bibitem#1\endcsname}%
\let\auto@bib@innerbib\@empty
\bibitem [{\citenamefont {von Kuk}\ \emph {et~al.}(2023)\citenamefont {von
  Kuk}, \citenamefont {Michel},\ and\ \citenamefont {Sun}}]{vonKuk:2023jfd}%
  \BibitemOpen
  \bibfield  {author} {\bibinfo {author} {\bibfnamefont {R.}~\bibnamefont {von
  Kuk}}, \bibinfo {author} {\bibfnamefont {J.~K.~L.}\ \bibnamefont {Michel}},\
  and\ \bibinfo {author} {\bibfnamefont {Z.}~\bibnamefont {Sun}},\ }\bibfield
  {title} {\bibinfo {title} {{Transverse momentum distributions of heavy
  hadrons and polarized heavy quarks}},\ }\href
  {https://doi.org/10.1007/JHEP09(2023)205} {\bibfield  {journal} {\bibinfo
  {journal} {JHEP}\ }\textbf {\bibinfo {volume} {09}},\ \bibinfo {pages}
  {205}},\ \Eprint {https://arxiv.org/abs/2305.15461} {arXiv:2305.15461
  [hep-ph]} \BibitemShut {NoStop}%
\bibitem [{\citenamefont {Mele}\ and\ \citenamefont
  {Nason}(1991)}]{Mele:1990cw}%
  \BibitemOpen
  \bibfield  {author} {\bibinfo {author} {\bibfnamefont {B.}~\bibnamefont
  {Mele}}\ and\ \bibinfo {author} {\bibfnamefont {P.}~\bibnamefont {Nason}},\
  }\bibfield  {title} {\bibinfo {title} {{The Fragmentation function for heavy
  quarks in QCD}},\ }\href {https://doi.org/10.1016/0550-3213(91)90597-Q}
  {\bibfield  {journal} {\bibinfo  {journal} {Nucl. Phys. B}\ }\textbf
  {\bibinfo {volume} {361}},\ \bibinfo {pages} {626} (\bibinfo {year}
  {1991})},\ \bibinfo {note} {[Erratum: Nucl.Phys.B 921, 841--842
  (2017)]}\BibitemShut {NoStop}%
\bibitem [{\citenamefont {Jaffe}\ and\ \citenamefont
  {Randall}(1994)}]{Jaffe:1993ie}%
  \BibitemOpen
  \bibfield  {author} {\bibinfo {author} {\bibfnamefont {R.~L.}\ \bibnamefont
  {Jaffe}}\ and\ \bibinfo {author} {\bibfnamefont {L.}~\bibnamefont
  {Randall}},\ }\bibfield  {title} {\bibinfo {title} {{Heavy quark
  fragmentation into heavy mesons}},\ }\href
  {https://doi.org/10.1016/0550-3213(94)90495-2} {\bibfield  {journal}
  {\bibinfo  {journal} {Nucl. Phys. B}\ }\textbf {\bibinfo {volume} {412}},\
  \bibinfo {pages} {79} (\bibinfo {year} {1994})},\ \Eprint
  {https://arxiv.org/abs/hep-ph/9306201} {arXiv:hep-ph/9306201} \BibitemShut
  {NoStop}%
\bibitem [{\citenamefont {Neubert}(2007)}]{Neubert:2007je}%
  \BibitemOpen
  \bibfield  {author} {\bibinfo {author} {\bibfnamefont {M.}~\bibnamefont
  {Neubert}},\ }\bibfield  {title} {\bibinfo {title} {{Factorization analysis
  for the fragmentation functions of hadrons containing a heavy quark}},\
  }\href@noop {} {\  (\bibinfo {year} {2007})},\ \Eprint
  {https://arxiv.org/abs/0706.2136} {arXiv:0706.2136 [hep-ph]} \BibitemShut
  {NoStop}%
\bibitem [{\citenamefont {Fickinger}\ \emph {et~al.}(2016)\citenamefont
  {Fickinger}, \citenamefont {Fleming}, \citenamefont {Kim},\ and\
  \citenamefont {Mereghetti}}]{Fickinger:2016rfd}%
  \BibitemOpen
  \bibfield  {author} {\bibinfo {author} {\bibfnamefont {M.}~\bibnamefont
  {Fickinger}}, \bibinfo {author} {\bibfnamefont {S.}~\bibnamefont {Fleming}},
  \bibinfo {author} {\bibfnamefont {C.}~\bibnamefont {Kim}},\ and\ \bibinfo
  {author} {\bibfnamefont {E.}~\bibnamefont {Mereghetti}},\ }\bibfield  {title}
  {\bibinfo {title} {{Effective field theory approach to heavy quark
  fragmentation}},\ }\href {https://doi.org/10.1007/JHEP11(2016)095} {\bibfield
   {journal} {\bibinfo  {journal} {JHEP}\ }\textbf {\bibinfo {volume} {11}},\
  \bibinfo {pages} {095}},\ \Eprint {https://arxiv.org/abs/1606.07737}
  {arXiv:1606.07737 [hep-ph]} \BibitemShut {NoStop}%
\bibitem [{\citenamefont {Bonino}\ \emph {et~al.}(2024)\citenamefont {Bonino},
  \citenamefont {Cacciari},\ and\ \citenamefont {Stagnitto}}]{Bonino:2023icn}%
  \BibitemOpen
  \bibfield  {author} {\bibinfo {author} {\bibfnamefont {L.}~\bibnamefont
  {Bonino}}, \bibinfo {author} {\bibfnamefont {M.}~\bibnamefont {Cacciari}},\
  and\ \bibinfo {author} {\bibfnamefont {G.}~\bibnamefont {Stagnitto}},\
  }\bibfield  {title} {\bibinfo {title} {{Heavy quark fragmentation in
  e$^{+}$e$^{-}$ collisions to NNLO+NNLL accuracy in perturbative QCD}},\
  }\href {https://doi.org/10.1007/JHEP06(2024)040} {\bibfield  {journal}
  {\bibinfo  {journal} {JHEP}\ }\textbf {\bibinfo {volume} {06}},\ \bibinfo
  {pages} {040}},\ \Eprint {https://arxiv.org/abs/2312.12519} {arXiv:2312.12519
  [hep-ph]} \BibitemShut {NoStop}%
\bibitem [{\citenamefont {Isgur}\ and\ \citenamefont
  {Wise}(1989)}]{Isgur:1989vq}%
  \BibitemOpen
  \bibfield  {author} {\bibinfo {author} {\bibfnamefont {N.}~\bibnamefont
  {Isgur}}\ and\ \bibinfo {author} {\bibfnamefont {M.~B.}\ \bibnamefont
  {Wise}},\ }\bibfield  {title} {\bibinfo {title} {{Weak Decays of Heavy Mesons
  in the Static Quark Approximation}},\ }\href
  {https://doi.org/10.1016/0370-2693(89)90566-2} {\bibfield  {journal}
  {\bibinfo  {journal} {Phys. Lett. B}\ }\textbf {\bibinfo {volume} {232}},\
  \bibinfo {pages} {113} (\bibinfo {year} {1989})}\BibitemShut {NoStop}%
\bibitem [{\citenamefont {Isgur}\ and\ \citenamefont
  {Wise}(1990)}]{Isgur:1990yhj}%
  \BibitemOpen
  \bibfield  {author} {\bibinfo {author} {\bibfnamefont {N.}~\bibnamefont
  {Isgur}}\ and\ \bibinfo {author} {\bibfnamefont {M.~B.}\ \bibnamefont
  {Wise}},\ }\bibfield  {title} {\bibinfo {title} {{Weak transition form
  factors between heavy mesons}},\ }\href
  {https://doi.org/10.1016/0370-2693(90)91219-2} {\bibfield  {journal}
  {\bibinfo  {journal} {Phys. Lett. B}\ }\textbf {\bibinfo {volume} {237}},\
  \bibinfo {pages} {527} (\bibinfo {year} {1990})}\BibitemShut {NoStop}%
\bibitem [{\citenamefont {Isgur}\ and\ \citenamefont
  {Wise}(1991)}]{Isgur:1991wq}%
  \BibitemOpen
  \bibfield  {author} {\bibinfo {author} {\bibfnamefont {N.}~\bibnamefont
  {Isgur}}\ and\ \bibinfo {author} {\bibfnamefont {M.~B.}\ \bibnamefont
  {Wise}},\ }\bibfield  {title} {\bibinfo {title} {{Spectroscopy with heavy
  quark symmetry}},\ }\href {https://doi.org/10.1103/PhysRevLett.66.1130}
  {\bibfield  {journal} {\bibinfo  {journal} {Phys. Rev. Lett.}\ }\textbf
  {\bibinfo {volume} {66}},\ \bibinfo {pages} {1130} (\bibinfo {year}
  {1991})}\BibitemShut {NoStop}%
\bibitem [{\citenamefont {Neubert}(1994)}]{Neubert:1993mb}%
  \BibitemOpen
  \bibfield  {author} {\bibinfo {author} {\bibfnamefont {M.}~\bibnamefont
  {Neubert}},\ }\bibfield  {title} {\bibinfo {title} {{Heavy quark symmetry}},\
  }\href {https://doi.org/10.1016/0370-1573(94)90091-4} {\bibfield  {journal}
  {\bibinfo  {journal} {Phys. Rept.}\ }\textbf {\bibinfo {volume} {245}},\
  \bibinfo {pages} {259} (\bibinfo {year} {1994})},\ \Eprint
  {https://arxiv.org/abs/hep-ph/9306320} {arXiv:hep-ph/9306320} \BibitemShut
  {NoStop}%
\bibitem [{\citenamefont {Manohar}\ and\ \citenamefont
  {Wise}(2000)}]{Manohar:2000dt}%
  \BibitemOpen
  \bibfield  {author} {\bibinfo {author} {\bibfnamefont {A.~V.}\ \bibnamefont
  {Manohar}}\ and\ \bibinfo {author} {\bibfnamefont {M.~B.}\ \bibnamefont
  {Wise}},\ }\href {https://doi.org/10.1017/9781009402125} {\emph {\bibinfo
  {title} {{Heavy quark physics}}}},\ Vol.~\bibinfo {volume} {10}\ (\bibinfo
  {year} {2000})\BibitemShut {NoStop}%
\bibitem [{\citenamefont {Falk}\ and\ \citenamefont
  {Peskin}(1994)}]{Falk:1993rf}%
  \BibitemOpen
  \bibfield  {author} {\bibinfo {author} {\bibfnamefont {A.~F.}\ \bibnamefont
  {Falk}}\ and\ \bibinfo {author} {\bibfnamefont {M.~E.}\ \bibnamefont
  {Peskin}},\ }\bibfield  {title} {\bibinfo {title} {{Production, decay, and
  polarization of excited heavy hadrons}},\ }\href
  {https://doi.org/10.1103/PhysRevD.49.3320} {\bibfield  {journal} {\bibinfo
  {journal} {Phys. Rev. D}\ }\textbf {\bibinfo {volume} {49}},\ \bibinfo
  {pages} {3320} (\bibinfo {year} {1994})},\ \Eprint
  {https://arxiv.org/abs/hep-ph/9308241} {arXiv:hep-ph/9308241} \BibitemShut
  {NoStop}%
\bibitem [{\citenamefont {Galanti}\ \emph {et~al.}(2015)\citenamefont
  {Galanti}, \citenamefont {Giammanco}, \citenamefont {Grossman}, \citenamefont
  {Kats}, \citenamefont {Stamou},\ and\ \citenamefont
  {Zupan}}]{Galanti:2015pqa}%
  \BibitemOpen
  \bibfield  {author} {\bibinfo {author} {\bibfnamefont {M.}~\bibnamefont
  {Galanti}}, \bibinfo {author} {\bibfnamefont {A.}~\bibnamefont {Giammanco}},
  \bibinfo {author} {\bibfnamefont {Y.}~\bibnamefont {Grossman}}, \bibinfo
  {author} {\bibfnamefont {Y.}~\bibnamefont {Kats}}, \bibinfo {author}
  {\bibfnamefont {E.}~\bibnamefont {Stamou}},\ and\ \bibinfo {author}
  {\bibfnamefont {J.}~\bibnamefont {Zupan}},\ }\bibfield  {title} {\bibinfo
  {title} {{Heavy baryons as polarimeters at colliders}},\ }\href
  {https://doi.org/10.1007/JHEP11(2015)067} {\bibfield  {journal} {\bibinfo
  {journal} {JHEP}\ }\textbf {\bibinfo {volume} {11}},\ \bibinfo {pages}
  {067}},\ \Eprint {https://arxiv.org/abs/1505.02771} {arXiv:1505.02771
  [hep-ph]} \BibitemShut {NoStop}%
\bibitem [{\citenamefont {Masouminia}\ and\ \citenamefont
  {Richardson}(2024)}]{Masouminia:2023zhb}%
  \BibitemOpen
  \bibfield  {author} {\bibinfo {author} {\bibfnamefont {M.~R.}\ \bibnamefont
  {Masouminia}}\ and\ \bibinfo {author} {\bibfnamefont {P.}~\bibnamefont
  {Richardson}},\ }\bibfield  {title} {\bibinfo {title} {{Hadronization and
  decay of excited heavy hadrons in Herwig 7}},\ }\href
  {https://doi.org/10.1007/JHEP07(2024)278} {\bibfield  {journal} {\bibinfo
  {journal} {JHEP}\ }\textbf {\bibinfo {volume} {07}},\ \bibinfo {pages}
  {278}},\ \Eprint {https://arxiv.org/abs/2312.02757} {arXiv:2312.02757
  [hep-ph]} \BibitemShut {NoStop}%
\bibitem [{\citenamefont {Mulders}\ and\ \citenamefont
  {Tangerman}(1996)}]{Mulders:1995dh}%
  \BibitemOpen
  \bibfield  {author} {\bibinfo {author} {\bibfnamefont {P.~J.}\ \bibnamefont
  {Mulders}}\ and\ \bibinfo {author} {\bibfnamefont {R.~D.}\ \bibnamefont
  {Tangerman}},\ }\bibfield  {title} {\bibinfo {title} {{The Complete tree
  level result up to order 1/Q for polarized deep inelastic leptoproduction}},\
  }\href {https://doi.org/10.1016/0550-3213(95)00632-X} {\bibfield  {journal}
  {\bibinfo  {journal} {Nucl. Phys. B}\ }\textbf {\bibinfo {volume} {461}},\
  \bibinfo {pages} {197} (\bibinfo {year} {1996})},\ \bibinfo {note} {[Erratum:
  Nucl.Phys.B 484, 538--540 (1997)]},\ \Eprint
  {https://arxiv.org/abs/hep-ph/9510301} {arXiv:hep-ph/9510301} \BibitemShut
  {NoStop}%
\bibitem [{\citenamefont {Boer}\ \emph {et~al.}(1997)\citenamefont {Boer},
  \citenamefont {Jakob},\ and\ \citenamefont {Mulders}}]{Boer:1997mf}%
  \BibitemOpen
  \bibfield  {author} {\bibinfo {author} {\bibfnamefont {D.}~\bibnamefont
  {Boer}}, \bibinfo {author} {\bibfnamefont {R.}~\bibnamefont {Jakob}},\ and\
  \bibinfo {author} {\bibfnamefont {P.~J.}\ \bibnamefont {Mulders}},\
  }\bibfield  {title} {\bibinfo {title} {{Asymmetries in polarized hadron
  production in e+ e- annihilation up to order 1/Q}},\ }\href
  {https://doi.org/10.1016/S0550-3213(97)00456-2} {\bibfield  {journal}
  {\bibinfo  {journal} {Nucl. Phys. B}\ }\textbf {\bibinfo {volume} {504}},\
  \bibinfo {pages} {345} (\bibinfo {year} {1997})},\ \Eprint
  {https://arxiv.org/abs/hep-ph/9702281} {arXiv:hep-ph/9702281} \BibitemShut
  {NoStop}%
\bibitem [{\citenamefont {Bacchetta}\ and\ \citenamefont
  {Mulders}(2000)}]{Bacchetta:2000jk}%
  \BibitemOpen
  \bibfield  {author} {\bibinfo {author} {\bibfnamefont {A.}~\bibnamefont
  {Bacchetta}}\ and\ \bibinfo {author} {\bibfnamefont {P.~J.}\ \bibnamefont
  {Mulders}},\ }\bibfield  {title} {\bibinfo {title} {{Deep inelastic
  leptoproduction of spin-one hadrons}},\ }\href
  {https://doi.org/10.1103/PhysRevD.62.114004} {\bibfield  {journal} {\bibinfo
  {journal} {Phys. Rev. D}\ }\textbf {\bibinfo {volume} {62}},\ \bibinfo
  {pages} {114004} (\bibinfo {year} {2000})},\ \Eprint
  {https://arxiv.org/abs/hep-ph/0007120} {arXiv:hep-ph/0007120} \BibitemShut
  {NoStop}%
\bibitem [{\citenamefont {Chen}\ \emph {et~al.}(2016)\citenamefont {Chen},
  \citenamefont {Yang}, \citenamefont {Wei},\ and\ \citenamefont
  {Liang}}]{Chen:2016moq}%
  \BibitemOpen
  \bibfield  {author} {\bibinfo {author} {\bibfnamefont {K.-b.}\ \bibnamefont
  {Chen}}, \bibinfo {author} {\bibfnamefont {W.-h.}\ \bibnamefont {Yang}},
  \bibinfo {author} {\bibfnamefont {S.-y.}\ \bibnamefont {Wei}},\ and\ \bibinfo
  {author} {\bibfnamefont {Z.-t.}\ \bibnamefont {Liang}},\ }\bibfield  {title}
  {\bibinfo {title} {{Tensor polarization dependent fragmentation functions and
  e+e-{\textbackslash}to V {\textbackslash}pi X at high energies}},\ }\href
  {https://doi.org/10.1103/PhysRevD.94.034003} {\bibfield  {journal} {\bibinfo
  {journal} {Phys. Rev. D}\ }\textbf {\bibinfo {volume} {94}},\ \bibinfo
  {pages} {034003} (\bibinfo {year} {2016})},\ \Eprint
  {https://arxiv.org/abs/1605.07790} {arXiv:1605.07790 [hep-ph]} \BibitemShut
  {NoStop}%
\bibitem [{\citenamefont {Zhao}\ \emph {et~al.}(2022)\citenamefont {Zhao},
  \citenamefont {Zhang}, \citenamefont {Liang}, \citenamefont {Liu},\ and\
  \citenamefont {Zhou}}]{Zhao:2022lbw}%
  \BibitemOpen
  \bibfield  {author} {\bibinfo {author} {\bibfnamefont {J.}~\bibnamefont
  {Zhao}}, \bibinfo {author} {\bibfnamefont {Z.}~\bibnamefont {Zhang}},
  \bibinfo {author} {\bibfnamefont {Z.-t.}\ \bibnamefont {Liang}}, \bibinfo
  {author} {\bibfnamefont {T.}~\bibnamefont {Liu}},\ and\ \bibinfo {author}
  {\bibfnamefont {Y.-j.}\ \bibnamefont {Zhou}},\ }\bibfield  {title} {\bibinfo
  {title} {{Inclusive and semi-inclusive production of spin-3/2 hadrons in e+e-
  annihilation}},\ }\href {https://doi.org/10.1103/PhysRevD.106.094006}
  {\bibfield  {journal} {\bibinfo  {journal} {Phys. Rev. D}\ }\textbf {\bibinfo
  {volume} {106}},\ \bibinfo {pages} {094006} (\bibinfo {year} {2022})},\
  \Eprint {https://arxiv.org/abs/2206.11742} {arXiv:2206.11742 [hep-ph]}
  \BibitemShut {NoStop}%
\bibitem [{\citenamefont {Metz}\ and\ \citenamefont
  {Vossen}(2016)}]{Metz:2016swz}%
  \BibitemOpen
  \bibfield  {author} {\bibinfo {author} {\bibfnamefont {A.}~\bibnamefont
  {Metz}}\ and\ \bibinfo {author} {\bibfnamefont {A.}~\bibnamefont {Vossen}},\
  }\bibfield  {title} {\bibinfo {title} {{Parton Fragmentation Functions}},\
  }\href {https://doi.org/10.1016/j.ppnp.2016.08.003} {\bibfield  {journal}
  {\bibinfo  {journal} {Prog. Part. Nucl. Phys.}\ }\textbf {\bibinfo {volume}
  {91}},\ \bibinfo {pages} {136} (\bibinfo {year} {2016})},\ \Eprint
  {https://arxiv.org/abs/1607.02521} {arXiv:1607.02521 [hep-ex]} \BibitemShut
  {NoStop}%
\bibitem [{\citenamefont {Anselmino}\ \emph {et~al.}(2020)\citenamefont
  {Anselmino}, \citenamefont {Mukherjee},\ and\ \citenamefont
  {Vossen}}]{Anselmino:2020vlp}%
  \BibitemOpen
  \bibfield  {author} {\bibinfo {author} {\bibfnamefont {M.}~\bibnamefont
  {Anselmino}}, \bibinfo {author} {\bibfnamefont {A.}~\bibnamefont
  {Mukherjee}},\ and\ \bibinfo {author} {\bibfnamefont {A.}~\bibnamefont
  {Vossen}},\ }\bibfield  {title} {\bibinfo {title} {{Transverse spin effects
  in hard semi-inclusive collisions}},\ }\href
  {https://doi.org/10.1016/j.ppnp.2020.103806} {\bibfield  {journal} {\bibinfo
  {journal} {Prog. Part. Nucl. Phys.}\ }\textbf {\bibinfo {volume} {114}},\
  \bibinfo {pages} {103806} (\bibinfo {year} {2020})},\ \Eprint
  {https://arxiv.org/abs/2001.05415} {arXiv:2001.05415 [hep-ph]} \BibitemShut
  {NoStop}%
\bibitem [{\citenamefont {Boussarie}\ \emph {et~al.}(2023)\citenamefont
  {Boussarie} \emph {et~al.}}]{Boussarie:2023izj}%
  \BibitemOpen
  \bibfield  {author} {\bibinfo {author} {\bibfnamefont {R.}~\bibnamefont
  {Boussarie}} \emph {et~al.},\ }\bibfield  {title} {\bibinfo {title} {{TMD
  Handbook}},\ }\href@noop {} {\  (\bibinfo {year} {2023})},\ \Eprint
  {https://arxiv.org/abs/2304.03302} {arXiv:2304.03302 [hep-ph]} \BibitemShut
  {NoStop}%
\bibitem [{\citenamefont {Chen}\ \emph {et~al.}(2023)\citenamefont {Chen},
  \citenamefont {Liu}, \citenamefont {Song},\ and\ \citenamefont
  {Wei}}]{Chen:2023kqw}%
  \BibitemOpen
  \bibfield  {author} {\bibinfo {author} {\bibfnamefont {K.-B.}\ \bibnamefont
  {Chen}}, \bibinfo {author} {\bibfnamefont {T.}~\bibnamefont {Liu}}, \bibinfo
  {author} {\bibfnamefont {Y.-K.}\ \bibnamefont {Song}},\ and\ \bibinfo
  {author} {\bibfnamefont {S.-Y.}\ \bibnamefont {Wei}},\ }\bibfield  {title}
  {\bibinfo {title} {{Several Topics on Transverse Momentum-Dependent
  Fragmentation Functions}},\ }\href {https://doi.org/10.3390/particles6020029}
  {\bibfield  {journal} {\bibinfo  {journal} {Particles}\ }\textbf {\bibinfo
  {volume} {6}},\ \bibinfo {pages} {515} (\bibinfo {year} {2023})},\ \Eprint
  {https://arxiv.org/abs/2307.02874} {arXiv:2307.02874 [hep-ph]} \BibitemShut
  {NoStop}%
\bibitem [{\citenamefont {Kumano}(2024)}]{Kumano:2024fpr}%
  \BibitemOpen
  \bibfield  {author} {\bibinfo {author} {\bibfnamefont {S.}~\bibnamefont
  {Kumano}},\ }\bibfield  {title} {\bibinfo {title} {{Parton distribution
  functions and fragmentation functions of spin-1 hadrons}},\ }\href
  {https://doi.org/10.1140/epja/s10050-024-01411-6} {\bibfield  {journal}
  {\bibinfo  {journal} {Eur. Phys. J. A}\ }\textbf {\bibinfo {volume} {60}},\
  \bibinfo {pages} {205} (\bibinfo {year} {2024})},\ \Eprint
  {https://arxiv.org/abs/2406.01180} {arXiv:2406.01180 [hep-ph]} \BibitemShut
  {NoStop}%
\bibitem [{\citenamefont {Accardi}\ \emph {et~al.}(2016)\citenamefont {Accardi}
  \emph {et~al.}}]{Accardi:2012qut}%
  \BibitemOpen
  \bibfield  {author} {\bibinfo {author} {\bibfnamefont {A.}~\bibnamefont
  {Accardi}} \emph {et~al.},\ }\bibfield  {title} {\bibinfo {title} {{Electron
  Ion Collider: The Next QCD Frontier}: {Understanding the glue that binds us
  all}},\ }\href {https://doi.org/10.1140/epja/i2016-16268-9} {\bibfield
  {journal} {\bibinfo  {journal} {Eur. Phys. J. A}\ }\textbf {\bibinfo {volume}
  {52}},\ \bibinfo {pages} {268} (\bibinfo {year} {2016})},\ \Eprint
  {https://arxiv.org/abs/1212.1701} {arXiv:1212.1701 [nucl-ex]} \BibitemShut
  {NoStop}%
\bibitem [{\citenamefont {Abdul~Khalek}\ \emph {et~al.}(2022)\citenamefont
  {Abdul~Khalek} \emph {et~al.}}]{AbdulKhalek:2021gbh}%
  \BibitemOpen
  \bibfield  {author} {\bibinfo {author} {\bibfnamefont {R.}~\bibnamefont
  {Abdul~Khalek}} \emph {et~al.},\ }\bibfield  {title} {\bibinfo {title}
  {{Science Requirements and Detector Concepts for the Electron-Ion Collider}:
  {EIC Yellow Report}},\ }\href
  {https://doi.org/10.1016/j.nuclphysa.2022.122447} {\bibfield  {journal}
  {\bibinfo  {journal} {Nucl. Phys. A}\ }\textbf {\bibinfo {volume} {1026}},\
  \bibinfo {pages} {122447} (\bibinfo {year} {2022})},\ \Eprint
  {https://arxiv.org/abs/2103.05419} {arXiv:2103.05419 [physics.ins-det]}
  \BibitemShut {NoStop}%
\bibitem [{\citenamefont {Bacchetta}\ \emph {et~al.}(2022)\citenamefont
  {Bacchetta}, \citenamefont {Bertone}, \citenamefont {Bissolotti},
  \citenamefont {Bozzi}, \citenamefont {Cerutti}, \citenamefont {Piacenza},
  \citenamefont {Radici},\ and\ \citenamefont {Signori}}]{Bacchetta:2022awv}%
  \BibitemOpen
  \bibfield  {author} {\bibinfo {author} {\bibfnamefont {A.}~\bibnamefont
  {Bacchetta}}, \bibinfo {author} {\bibfnamefont {V.}~\bibnamefont {Bertone}},
  \bibinfo {author} {\bibfnamefont {C.}~\bibnamefont {Bissolotti}}, \bibinfo
  {author} {\bibfnamefont {G.}~\bibnamefont {Bozzi}}, \bibinfo {author}
  {\bibfnamefont {M.}~\bibnamefont {Cerutti}}, \bibinfo {author} {\bibfnamefont
  {F.}~\bibnamefont {Piacenza}}, \bibinfo {author} {\bibfnamefont
  {M.}~\bibnamefont {Radici}},\ and\ \bibinfo {author} {\bibfnamefont
  {A.}~\bibnamefont {Signori}} (\bibinfo {collaboration} {MAP
  (Multi-dimensional Analyses of Partonic distributions)}),\ }\bibfield
  {title} {\bibinfo {title} {{Unpolarized transverse momentum distributions
  from a global fit of Drell-Yan and semi-inclusive deep-inelastic scattering
  data}},\ }\href {https://doi.org/10.1007/JHEP10(2022)127} {\bibfield
  {journal} {\bibinfo  {journal} {JHEP}\ }\textbf {\bibinfo {volume} {10}},\
  \bibinfo {pages} {127}},\ \Eprint {https://arxiv.org/abs/2206.07598}
  {arXiv:2206.07598 [hep-ph]} \BibitemShut {NoStop}%
\bibitem [{\citenamefont {Moos}\ \emph {et~al.}(2024)\citenamefont {Moos},
  \citenamefont {Scimemi}, \citenamefont {Vladimirov},\ and\ \citenamefont
  {Zurita}}]{Moos:2023yfa}%
  \BibitemOpen
  \bibfield  {author} {\bibinfo {author} {\bibfnamefont {V.}~\bibnamefont
  {Moos}}, \bibinfo {author} {\bibfnamefont {I.}~\bibnamefont {Scimemi}},
  \bibinfo {author} {\bibfnamefont {A.}~\bibnamefont {Vladimirov}},\ and\
  \bibinfo {author} {\bibfnamefont {P.}~\bibnamefont {Zurita}},\ }\bibfield
  {title} {\bibinfo {title} {{Extraction of unpolarized transverse momentum
  distributions from the fit of Drell-Yan data at N$^{4}$LL}},\ }\href
  {https://doi.org/10.1007/JHEP05(2024)036} {\bibfield  {journal} {\bibinfo
  {journal} {JHEP}\ }\textbf {\bibinfo {volume} {05}},\ \bibinfo {pages}
  {036}},\ \Eprint {https://arxiv.org/abs/2305.07473} {arXiv:2305.07473
  [hep-ph]} \BibitemShut {NoStop}%
\bibitem [{\citenamefont {Bacchetta}\ \emph {et~al.}(2024)\citenamefont
  {Bacchetta}, \citenamefont {Bertone}, \citenamefont {Bissolotti},
  \citenamefont {Bozzi}, \citenamefont {Cerutti}, \citenamefont {Delcarro},
  \citenamefont {Radici}, \citenamefont {Rossi},\ and\ \citenamefont
  {Signori}}]{Bacchetta:2024qre}%
  \BibitemOpen
  \bibfield  {author} {\bibinfo {author} {\bibfnamefont {A.}~\bibnamefont
  {Bacchetta}}, \bibinfo {author} {\bibfnamefont {V.}~\bibnamefont {Bertone}},
  \bibinfo {author} {\bibfnamefont {C.}~\bibnamefont {Bissolotti}}, \bibinfo
  {author} {\bibfnamefont {G.}~\bibnamefont {Bozzi}}, \bibinfo {author}
  {\bibfnamefont {M.}~\bibnamefont {Cerutti}}, \bibinfo {author} {\bibfnamefont
  {F.}~\bibnamefont {Delcarro}}, \bibinfo {author} {\bibfnamefont
  {M.}~\bibnamefont {Radici}}, \bibinfo {author} {\bibfnamefont
  {L.}~\bibnamefont {Rossi}},\ and\ \bibinfo {author} {\bibfnamefont
  {A.}~\bibnamefont {Signori}} (\bibinfo {collaboration} {MAP
  (Multi-dimensional Analyses of Partonic distributions)}),\ }\bibfield
  {title} {\bibinfo {title} {{Flavor dependence of unpolarized quark transverse
  momentum distributions from a global fit}},\ }\href
  {https://doi.org/10.1007/JHEP08(2024)232} {\bibfield  {journal} {\bibinfo
  {journal} {JHEP}\ }\textbf {\bibinfo {volume} {08}},\ \bibinfo {pages}
  {232}},\ \Eprint {https://arxiv.org/abs/2405.13833} {arXiv:2405.13833
  [hep-ph]} \BibitemShut {NoStop}%
\bibitem [{\citenamefont {Anselmino}\ \emph {et~al.}(2013)\citenamefont
  {Anselmino}, \citenamefont {Boglione}, \citenamefont {D'Alesio},
  \citenamefont {Melis}, \citenamefont {Murgia},\ and\ \citenamefont
  {Prokudin}}]{Anselmino:2013vqa}%
  \BibitemOpen
  \bibfield  {author} {\bibinfo {author} {\bibfnamefont {M.}~\bibnamefont
  {Anselmino}}, \bibinfo {author} {\bibfnamefont {M.}~\bibnamefont {Boglione}},
  \bibinfo {author} {\bibfnamefont {U.}~\bibnamefont {D'Alesio}}, \bibinfo
  {author} {\bibfnamefont {S.}~\bibnamefont {Melis}}, \bibinfo {author}
  {\bibfnamefont {F.}~\bibnamefont {Murgia}},\ and\ \bibinfo {author}
  {\bibfnamefont {A.}~\bibnamefont {Prokudin}},\ }\bibfield  {title} {\bibinfo
  {title} {{Simultaneous extraction of transversity and Collins functions from
  new SIDIS and e+e- data}},\ }\href
  {https://doi.org/10.1103/PhysRevD.87.094019} {\bibfield  {journal} {\bibinfo
  {journal} {Phys. Rev. D}\ }\textbf {\bibinfo {volume} {87}},\ \bibinfo
  {pages} {094019} (\bibinfo {year} {2013})},\ \Eprint
  {https://arxiv.org/abs/1303.3822} {arXiv:1303.3822 [hep-ph]} \BibitemShut
  {NoStop}%
\bibitem [{\citenamefont {Kang}\ \emph {et~al.}(2016)\citenamefont {Kang},
  \citenamefont {Prokudin}, \citenamefont {Sun},\ and\ \citenamefont
  {Yuan}}]{Kang:2015msa}%
  \BibitemOpen
  \bibfield  {author} {\bibinfo {author} {\bibfnamefont {Z.-B.}\ \bibnamefont
  {Kang}}, \bibinfo {author} {\bibfnamefont {A.}~\bibnamefont {Prokudin}},
  \bibinfo {author} {\bibfnamefont {P.}~\bibnamefont {Sun}},\ and\ \bibinfo
  {author} {\bibfnamefont {F.}~\bibnamefont {Yuan}},\ }\bibfield  {title}
  {\bibinfo {title} {{Extraction of Quark Transversity Distribution and Collins
  Fragmentation Functions with QCD Evolution}},\ }\href
  {https://doi.org/10.1103/PhysRevD.93.014009} {\bibfield  {journal} {\bibinfo
  {journal} {Phys. Rev. D}\ }\textbf {\bibinfo {volume} {93}},\ \bibinfo
  {pages} {014009} (\bibinfo {year} {2016})},\ \Eprint
  {https://arxiv.org/abs/1505.05589} {arXiv:1505.05589 [hep-ph]} \BibitemShut
  {NoStop}%
\bibitem [{\citenamefont {Cammarota}\ \emph {et~al.}(2020)\citenamefont
  {Cammarota}, \citenamefont {Gamberg}, \citenamefont {Kang}, \citenamefont
  {Miller}, \citenamefont {Pitonyak}, \citenamefont {Prokudin}, \citenamefont
  {Rogers},\ and\ \citenamefont {Sato}}]{Cammarota:2020qcw}%
  \BibitemOpen
  \bibfield  {author} {\bibinfo {author} {\bibfnamefont {J.}~\bibnamefont
  {Cammarota}}, \bibinfo {author} {\bibfnamefont {L.}~\bibnamefont {Gamberg}},
  \bibinfo {author} {\bibfnamefont {Z.-B.}\ \bibnamefont {Kang}}, \bibinfo
  {author} {\bibfnamefont {J.~A.}\ \bibnamefont {Miller}}, \bibinfo {author}
  {\bibfnamefont {D.}~\bibnamefont {Pitonyak}}, \bibinfo {author}
  {\bibfnamefont {A.}~\bibnamefont {Prokudin}}, \bibinfo {author}
  {\bibfnamefont {T.~C.}\ \bibnamefont {Rogers}},\ and\ \bibinfo {author}
  {\bibfnamefont {N.}~\bibnamefont {Sato}} (\bibinfo {collaboration} {Jefferson
  Lab Angular Momentum}),\ }\bibfield  {title} {\bibinfo {title} {{Origin of
  single transverse-spin asymmetries in high-energy collisions}},\ }\href
  {https://doi.org/10.1103/PhysRevD.102.054002} {\bibfield  {journal} {\bibinfo
   {journal} {Phys. Rev. D}\ }\textbf {\bibinfo {volume} {102}},\ \bibinfo
  {pages} {054002} (\bibinfo {year} {2020})},\ \Eprint
  {https://arxiv.org/abs/2002.08384} {arXiv:2002.08384 [hep-ph]} \BibitemShut
  {NoStop}%
\bibitem [{\citenamefont {Gamberg}\ \emph {et~al.}(2022)\citenamefont
  {Gamberg}, \citenamefont {Malda}, \citenamefont {Miller}, \citenamefont
  {Pitonyak}, \citenamefont {Prokudin},\ and\ \citenamefont
  {Sato}}]{Gamberg:2022kdb}%
  \BibitemOpen
  \bibfield  {author} {\bibinfo {author} {\bibfnamefont {L.}~\bibnamefont
  {Gamberg}}, \bibinfo {author} {\bibfnamefont {M.}~\bibnamefont {Malda}},
  \bibinfo {author} {\bibfnamefont {J.~A.}\ \bibnamefont {Miller}}, \bibinfo
  {author} {\bibfnamefont {D.}~\bibnamefont {Pitonyak}}, \bibinfo {author}
  {\bibfnamefont {A.}~\bibnamefont {Prokudin}},\ and\ \bibinfo {author}
  {\bibfnamefont {N.}~\bibnamefont {Sato}} (\bibinfo {collaboration} {Jefferson
  Lab Angular Momentum}),\ }\bibfield  {title} {\bibinfo {title} {{Updated QCD
  global analysis of single transverse-spin asymmetries: Extracting
  H{\textasciitilde}, and the role of the Soffer bound and lattice QCD}},\
  }\href {https://doi.org/10.1103/PhysRevD.106.034014} {\bibfield  {journal}
  {\bibinfo  {journal} {Phys. Rev. D}\ }\textbf {\bibinfo {volume} {106}},\
  \bibinfo {pages} {034014} (\bibinfo {year} {2022})},\ \Eprint
  {https://arxiv.org/abs/2205.00999} {arXiv:2205.00999 [hep-ph]} \BibitemShut
  {NoStop}%
\bibitem [{\citenamefont {Bunce}\ \emph {et~al.}(1976)\citenamefont {Bunce}
  \emph {et~al.}}]{Bunce:1976yb}%
  \BibitemOpen
  \bibfield  {author} {\bibinfo {author} {\bibfnamefont {G.}~\bibnamefont
  {Bunce}} \emph {et~al.},\ }\bibfield  {title} {\bibinfo {title} {{Lambda0
  Hyperon Polarization in Inclusive Production by 300-GeV Protons on
  Beryllium.}},\ }\href {https://doi.org/10.1103/PhysRevLett.36.1113}
  {\bibfield  {journal} {\bibinfo  {journal} {Phys. Rev. Lett.}\ }\textbf
  {\bibinfo {volume} {36}},\ \bibinfo {pages} {1113} (\bibinfo {year}
  {1976})}\BibitemShut {NoStop}%
\bibitem [{\citenamefont {Anselmino}\ \emph {et~al.}(2001)\citenamefont
  {Anselmino}, \citenamefont {Boer}, \citenamefont {D'Alesio},\ and\
  \citenamefont {Murgia}}]{Anselmino:2000vs}%
  \BibitemOpen
  \bibfield  {author} {\bibinfo {author} {\bibfnamefont {M.}~\bibnamefont
  {Anselmino}}, \bibinfo {author} {\bibfnamefont {D.}~\bibnamefont {Boer}},
  \bibinfo {author} {\bibfnamefont {U.}~\bibnamefont {D'Alesio}},\ and\
  \bibinfo {author} {\bibfnamefont {F.}~\bibnamefont {Murgia}},\ }\bibfield
  {title} {\bibinfo {title} {{Lambda polarization from unpolarized quark
  fragmentation}},\ }\href {https://doi.org/10.1103/PhysRevD.63.054029}
  {\bibfield  {journal} {\bibinfo  {journal} {Phys. Rev. D}\ }\textbf {\bibinfo
  {volume} {63}},\ \bibinfo {pages} {054029} (\bibinfo {year} {2001})},\
  \Eprint {https://arxiv.org/abs/hep-ph/0008186} {arXiv:hep-ph/0008186}
  \BibitemShut {NoStop}%
\bibitem [{\citenamefont {Boer}\ \emph {et~al.}(2010)\citenamefont {Boer},
  \citenamefont {Kang}, \citenamefont {Vogelsang},\ and\ \citenamefont
  {Yuan}}]{Boer:2010ya}%
  \BibitemOpen
  \bibfield  {author} {\bibinfo {author} {\bibfnamefont {D.}~\bibnamefont
  {Boer}}, \bibinfo {author} {\bibfnamefont {Z.-B.}\ \bibnamefont {Kang}},
  \bibinfo {author} {\bibfnamefont {W.}~\bibnamefont {Vogelsang}},\ and\
  \bibinfo {author} {\bibfnamefont {F.}~\bibnamefont {Yuan}},\ }\bibfield
  {title} {\bibinfo {title} {{Test of the Universality of
  Naive-time-reversal-odd Fragmentation Functions}},\ }\href
  {https://doi.org/10.1103/PhysRevLett.105.202001} {\bibfield  {journal}
  {\bibinfo  {journal} {Phys. Rev. Lett.}\ }\textbf {\bibinfo {volume} {105}},\
  \bibinfo {pages} {202001} (\bibinfo {year} {2010})},\ \Eprint
  {https://arxiv.org/abs/1008.3543} {arXiv:1008.3543 [hep-ph]} \BibitemShut
  {NoStop}%
\bibitem [{\citenamefont {Guan}\ \emph {et~al.}(2019)\citenamefont {Guan} \emph
  {et~al.}}]{Belle:2018ttu}%
  \BibitemOpen
  \bibfield  {author} {\bibinfo {author} {\bibfnamefont {Y.}~\bibnamefont
  {Guan}} \emph {et~al.} (\bibinfo {collaboration} {Belle}),\ }\bibfield
  {title} {\bibinfo {title} {{Observation of Transverse $\Lambda/\bar{\Lambda}$
  Hyperon Polarization in $e^+e^-$ Annihilation at Belle}},\ }\href
  {https://doi.org/10.1103/PhysRevLett.122.042001} {\bibfield  {journal}
  {\bibinfo  {journal} {Phys. Rev. Lett.}\ }\textbf {\bibinfo {volume} {122}},\
  \bibinfo {pages} {042001} (\bibinfo {year} {2019})},\ \Eprint
  {https://arxiv.org/abs/1808.05000} {arXiv:1808.05000 [hep-ex]} \BibitemShut
  {NoStop}%
\bibitem [{\citenamefont {Anselmino}\ \emph {et~al.}(2019)\citenamefont
  {Anselmino}, \citenamefont {Kishore},\ and\ \citenamefont
  {Mukherjee}}]{Anselmino:2019cqd}%
  \BibitemOpen
  \bibfield  {author} {\bibinfo {author} {\bibfnamefont {M.}~\bibnamefont
  {Anselmino}}, \bibinfo {author} {\bibfnamefont {R.}~\bibnamefont {Kishore}},\
  and\ \bibinfo {author} {\bibfnamefont {A.}~\bibnamefont {Mukherjee}},\
  }\bibfield  {title} {\bibinfo {title} {{Polarizing fragmentation function and
  the {\ensuremath{\Lambda}} polarization in $e^+e^-$ processes}},\ }\href
  {https://doi.org/10.1103/PhysRevD.100.014029} {\bibfield  {journal} {\bibinfo
   {journal} {Phys. Rev. D}\ }\textbf {\bibinfo {volume} {100}},\ \bibinfo
  {pages} {014029} (\bibinfo {year} {2019})},\ \Eprint
  {https://arxiv.org/abs/1905.02777} {arXiv:1905.02777 [hep-ph]} \BibitemShut
  {NoStop}%
\bibitem [{\citenamefont {D'Alesio}\ \emph {et~al.}(2020)\citenamefont
  {D'Alesio}, \citenamefont {Murgia},\ and\ \citenamefont
  {Zaccheddu}}]{DAlesio:2020wjq}%
  \BibitemOpen
  \bibfield  {author} {\bibinfo {author} {\bibfnamefont {U.}~\bibnamefont
  {D'Alesio}}, \bibinfo {author} {\bibfnamefont {F.}~\bibnamefont {Murgia}},\
  and\ \bibinfo {author} {\bibfnamefont {M.}~\bibnamefont {Zaccheddu}},\
  }\bibfield  {title} {\bibinfo {title} {{First extraction of the $\Lambda$
  polarizing fragmentation function from Belle $e^+e^-$ data}},\ }\href
  {https://doi.org/10.1103/PhysRevD.102.054001} {\bibfield  {journal} {\bibinfo
   {journal} {Phys. Rev. D}\ }\textbf {\bibinfo {volume} {102}},\ \bibinfo
  {pages} {054001} (\bibinfo {year} {2020})},\ \Eprint
  {https://arxiv.org/abs/2003.01128} {arXiv:2003.01128 [hep-ph]} \BibitemShut
  {NoStop}%
\bibitem [{\citenamefont {Callos}\ \emph {et~al.}(2020)\citenamefont {Callos},
  \citenamefont {Kang},\ and\ \citenamefont {Terry}}]{Callos:2020qtu}%
  \BibitemOpen
  \bibfield  {author} {\bibinfo {author} {\bibfnamefont {D.}~\bibnamefont
  {Callos}}, \bibinfo {author} {\bibfnamefont {Z.-B.}\ \bibnamefont {Kang}},\
  and\ \bibinfo {author} {\bibfnamefont {J.}~\bibnamefont {Terry}},\ }\bibfield
   {title} {\bibinfo {title} {{Extracting the transverse momentum dependent
  polarizing fragmentation functions}},\ }\href
  {https://doi.org/10.1103/PhysRevD.102.096007} {\bibfield  {journal} {\bibinfo
   {journal} {Phys. Rev. D}\ }\textbf {\bibinfo {volume} {102}},\ \bibinfo
  {pages} {096007} (\bibinfo {year} {2020})},\ \Eprint
  {https://arxiv.org/abs/2003.04828} {arXiv:2003.04828 [hep-ph]} \BibitemShut
  {NoStop}%
\bibitem [{\citenamefont {Li}\ \emph {et~al.}(2021)\citenamefont {Li},
  \citenamefont {Wang}, \citenamefont {Yang},\ and\ \citenamefont
  {Lu}}]{Li:2020oto}%
  \BibitemOpen
  \bibfield  {author} {\bibinfo {author} {\bibfnamefont {H.}~\bibnamefont
  {Li}}, \bibinfo {author} {\bibfnamefont {X.}~\bibnamefont {Wang}}, \bibinfo
  {author} {\bibfnamefont {Y.}~\bibnamefont {Yang}},\ and\ \bibinfo {author}
  {\bibfnamefont {Z.}~\bibnamefont {Lu}},\ }\bibfield  {title} {\bibinfo
  {title} {{The transverse polarization of $\Lambda $ hyperons in
  $e^+e^-\rightarrow \Lambda ^\uparrow h X$ processes within TMD
  factorization}},\ }\href {https://doi.org/10.1140/epjc/s10052-021-09064-1}
  {\bibfield  {journal} {\bibinfo  {journal} {Eur. Phys. J. C}\ }\textbf
  {\bibinfo {volume} {81}},\ \bibinfo {pages} {289} (\bibinfo {year} {2021})},\
  \Eprint {https://arxiv.org/abs/2009.07193} {arXiv:2009.07193 [hep-ph]}
  \BibitemShut {NoStop}%
\bibitem [{\citenamefont {Chen}\ \emph {et~al.}(2021)\citenamefont {Chen},
  \citenamefont {Liang}, \citenamefont {Pan}, \citenamefont {Song},\ and\
  \citenamefont {Wei}}]{Chen:2021hdn}%
  \BibitemOpen
  \bibfield  {author} {\bibinfo {author} {\bibfnamefont {K.-B.}\ \bibnamefont
  {Chen}}, \bibinfo {author} {\bibfnamefont {Z.-T.}\ \bibnamefont {Liang}},
  \bibinfo {author} {\bibfnamefont {Y.-L.}\ \bibnamefont {Pan}}, \bibinfo
  {author} {\bibfnamefont {Y.-K.}\ \bibnamefont {Song}},\ and\ \bibinfo
  {author} {\bibfnamefont {S.-Y.}\ \bibnamefont {Wei}},\ }\bibfield  {title}
  {\bibinfo {title} {{Isospin Symmetry of Fragmentation Functions}},\ }\href
  {https://doi.org/10.1016/j.physletb.2021.136217} {\bibfield  {journal}
  {\bibinfo  {journal} {Phys. Lett. B}\ }\textbf {\bibinfo {volume} {816}},\
  \bibinfo {pages} {136217} (\bibinfo {year} {2021})},\ \Eprint
  {https://arxiv.org/abs/2102.00658} {arXiv:2102.00658 [hep-ph]} \BibitemShut
  {NoStop}%
\bibitem [{\citenamefont {Gamberg}\ \emph {et~al.}(2021)\citenamefont
  {Gamberg}, \citenamefont {Kang}, \citenamefont {Shao}, \citenamefont
  {Terry},\ and\ \citenamefont {Zhao}}]{Gamberg:2021iat}%
  \BibitemOpen
  \bibfield  {author} {\bibinfo {author} {\bibfnamefont {L.}~\bibnamefont
  {Gamberg}}, \bibinfo {author} {\bibfnamefont {Z.-B.}\ \bibnamefont {Kang}},
  \bibinfo {author} {\bibfnamefont {D.~Y.}\ \bibnamefont {Shao}}, \bibinfo
  {author} {\bibfnamefont {J.}~\bibnamefont {Terry}},\ and\ \bibinfo {author}
  {\bibfnamefont {F.}~\bibnamefont {Zhao}},\ }\bibfield  {title} {\bibinfo
  {title} {{Transverse $\Lambda$ polarization in $e^+ e^-$ collisions}},\
  }\href {https://doi.org/10.1016/j.physletb.2021.136371} {\bibfield  {journal}
  {\bibinfo  {journal} {Phys. Lett. B}\ }\textbf {\bibinfo {volume} {818}},\
  \bibinfo {pages} {136371} (\bibinfo {year} {2021})},\ \Eprint
  {https://arxiv.org/abs/2102.05553} {arXiv:2102.05553 [hep-ph]} \BibitemShut
  {NoStop}%
\bibitem [{\citenamefont {D'Alesio}\ \emph {et~al.}(2022)\citenamefont
  {D'Alesio}, \citenamefont {Gamberg}, \citenamefont {Murgia},\ and\
  \citenamefont {Zaccheddu}}]{DAlesio:2022brl}%
  \BibitemOpen
  \bibfield  {author} {\bibinfo {author} {\bibfnamefont {U.}~\bibnamefont
  {D'Alesio}}, \bibinfo {author} {\bibfnamefont {L.}~\bibnamefont {Gamberg}},
  \bibinfo {author} {\bibfnamefont {F.}~\bibnamefont {Murgia}},\ and\ \bibinfo
  {author} {\bibfnamefont {M.}~\bibnamefont {Zaccheddu}},\ }\bibfield  {title}
  {\bibinfo {title} {{Transverse {\ensuremath{\Lambda}} polarization in
  e$^{+}$e$^{-}$ processes within a TMD factorization approach and the
  polarizing fragmentation function}},\ }\href
  {https://doi.org/10.1007/JHEP12(2022)074} {\bibfield  {journal} {\bibinfo
  {journal} {JHEP}\ }\textbf {\bibinfo {volume} {12}},\ \bibinfo {pages}
  {074}},\ \Eprint {https://arxiv.org/abs/2209.11670} {arXiv:2209.11670
  [hep-ph]} \BibitemShut {NoStop}%
\bibitem [{\citenamefont {D'Alesio}\ \emph {et~al.}(2023)\citenamefont
  {D'Alesio}, \citenamefont {Gamberg}, \citenamefont {Murgia},\ and\
  \citenamefont {Zaccheddu}}]{DAlesio:2023ozw}%
  \BibitemOpen
  \bibfield  {author} {\bibinfo {author} {\bibfnamefont {U.}~\bibnamefont
  {D'Alesio}}, \bibinfo {author} {\bibfnamefont {L.}~\bibnamefont {Gamberg}},
  \bibinfo {author} {\bibfnamefont {F.}~\bibnamefont {Murgia}},\ and\ \bibinfo
  {author} {\bibfnamefont {M.}~\bibnamefont {Zaccheddu}},\ }\bibfield  {title}
  {\bibinfo {title} {{Transverse {\ensuremath{\Lambda}} polarization in e+e-
  annihilations and in SIDIS processes at the EIC within TMD factorization}},\
  }\href {https://doi.org/10.1103/PhysRevD.108.094004} {\bibfield  {journal}
  {\bibinfo  {journal} {Phys. Rev. D}\ }\textbf {\bibinfo {volume} {108}},\
  \bibinfo {pages} {094004} (\bibinfo {year} {2023})},\ \Eprint
  {https://arxiv.org/abs/2307.02359} {arXiv:2307.02359 [hep-ph]} \BibitemShut
  {NoStop}%
\bibitem [{\citenamefont {Aboona}\ \emph {et~al.}(2026)\citenamefont {Aboona}
  \emph {et~al.}}]{STAR:2025jwc}%
  \BibitemOpen
  \bibfield  {author} {\bibinfo {author} {\bibfnamefont {B.~E.}\ \bibnamefont
  {Aboona}} \emph {et~al.} (\bibinfo {collaboration} {STAR}),\ }\bibfield
  {title} {\bibinfo {title} {{Measurement of transverse polarization of
  {\ensuremath{\Lambda}} and $\overline{\Lambda }$ hyperons inside jets in
  unpolarized proton-proton collisions at $\sqrt{s}=200$ GeV}},\ }\href
  {https://doi.org/10.1007/JHEP06(2026)050} {\bibfield  {journal} {\bibinfo
  {journal} {JHEP}\ }\textbf {\bibinfo {volume} {06}},\ \bibinfo {pages}
  {050}},\ \Eprint {https://arxiv.org/abs/2509.17487} {arXiv:2509.17487
  [hep-ex]} \BibitemShut {NoStop}%
\bibitem [{\citenamefont {Kang}\ \emph {et~al.}(2020)\citenamefont {Kang},
  \citenamefont {Lee},\ and\ \citenamefont {Zhao}}]{Kang:2020xyq}%
  \BibitemOpen
  \bibfield  {author} {\bibinfo {author} {\bibfnamefont {Z.-B.}\ \bibnamefont
  {Kang}}, \bibinfo {author} {\bibfnamefont {K.}~\bibnamefont {Lee}},\ and\
  \bibinfo {author} {\bibfnamefont {F.}~\bibnamefont {Zhao}},\ }\bibfield
  {title} {\bibinfo {title} {{Polarized jet fragmentation functions}},\ }\href
  {https://doi.org/10.1016/j.physletb.2020.135756} {\bibfield  {journal}
  {\bibinfo  {journal} {Phys. Lett. B}\ }\textbf {\bibinfo {volume} {809}},\
  \bibinfo {pages} {135756} (\bibinfo {year} {2020})},\ \Eprint
  {https://arxiv.org/abs/2005.02398} {arXiv:2005.02398 [hep-ph]} \BibitemShut
  {NoStop}%
\bibitem [{\citenamefont {Kang}\ \emph {et~al.}(2021)\citenamefont {Kang},
  \citenamefont {Lee}, \citenamefont {Shao},\ and\ \citenamefont
  {Zhao}}]{Kang:2021ffh}%
  \BibitemOpen
  \bibfield  {author} {\bibinfo {author} {\bibfnamefont {Z.-B.}\ \bibnamefont
  {Kang}}, \bibinfo {author} {\bibfnamefont {K.}~\bibnamefont {Lee}}, \bibinfo
  {author} {\bibfnamefont {D.~Y.}\ \bibnamefont {Shao}},\ and\ \bibinfo
  {author} {\bibfnamefont {F.}~\bibnamefont {Zhao}},\ }\bibfield  {title}
  {\bibinfo {title} {{Spin asymmetries in electron-jet production at the future
  electron ion collider}},\ }\href {https://doi.org/10.1007/JHEP11(2021)005}
  {\bibfield  {journal} {\bibinfo  {journal} {JHEP}\ }\textbf {\bibinfo
  {volume} {11}},\ \bibinfo {pages} {005}},\ \Eprint
  {https://arxiv.org/abs/2106.15624} {arXiv:2106.15624 [hep-ph]} \BibitemShut
  {NoStop}%
\bibitem [{\citenamefont {Kang}\ \emph {et~al.}(2022)\citenamefont {Kang},
  \citenamefont {Terry}, \citenamefont {Vossen}, \citenamefont {Xu},\ and\
  \citenamefont {Zhang}}]{Kang:2021kpt}%
  \BibitemOpen
  \bibfield  {author} {\bibinfo {author} {\bibfnamefont {Z.-B.}\ \bibnamefont
  {Kang}}, \bibinfo {author} {\bibfnamefont {J.}~\bibnamefont {Terry}},
  \bibinfo {author} {\bibfnamefont {A.}~\bibnamefont {Vossen}}, \bibinfo
  {author} {\bibfnamefont {Q.}~\bibnamefont {Xu}},\ and\ \bibinfo {author}
  {\bibfnamefont {J.}~\bibnamefont {Zhang}},\ }\bibfield  {title} {\bibinfo
  {title} {{Transverse Lambda production at the future Electron-Ion
  Collider}},\ }\href {https://doi.org/10.1103/PhysRevD.105.094033} {\bibfield
  {journal} {\bibinfo  {journal} {Phys. Rev. D}\ }\textbf {\bibinfo {volume}
  {105}},\ \bibinfo {pages} {094033} (\bibinfo {year} {2022})},\ \Eprint
  {https://arxiv.org/abs/2108.05383} {arXiv:2108.05383 [hep-ph]} \BibitemShut
  {NoStop}%
\bibitem [{\citenamefont {Chen}\ \emph {et~al.}(2022)\citenamefont {Chen},
  \citenamefont {Liang}, \citenamefont {Song},\ and\ \citenamefont
  {Wei}}]{Chen:2021zrr}%
  \BibitemOpen
  \bibfield  {author} {\bibinfo {author} {\bibfnamefont {K.-b.}\ \bibnamefont
  {Chen}}, \bibinfo {author} {\bibfnamefont {Z.-t.}\ \bibnamefont {Liang}},
  \bibinfo {author} {\bibfnamefont {Y.-k.}\ \bibnamefont {Song}},\ and\
  \bibinfo {author} {\bibfnamefont {S.-y.}\ \bibnamefont {Wei}},\ }\bibfield
  {title} {\bibinfo {title} {{Longitudinal and transverse polarizations of
  {\ensuremath{\Lambda}} hyperon in unpolarized SIDIS and e+e- annihilation}},\
  }\href {https://doi.org/10.1103/PhysRevD.105.034027} {\bibfield  {journal}
  {\bibinfo  {journal} {Phys. Rev. D}\ }\textbf {\bibinfo {volume} {105}},\
  \bibinfo {pages} {034027} (\bibinfo {year} {2022})},\ \Eprint
  {https://arxiv.org/abs/2108.07740} {arXiv:2108.07740 [hep-ph]} \BibitemShut
  {NoStop}%
\bibitem [{\citenamefont {Gao}\ \emph {et~al.}(2024)\citenamefont {Gao},
  \citenamefont {Chen}, \citenamefont {Song},\ and\ \citenamefont
  {Wei}}]{Gao:2024bfp}%
  \BibitemOpen
  \bibfield  {author} {\bibinfo {author} {\bibfnamefont {Y.}~\bibnamefont
  {Gao}}, \bibinfo {author} {\bibfnamefont {K.-B.}\ \bibnamefont {Chen}},
  \bibinfo {author} {\bibfnamefont {Y.-K.}\ \bibnamefont {Song}},\ and\
  \bibinfo {author} {\bibfnamefont {S.-Y.}\ \bibnamefont {Wei}},\ }\bibfield
  {title} {\bibinfo {title} {{Transverse polarization of Lambda hyperons in
  hadronic collisions}},\ }\href
  {https://doi.org/10.1016/j.physletb.2024.139026} {\bibfield  {journal}
  {\bibinfo  {journal} {Phys. Lett. B}\ }\textbf {\bibinfo {volume} {858}},\
  \bibinfo {pages} {139026} (\bibinfo {year} {2024})},\ \Eprint
  {https://arxiv.org/abs/2403.06133} {arXiv:2403.06133 [hep-ph]} \BibitemShut
  {NoStop}%
\bibitem [{\citenamefont {Yang}\ \emph {et~al.}(2025)\citenamefont {Yang},
  \citenamefont {Song},\ and\ \citenamefont {Wei}}]{Yang:2024kjn}%
  \BibitemOpen
  \bibfield  {author} {\bibinfo {author} {\bibfnamefont {L.}~\bibnamefont
  {Yang}}, \bibinfo {author} {\bibfnamefont {Y.-K.}\ \bibnamefont {Song}},\
  and\ \bibinfo {author} {\bibfnamefont {S.-Y.}\ \bibnamefont {Wei}},\
  }\bibfield  {title} {\bibinfo {title} {{Transverse spin correlation of
  back-to-back dihadrons in unpolarized collisions}},\ }\href
  {https://doi.org/10.1103/PhysRevD.111.054035} {\bibfield  {journal} {\bibinfo
   {journal} {Phys. Rev. D}\ }\textbf {\bibinfo {volume} {111}},\ \bibinfo
  {pages} {054035} (\bibinfo {year} {2025})},\ \Eprint
  {https://arxiv.org/abs/2410.20917} {arXiv:2410.20917 [hep-ph]} \BibitemShut
  {NoStop}%
\bibitem [{\citenamefont {Buskulic}\ \emph
  {et~al.}(1996{\natexlab{a}})\citenamefont {Buskulic} \emph
  {et~al.}}]{ALEPH:1996oew}%
  \BibitemOpen
  \bibfield  {author} {\bibinfo {author} {\bibfnamefont {D.}~\bibnamefont
  {Buskulic}} \emph {et~al.} (\bibinfo {collaboration} {ALEPH}),\ }\bibfield
  {title} {\bibinfo {title} {{Measurement of Lambda polarization from Z
  decays}},\ }\href {https://doi.org/10.1016/0370-2693(96)00300-0} {\bibfield
  {journal} {\bibinfo  {journal} {Phys. Lett. B}\ }\textbf {\bibinfo {volume}
  {374}},\ \bibinfo {pages} {319} (\bibinfo {year}
  {1996}{\natexlab{a}})}\BibitemShut {NoStop}%
\bibitem [{\citenamefont {Ackerstaff}\ \emph {et~al.}(1998)\citenamefont
  {Ackerstaff} \emph {et~al.}}]{OPAL:1997oem}%
  \BibitemOpen
  \bibfield  {author} {\bibinfo {author} {\bibfnamefont {K.}~\bibnamefont
  {Ackerstaff}} \emph {et~al.} (\bibinfo {collaboration} {OPAL}),\ }\bibfield
  {title} {\bibinfo {title} {{Polarization and forward - backward asymmetry of
  Lambda baryons in hadronic Z0 decays}},\ }\href
  {https://doi.org/10.1007/s100520050123} {\bibfield  {journal} {\bibinfo
  {journal} {Eur. Phys. J. C}\ }\textbf {\bibinfo {volume} {2}},\ \bibinfo
  {pages} {49} (\bibinfo {year} {1998})},\ \Eprint
  {https://arxiv.org/abs/hep-ex/9708027} {arXiv:hep-ex/9708027} \BibitemShut
  {NoStop}%
\bibitem [{\citenamefont {Ackerstaff}\ \emph
  {et~al.}(1997{\natexlab{a}})\citenamefont {Ackerstaff} \emph
  {et~al.}}]{OPAL:1997vmw}%
  \BibitemOpen
  \bibfield  {author} {\bibinfo {author} {\bibfnamefont {K.}~\bibnamefont
  {Ackerstaff}} \emph {et~al.} (\bibinfo {collaboration} {OPAL}),\ }\bibfield
  {title} {\bibinfo {title} {{Spin alignment of leading K*(892)0 mesons in
  hadronic Z0 decays}},\ }\href {https://doi.org/10.1016/S0370-2693(97)01077-0}
  {\bibfield  {journal} {\bibinfo  {journal} {Phys. Lett. B}\ }\textbf
  {\bibinfo {volume} {412}},\ \bibinfo {pages} {210} (\bibinfo {year}
  {1997}{\natexlab{a}})},\ \Eprint {https://arxiv.org/abs/hep-ex/9708022}
  {arXiv:hep-ex/9708022} \BibitemShut {NoStop}%
\bibitem [{\citenamefont {Chen}\ \emph {et~al.}(2020)\citenamefont {Chen},
  \citenamefont {Liang}, \citenamefont {Song},\ and\ \citenamefont
  {Wei}}]{Chen:2020pty}%
  \BibitemOpen
  \bibfield  {author} {\bibinfo {author} {\bibfnamefont {K.-b.}\ \bibnamefont
  {Chen}}, \bibinfo {author} {\bibfnamefont {Z.-t.}\ \bibnamefont {Liang}},
  \bibinfo {author} {\bibfnamefont {Y.-k.}\ \bibnamefont {Song}},\ and\
  \bibinfo {author} {\bibfnamefont {S.-y.}\ \bibnamefont {Wei}},\ }\bibfield
  {title} {\bibinfo {title} {{Spin alignment of vector mesons in high energy
  $pp$ collisions}},\ }\href {https://doi.org/10.1103/PhysRevD.102.034001}
  {\bibfield  {journal} {\bibinfo  {journal} {Phys. Rev. D}\ }\textbf {\bibinfo
  {volume} {102}},\ \bibinfo {pages} {034001} (\bibinfo {year} {2020})},\
  \Eprint {https://arxiv.org/abs/2002.09890} {arXiv:2002.09890 [hep-ph]}
  \BibitemShut {NoStop}%
\bibitem [{\citenamefont {Dai}\ \emph {et~al.}(2024)\citenamefont {Dai},
  \citenamefont {Kim},\ and\ \citenamefont {Leibovich}}]{Dai:2023rvd}%
  \BibitemOpen
  \bibfield  {author} {\bibinfo {author} {\bibfnamefont {L.}~\bibnamefont
  {Dai}}, \bibinfo {author} {\bibfnamefont {C.}~\bibnamefont {Kim}},\ and\
  \bibinfo {author} {\bibfnamefont {A.~K.}\ \bibnamefont {Leibovich}},\
  }\bibfield  {title} {\bibinfo {title} {{Heavy quark transverse momentum
  dependent fragmentation}},\ }\href {https://doi.org/10.1007/JHEP05(2024)002}
  {\bibfield  {journal} {\bibinfo  {journal} {JHEP}\ }\textbf {\bibinfo
  {volume} {05}},\ \bibinfo {pages} {002}},\ \Eprint
  {https://arxiv.org/abs/2310.19207} {arXiv:2310.19207 [hep-ph]} \BibitemShut
  {NoStop}%
\bibitem [{\citenamefont {Fleming}\ \emph
  {et~al.}(2008{\natexlab{a}})\citenamefont {Fleming}, \citenamefont {Hoang},
  \citenamefont {Mantry},\ and\ \citenamefont {Stewart}}]{Fleming:2007qr}%
  \BibitemOpen
  \bibfield  {author} {\bibinfo {author} {\bibfnamefont {S.}~\bibnamefont
  {Fleming}}, \bibinfo {author} {\bibfnamefont {A.~H.}\ \bibnamefont {Hoang}},
  \bibinfo {author} {\bibfnamefont {S.}~\bibnamefont {Mantry}},\ and\ \bibinfo
  {author} {\bibfnamefont {I.~W.}\ \bibnamefont {Stewart}},\ }\bibfield
  {title} {\bibinfo {title} {{Jets from massive unstable particles: Top-mass
  determination}},\ }\href {https://doi.org/10.1103/PhysRevD.77.074010}
  {\bibfield  {journal} {\bibinfo  {journal} {Phys. Rev. D}\ }\textbf {\bibinfo
  {volume} {77}},\ \bibinfo {pages} {074010} (\bibinfo {year}
  {2008}{\natexlab{a}})},\ \Eprint {https://arxiv.org/abs/hep-ph/0703207}
  {arXiv:hep-ph/0703207} \BibitemShut {NoStop}%
\bibitem [{\citenamefont {Fleming}\ \emph
  {et~al.}(2008{\natexlab{b}})\citenamefont {Fleming}, \citenamefont {Hoang},
  \citenamefont {Mantry},\ and\ \citenamefont {Stewart}}]{Fleming:2007xt}%
  \BibitemOpen
  \bibfield  {author} {\bibinfo {author} {\bibfnamefont {S.}~\bibnamefont
  {Fleming}}, \bibinfo {author} {\bibfnamefont {A.~H.}\ \bibnamefont {Hoang}},
  \bibinfo {author} {\bibfnamefont {S.}~\bibnamefont {Mantry}},\ and\ \bibinfo
  {author} {\bibfnamefont {I.~W.}\ \bibnamefont {Stewart}},\ }\bibfield
  {title} {\bibinfo {title} {{Top Jets in the Peak Region: Factorization
  Analysis with NLL Resummation}},\ }\href
  {https://doi.org/10.1103/PhysRevD.77.114003} {\bibfield  {journal} {\bibinfo
  {journal} {Phys. Rev. D}\ }\textbf {\bibinfo {volume} {77}},\ \bibinfo
  {pages} {114003} (\bibinfo {year} {2008}{\natexlab{b}})},\ \Eprint
  {https://arxiv.org/abs/0711.2079} {arXiv:0711.2079 [hep-ph]} \BibitemShut
  {NoStop}%
\bibitem [{\citenamefont {von Kuk}\ \emph {et~al.}(2024)\citenamefont {von
  Kuk}, \citenamefont {Michel},\ and\ \citenamefont {Sun}}]{vonKuk:2024uxe}%
  \BibitemOpen
  \bibfield  {author} {\bibinfo {author} {\bibfnamefont {R.}~\bibnamefont {von
  Kuk}}, \bibinfo {author} {\bibfnamefont {J.~K.~L.}\ \bibnamefont {Michel}},\
  and\ \bibinfo {author} {\bibfnamefont {Z.}~\bibnamefont {Sun}},\ }\bibfield
  {title} {\bibinfo {title} {{Transverse momentum-dependent heavy-quark
  fragmentation at next-to-leading order}},\ }\href
  {https://doi.org/10.1007/JHEP07(2024)129} {\bibfield  {journal} {\bibinfo
  {journal} {JHEP}\ }\textbf {\bibinfo {volume} {07}},\ \bibinfo {pages}
  {129}},\ \Eprint {https://arxiv.org/abs/2404.08622} {arXiv:2404.08622
  [hep-ph]} \BibitemShut {NoStop}%
\bibitem [{\citenamefont {von Kuk}\ \emph {et~al.}(2025)\citenamefont {von
  Kuk}, \citenamefont {Lee}, \citenamefont {Michel},\ and\ \citenamefont
  {Sun}}]{vonKuk:2025kbv}%
  \BibitemOpen
  \bibfield  {author} {\bibinfo {author} {\bibfnamefont {R.}~\bibnamefont {von
  Kuk}}, \bibinfo {author} {\bibfnamefont {K.}~\bibnamefont {Lee}}, \bibinfo
  {author} {\bibfnamefont {J.~K.~L.}\ \bibnamefont {Michel}},\ and\ \bibinfo
  {author} {\bibfnamefont {Z.}~\bibnamefont {Sun}},\ }\bibfield  {title}
  {\bibinfo {title} {{Towards a Quantum Information Theory of Hadronization:
  Dihadron Fragmentation and Neutral Polarization in Heavy Baryons}},\
  }\href@noop {} {\  (\bibinfo {year} {2025})},\ \Eprint
  {https://arxiv.org/abs/2503.22607} {arXiv:2503.22607 [hep-ph]} \BibitemShut
  {NoStop}%
\bibitem [{\citenamefont {Aad}\ \emph {et~al.}(2021)\citenamefont {Aad} \emph
  {et~al.}}]{ATLAS:2021agf}%
  \BibitemOpen
  \bibfield  {author} {\bibinfo {author} {\bibfnamefont {G.}~\bibnamefont
  {Aad}} \emph {et~al.} (\bibinfo {collaboration} {ATLAS}),\ }\bibfield
  {title} {\bibinfo {title} {{Measurement of $b$-quark fragmentation properties
  in jets using the decay $B^{\pm} \to J/\psi K^{\pm}$ in $pp$ collisions at $
  \sqrt{s} $ = 13 TeV with the ATLAS detector}},\ }\href
  {https://doi.org/10.1007/JHEP12(2021)131} {\bibfield  {journal} {\bibinfo
  {journal} {JHEP}\ }\textbf {\bibinfo {volume} {12}},\ \bibinfo {pages}
  {131}},\ \Eprint {https://arxiv.org/abs/2108.11650} {arXiv:2108.11650
  [hep-ex]} \BibitemShut {NoStop}%
\bibitem [{\citenamefont {Aaij}\ \emph {et~al.}(2026)\citenamefont {Aaij} \emph
  {et~al.}}]{LHCb:2026mbl}%
  \BibitemOpen
  \bibfield  {author} {\bibinfo {author} {\bibfnamefont {R.}~\bibnamefont
  {Aaij}} \emph {et~al.} (\bibinfo {collaboration} {LHCb}),\ }\bibfield
  {title} {\bibinfo {title} {{B-jet fragmentation with
  B{\ensuremath{\pm}}{\textrightarrow}J/{\ensuremath{\psi}}K{\ensuremath{\pm}}
  decays in s=13{\,}{\,}TeV pp collisions at LHCb}},\ }\href
  {https://doi.org/10.1103/3d7r-c3y4} {\bibfield  {journal} {\bibinfo
  {journal} {Phys. Rev. D}\ }\textbf {\bibinfo {volume} {113}},\ \bibinfo
  {pages} {112006} (\bibinfo {year} {2026})},\ \Eprint
  {https://arxiv.org/abs/2603.23663} {arXiv:2603.23663 [hep-ex]} \BibitemShut
  {NoStop}%
\bibitem [{\citenamefont {Acharya}\ \emph {et~al.}(2019)\citenamefont {Acharya}
  \emph {et~al.}}]{ALICE:2019cbr}%
  \BibitemOpen
  \bibfield  {author} {\bibinfo {author} {\bibfnamefont {S.}~\bibnamefont
  {Acharya}} \emph {et~al.} (\bibinfo {collaboration} {ALICE}),\ }\bibfield
  {title} {\bibinfo {title} {{Measurement of the production of charm jets
  tagged with D$^{0}$ mesons in pp collisions at $ \sqrt{\mathrm{s}}=7 $
  TeV}},\ }\href {https://doi.org/10.1007/JHEP08(2019)133} {\bibfield
  {journal} {\bibinfo  {journal} {JHEP}\ }\textbf {\bibinfo {volume} {08}},\
  \bibinfo {pages} {133}},\ \Eprint {https://arxiv.org/abs/1905.02510}
  {arXiv:1905.02510 [nucl-ex]} \BibitemShut {NoStop}%
\bibitem [{\citenamefont {Acharya}\ \emph
  {et~al.}(2023{\natexlab{a}})\citenamefont {Acharya} \emph
  {et~al.}}]{ALICE:2022mur}%
  \BibitemOpen
  \bibfield  {author} {\bibinfo {author} {\bibfnamefont {S.}~\bibnamefont
  {Acharya}} \emph {et~al.} (\bibinfo {collaboration} {ALICE}),\ }\bibfield
  {title} {\bibinfo {title} {{Measurement of the production of charm jets
  tagged with D$^{0}$ mesons in pp collisions at $ \sqrt{s} $ = 5.02 and 13
  TeV}},\ }\href {https://doi.org/10.1007/JHEP06(2023)133} {\bibfield
  {journal} {\bibinfo  {journal} {JHEP}\ }\textbf {\bibinfo {volume} {06}},\
  \bibinfo {pages} {133}},\ \Eprint {https://arxiv.org/abs/2204.10167}
  {arXiv:2204.10167 [nucl-ex]} \BibitemShut {NoStop}%
\bibitem [{\citenamefont {Acharya}\ \emph {et~al.}(2024)\citenamefont {Acharya}
  \emph {et~al.}}]{ALICE:2023jgm}%
  \BibitemOpen
  \bibfield  {author} {\bibinfo {author} {\bibfnamefont {S.}~\bibnamefont
  {Acharya}} \emph {et~al.} (\bibinfo {collaboration} {ALICE}),\ }\bibfield
  {title} {\bibinfo {title} {{Measurement of the fraction of jet longitudinal
  momentum carried by {\ensuremath{\Lambda}}c+ baryons in pp collisions}},\
  }\href {https://doi.org/10.1103/PhysRevD.109.072005} {\bibfield  {journal}
  {\bibinfo  {journal} {Phys. Rev. D}\ }\textbf {\bibinfo {volume} {109}},\
  \bibinfo {pages} {072005} (\bibinfo {year} {2024})},\ \Eprint
  {https://arxiv.org/abs/2301.13798} {arXiv:2301.13798 [nucl-ex]} \BibitemShut
  {NoStop}%
\bibitem [{\citenamefont {Sirunyan}\ \emph {et~al.}(2020)\citenamefont
  {Sirunyan} \emph {et~al.}}]{CMS:2019jis}%
  \BibitemOpen
  \bibfield  {author} {\bibinfo {author} {\bibfnamefont {A.~M.}\ \bibnamefont
  {Sirunyan}} \emph {et~al.} (\bibinfo {collaboration} {CMS}),\ }\bibfield
  {title} {\bibinfo {title} {{Studies of charm quark diffusion inside jets
  using PbPb and pp collisions at $\sqrt{s_\mathrm{NN}} =$ 5.02 TeV}},\ }\href
  {https://doi.org/10.1103/PhysRevLett.125.102001} {\bibfield  {journal}
  {\bibinfo  {journal} {Phys. Rev. Lett.}\ }\textbf {\bibinfo {volume} {125}},\
  \bibinfo {pages} {102001} (\bibinfo {year} {2020})},\ \Eprint
  {https://arxiv.org/abs/1911.01461} {arXiv:1911.01461 [hep-ex]} \BibitemShut
  {NoStop}%
\bibitem [{\citenamefont {Buskulic}\ \emph
  {et~al.}(1996{\natexlab{b}})\citenamefont {Buskulic} \emph
  {et~al.}}]{ALEPH:1995aqx}%
  \BibitemOpen
  \bibfield  {author} {\bibinfo {author} {\bibfnamefont {D.}~\bibnamefont
  {Buskulic}} \emph {et~al.} (\bibinfo {collaboration} {ALEPH}),\ }\bibfield
  {title} {\bibinfo {title} {{Measurement of Lambda(b) polarization in Z
  decays}},\ }\href {https://doi.org/10.1016/0370-2693(95)01433-0} {\bibfield
  {journal} {\bibinfo  {journal} {Phys. Lett. B}\ }\textbf {\bibinfo {volume}
  {365}},\ \bibinfo {pages} {437} (\bibinfo {year}
  {1996}{\natexlab{b}})}\BibitemShut {NoStop}%
\bibitem [{\citenamefont {Abbiendi}\ \emph {et~al.}(1998)\citenamefont
  {Abbiendi} \emph {et~al.}}]{OPAL:1998wmk}%
  \BibitemOpen
  \bibfield  {author} {\bibinfo {author} {\bibfnamefont {G.}~\bibnamefont
  {Abbiendi}} \emph {et~al.} (\bibinfo {collaboration} {OPAL}),\ }\bibfield
  {title} {\bibinfo {title} {{Measurement of the average polarization of b
  baryons in hadronic Z0 decays}},\ }\href
  {https://doi.org/10.1016/S0370-2693(98)01387-2} {\bibfield  {journal}
  {\bibinfo  {journal} {Phys. Lett. B}\ }\textbf {\bibinfo {volume} {444}},\
  \bibinfo {pages} {539} (\bibinfo {year} {1998})},\ \Eprint
  {https://arxiv.org/abs/hep-ex/9808006} {arXiv:hep-ex/9808006} \BibitemShut
  {NoStop}%
\bibitem [{\citenamefont {Abreu}\ \emph {et~al.}(2000)\citenamefont {Abreu}
  \emph {et~al.}}]{DELPHI:1999hkl}%
  \BibitemOpen
  \bibfield  {author} {\bibinfo {author} {\bibfnamefont {P.}~\bibnamefont
  {Abreu}} \emph {et~al.} (\bibinfo {collaboration} {DELPHI}),\ }\bibfield
  {title} {\bibinfo {title} {{Lambda(b) polarization in Z0 decays at LEP}},\
  }\href {https://doi.org/10.1016/S0370-2693(99)01431-8} {\bibfield  {journal}
  {\bibinfo  {journal} {Phys. Lett. B}\ }\textbf {\bibinfo {volume} {474}},\
  \bibinfo {pages} {205} (\bibinfo {year} {2000})}\BibitemShut {NoStop}%
\bibitem [{\citenamefont {Aaij}\ \emph {et~al.}(2013)\citenamefont {Aaij} \emph
  {et~al.}}]{LHCb:2013hzx}%
  \BibitemOpen
  \bibfield  {author} {\bibinfo {author} {\bibfnamefont {R.}~\bibnamefont
  {Aaij}} \emph {et~al.} (\bibinfo {collaboration} {LHCb}),\ }\bibfield
  {title} {\bibinfo {title} {{Measurements of the $\Lambda_b^0 \to J/\psi
  \Lambda$ decay amplitudes and the $\Lambda_b^0$ polarisation in $pp$
  collisions at $\sqrt{s} = 7$ TeV}},\ }\href
  {https://doi.org/10.1016/j.physletb.2013.05.041} {\bibfield  {journal}
  {\bibinfo  {journal} {Phys. Lett. B}\ }\textbf {\bibinfo {volume} {724}},\
  \bibinfo {pages} {27} (\bibinfo {year} {2013})},\ \Eprint
  {https://arxiv.org/abs/1302.5578} {arXiv:1302.5578 [hep-ex]} \BibitemShut
  {NoStop}%
\bibitem [{\citenamefont {Aad}\ \emph {et~al.}(2014)\citenamefont {Aad} \emph
  {et~al.}}]{ATLAS:2014swk}%
  \BibitemOpen
  \bibfield  {author} {\bibinfo {author} {\bibfnamefont {G.}~\bibnamefont
  {Aad}} \emph {et~al.} (\bibinfo {collaboration} {ATLAS}),\ }\bibfield
  {title} {\bibinfo {title} {{Measurement of the parity-violating asymmetry
  parameter $\alpha_b$ and the helicity amplitudes for the decay
  $\Lambda_b^0\to J/\psi+\Lambda^0$ with the ATLAS detector}},\ }\href
  {https://doi.org/10.1103/PhysRevD.89.092009} {\bibfield  {journal} {\bibinfo
  {journal} {Phys. Rev. D}\ }\textbf {\bibinfo {volume} {89}},\ \bibinfo
  {pages} {092009} (\bibinfo {year} {2014})},\ \Eprint
  {https://arxiv.org/abs/1404.1071} {arXiv:1404.1071 [hep-ex]} \BibitemShut
  {NoStop}%
\bibitem [{\citenamefont {Sirunyan}\ \emph {et~al.}(2018)\citenamefont
  {Sirunyan} \emph {et~al.}}]{CMS:2018wjk}%
  \BibitemOpen
  \bibfield  {author} {\bibinfo {author} {\bibfnamefont {A.~M.}\ \bibnamefont
  {Sirunyan}} \emph {et~al.} (\bibinfo {collaboration} {CMS}),\ }\bibfield
  {title} {\bibinfo {title} {{Measurement of the $\Lambda_b$ polarization and
  angular parameters in $\Lambda_b\to J/\psi\, \Lambda$ decays from pp
  collisions at $\sqrt{s}=$ 7 and 8 TeV}},\ }\href
  {https://doi.org/10.1103/PhysRevD.97.072010} {\bibfield  {journal} {\bibinfo
  {journal} {Phys. Rev. D}\ }\textbf {\bibinfo {volume} {97}},\ \bibinfo
  {pages} {072010} (\bibinfo {year} {2018})},\ \Eprint
  {https://arxiv.org/abs/1802.04867} {arXiv:1802.04867 [hep-ex]} \BibitemShut
  {NoStop}%
\bibitem [{\citenamefont {Aaij}\ \emph {et~al.}(2020)\citenamefont {Aaij} \emph
  {et~al.}}]{LHCb:2020iux}%
  \BibitemOpen
  \bibfield  {author} {\bibinfo {author} {\bibfnamefont {R.}~\bibnamefont
  {Aaij}} \emph {et~al.} (\bibinfo {collaboration} {LHCb}),\ }\bibfield
  {title} {\bibinfo {title} {{Measurement of the $\Lambda^0_b\rightarrow
  J/\psi\Lambda$ angular distribution and the $\Lambda^0_b$ polarisation in
  $pp$ collisions}},\ }\href {https://doi.org/10.1007/JHEP06(2020)110}
  {\bibfield  {journal} {\bibinfo  {journal} {JHEP}\ }\textbf {\bibinfo
  {volume} {06}},\ \bibinfo {pages} {110}},\ \Eprint
  {https://arxiv.org/abs/2004.10563} {arXiv:2004.10563 [hep-ex]} \BibitemShut
  {NoStop}%
\bibitem [{\citenamefont {Aaij}\ \emph
  {et~al.}(2023{\natexlab{a}})\citenamefont {Aaij} \emph
  {et~al.}}]{LHCb:2022sck}%
  \BibitemOpen
  \bibfield  {author} {\bibinfo {author} {\bibfnamefont {R.}~\bibnamefont
  {Aaij}} \emph {et~al.} (\bibinfo {collaboration} {LHCb}),\ }\bibfield
  {title} {\bibinfo {title} {{Amplitude analysis of the
  {\ensuremath{\Lambda}}c+{\textrightarrow}pK-{\ensuremath{\pi}}+ decay and
  {\ensuremath{\Lambda}}c+ baryon polarization measurement in semileptonic
  beauty hadron decays}},\ }\href {https://doi.org/10.1103/PhysRevD.108.012023}
  {\bibfield  {journal} {\bibinfo  {journal} {Phys. Rev. D}\ }\textbf {\bibinfo
  {volume} {108}},\ \bibinfo {pages} {012023} (\bibinfo {year}
  {2023}{\natexlab{a}})},\ \Eprint {https://arxiv.org/abs/2208.03262}
  {arXiv:2208.03262 [hep-ex]} \BibitemShut {NoStop}%
\bibitem [{\citenamefont {Aaij}\ \emph
  {et~al.}(2023{\natexlab{b}})\citenamefont {Aaij} \emph
  {et~al.}}]{LHCb:2023crj}%
  \BibitemOpen
  \bibfield  {author} {\bibinfo {author} {\bibfnamefont {R.}~\bibnamefont
  {Aaij}} \emph {et~al.} (\bibinfo {collaboration} {LHCb}),\ }\bibfield
  {title} {\bibinfo {title} {{$ {\Lambda}_c^{+} $ polarimetry using the
  dominant hadronic mode}},\ }\href {https://doi.org/10.1007/JHEP07(2023)228}
  {\bibfield  {journal} {\bibinfo  {journal} {JHEP}\ }\textbf {\bibinfo
  {volume} {07}},\ \bibinfo {pages} {228}},\ \Eprint
  {https://arxiv.org/abs/2301.07010} {arXiv:2301.07010 [hep-ex]} \BibitemShut
  {NoStop}%
\bibitem [{\citenamefont {Ackerstaff}\ \emph
  {et~al.}(1997{\natexlab{b}})\citenamefont {Ackerstaff} \emph
  {et~al.}}]{OPAL:1997nwj}%
  \BibitemOpen
  \bibfield  {author} {\bibinfo {author} {\bibfnamefont {K.}~\bibnamefont
  {Ackerstaff}} \emph {et~al.} (\bibinfo {collaboration} {OPAL}),\ }\bibfield
  {title} {\bibinfo {title} {{Study of phi (1020), D*+- and B* spin alignment
  in hadronic Z0 decays}},\ }\href {https://doi.org/10.1007/s002880050406}
  {\bibfield  {journal} {\bibinfo  {journal} {Z. Phys. C}\ }\textbf {\bibinfo
  {volume} {74}},\ \bibinfo {pages} {437} (\bibinfo {year}
  {1997}{\natexlab{b}})}\BibitemShut {NoStop}%
\bibitem [{\citenamefont {Acharya}\ \emph
  {et~al.}(2023{\natexlab{b}})\citenamefont {Acharya} \emph
  {et~al.}}]{ALICE:2022byg}%
  \BibitemOpen
  \bibfield  {author} {\bibinfo {author} {\bibfnamefont {S.}~\bibnamefont
  {Acharya}} \emph {et~al.} (\bibinfo {collaboration} {ALICE}),\ }\bibfield
  {title} {\bibinfo {title} {{First measurement of prompt and non-prompt
  D{\textasteriskcentered}+ vector meson spin alignment in pp collisions at
  s=13 TeV}},\ }\href {https://doi.org/10.1016/j.physletb.2023.137920}
  {\bibfield  {journal} {\bibinfo  {journal} {Phys. Lett. B}\ }\textbf
  {\bibinfo {volume} {846}},\ \bibinfo {pages} {137920} (\bibinfo {year}
  {2023}{\natexlab{b}})},\ \Eprint {https://arxiv.org/abs/2212.06588}
  {arXiv:2212.06588 [nucl-ex]} \BibitemShut {NoStop}%
\bibitem [{\citenamefont {Kats}\ and\ \citenamefont
  {Uzan}(2024)}]{Kats:2023zxb}%
  \BibitemOpen
  \bibfield  {author} {\bibinfo {author} {\bibfnamefont {Y.}~\bibnamefont
  {Kats}}\ and\ \bibinfo {author} {\bibfnamefont {D.}~\bibnamefont {Uzan}},\
  }\bibfield  {title} {\bibinfo {title} {{Prospects for measuring quark
  polarization and spin correlations in $b\overline{b }$ and $c\overline{c }$
  samples at the LHC}},\ }\href {https://doi.org/10.1007/JHEP03(2024)063}
  {\bibfield  {journal} {\bibinfo  {journal} {JHEP}\ }\textbf {\bibinfo
  {volume} {03}},\ \bibinfo {pages} {063}},\ \Eprint
  {https://arxiv.org/abs/2311.08226} {arXiv:2311.08226 [hep-ph]} \BibitemShut
  {NoStop}%
\bibitem [{\citenamefont {Afik}\ \emph {et~al.}(2025)\citenamefont {Afik},
  \citenamefont {Kats}, \citenamefont {de~Nova}, \citenamefont {Soffer},\ and\
  \citenamefont {Uzan}}]{Afik:2025grr}%
  \BibitemOpen
  \bibfield  {author} {\bibinfo {author} {\bibfnamefont {Y.}~\bibnamefont
  {Afik}}, \bibinfo {author} {\bibfnamefont {Y.}~\bibnamefont {Kats}}, \bibinfo
  {author} {\bibfnamefont {J.~R.~M.}\ \bibnamefont {de~Nova}}, \bibinfo
  {author} {\bibfnamefont {A.}~\bibnamefont {Soffer}},\ and\ \bibinfo {author}
  {\bibfnamefont {D.}~\bibnamefont {Uzan}},\ }\bibfield  {title} {\bibinfo
  {title} {{Entanglement and Bell nonlocality with bottom-quark pairs at hadron
  colliders}},\ }\href {https://doi.org/10.1103/fhkc-kfhr} {\bibfield
  {journal} {\bibinfo  {journal} {Phys. Rev. D}\ }\textbf {\bibinfo {volume}
  {111}},\ \bibinfo {pages} {L111902} (\bibinfo {year} {2025})},\ \Eprint
  {https://arxiv.org/abs/2406.04402} {arXiv:2406.04402 [hep-ph]} \BibitemShut
  {NoStop}%
\bibitem [{\citenamefont {Aad}\ \emph {et~al.}(2026)\citenamefont {Aad} \emph
  {et~al.}}]{ATLAS:2025dkv}%
  \BibitemOpen
  \bibfield  {author} {\bibinfo {author} {\bibfnamefont {G.}~\bibnamefont
  {Aad}} \emph {et~al.} (\bibinfo {collaboration} {ATLAS}),\ }\bibfield
  {title} {\bibinfo {title} {{Transforming jet flavour tagging at ATLAS}},\
  }\href {https://doi.org/10.1038/s41467-025-65059-6} {\bibfield  {journal}
  {\bibinfo  {journal} {Nature Commun.}\ }\textbf {\bibinfo {volume} {17}},\
  \bibinfo {pages} {541} (\bibinfo {year} {2026})},\ \Eprint
  {https://arxiv.org/abs/2505.19689} {arXiv:2505.19689 [hep-ex]} \BibitemShut
  {NoStop}%
\bibitem [{\citenamefont {Seuster}\ \emph {et~al.}(2006)\citenamefont {Seuster}
  \emph {et~al.}}]{Belle:2005mtx}%
  \BibitemOpen
  \bibfield  {author} {\bibinfo {author} {\bibfnamefont {R.}~\bibnamefont
  {Seuster}} \emph {et~al.} (\bibinfo {collaboration} {Belle}),\ }\bibfield
  {title} {\bibinfo {title} {{Charm hadrons from fragmentation and B decays in
  e+ e- annihilation at s**(1/2) = 10.6-GeV}},\ }\href
  {https://doi.org/10.1103/PhysRevD.73.032002} {\bibfield  {journal} {\bibinfo
  {journal} {Phys. Rev. D}\ }\textbf {\bibinfo {volume} {73}},\ \bibinfo
  {pages} {032002} (\bibinfo {year} {2006})},\ \Eprint
  {https://arxiv.org/abs/hep-ex/0506068} {arXiv:hep-ex/0506068} \BibitemShut
  {NoStop}%
\bibitem [{\citenamefont {Seidl}\ \emph {et~al.}(2006)\citenamefont {Seidl}
  \emph {et~al.}}]{Belle:2005dmx}%
  \BibitemOpen
  \bibfield  {author} {\bibinfo {author} {\bibfnamefont {R.}~\bibnamefont
  {Seidl}} \emph {et~al.} (\bibinfo {collaboration} {Belle}),\ }\bibfield
  {title} {\bibinfo {title} {{Measurement of azimuthal asymmetries in inclusive
  production of hadron pairs in e+ e- annihilation at Belle}},\ }\href
  {https://doi.org/10.1103/PhysRevLett.96.232002} {\bibfield  {journal}
  {\bibinfo  {journal} {Phys. Rev. Lett.}\ }\textbf {\bibinfo {volume} {96}},\
  \bibinfo {pages} {232002} (\bibinfo {year} {2006})},\ \Eprint
  {https://arxiv.org/abs/hep-ex/0507063} {arXiv:hep-ex/0507063} \BibitemShut
  {NoStop}%
\bibitem [{\citenamefont {Lees}\ \emph {et~al.}(2014)\citenamefont {Lees} \emph
  {et~al.}}]{BaBar:2013jdt}%
  \BibitemOpen
  \bibfield  {author} {\bibinfo {author} {\bibfnamefont {J.~P.}\ \bibnamefont
  {Lees}} \emph {et~al.} (\bibinfo {collaboration} {BaBar}),\ }\bibfield
  {title} {\bibinfo {title} {{Measurement of Collins asymmetries in inclusive
  production of charged pion pairs in $e^+e^-$ annihilation at BABAR}},\ }\href
  {https://doi.org/10.1103/PhysRevD.90.052003} {\bibfield  {journal} {\bibinfo
  {journal} {Phys. Rev. D}\ }\textbf {\bibinfo {volume} {90}},\ \bibinfo
  {pages} {052003} (\bibinfo {year} {2014})},\ \Eprint
  {https://arxiv.org/abs/1309.5278} {arXiv:1309.5278 [hep-ex]} \BibitemShut
  {NoStop}%
\bibitem [{\citenamefont {Accardi}\ \emph {et~al.}(2022)\citenamefont {Accardi}
  \emph {et~al.}}]{Accardi:2022oog}%
  \BibitemOpen
  \bibfield  {author} {\bibinfo {author} {\bibfnamefont {A.}~\bibnamefont
  {Accardi}} \emph {et~al.},\ }\bibfield  {title} {\bibinfo {title}
  {{Opportunities for precision QCD physics in hadronization at Belle II -- a
  snowmass whitepaper}},\ }in\ \href@noop {} {\emph {\bibinfo {booktitle}
  {{Snowmass 2021}}}}\ (\bibinfo {year} {2022})\ \Eprint
  {https://arxiv.org/abs/2204.02280} {arXiv:2204.02280 [hep-ex]} \BibitemShut
  {NoStop}%
\bibitem [{\citenamefont {Abe}\ \emph {et~al.}(2010)\citenamefont {Abe} \emph
  {et~al.}}]{Belle-II:2010dht}%
  \BibitemOpen
  \bibfield  {author} {\bibinfo {author} {\bibfnamefont {T.}~\bibnamefont
  {Abe}} \emph {et~al.} (\bibinfo {collaboration} {Belle-II}),\ }\bibfield
  {title} {\bibinfo {title} {{Belle II Technical Design Report}},\ }\href@noop
  {} {\  (\bibinfo {year} {2010})},\ \Eprint {https://arxiv.org/abs/1011.0352}
  {arXiv:1011.0352 [physics.ins-det]} \BibitemShut {NoStop}%
\bibitem [{\citenamefont {Collins}(2011)}]{Collins:2011zzd}%
  \BibitemOpen
  \bibfield  {author} {\bibinfo {author} {\bibfnamefont {J.}~\bibnamefont
  {Collins}},\ }\href {https://doi.org/10.1017/9781009401845} {\emph {\bibinfo
  {title} {{Foundations of Perturbative QCD}}}},\ Vol.~\bibinfo {volume} {32}\
  (\bibinfo  {publisher} {Cambridge University Press},\ \bibinfo {year}
  {2011})\BibitemShut {NoStop}%
\bibitem [{\citenamefont {Hoang}\ \emph {et~al.}(2015)\citenamefont {Hoang},
  \citenamefont {Pathak}, \citenamefont {Pietrulewicz},\ and\ \citenamefont
  {Stewart}}]{Hoang:2015vua}%
  \BibitemOpen
  \bibfield  {author} {\bibinfo {author} {\bibfnamefont {A.~H.}\ \bibnamefont
  {Hoang}}, \bibinfo {author} {\bibfnamefont {A.}~\bibnamefont {Pathak}},
  \bibinfo {author} {\bibfnamefont {P.}~\bibnamefont {Pietrulewicz}},\ and\
  \bibinfo {author} {\bibfnamefont {I.~W.}\ \bibnamefont {Stewart}},\
  }\bibfield  {title} {\bibinfo {title} {{Hard Matching for Boosted Tops at Two
  Loops}},\ }\href {https://doi.org/10.1007/JHEP12(2015)059} {\bibfield
  {journal} {\bibinfo  {journal} {JHEP}\ }\textbf {\bibinfo {volume} {12}},\
  \bibinfo {pages} {059}},\ \Eprint {https://arxiv.org/abs/1508.04137}
  {arXiv:1508.04137 [hep-ph]} \BibitemShut {NoStop}%
\bibitem [{\citenamefont {Hoang}\ \emph {et~al.}(2019)\citenamefont {Hoang},
  \citenamefont {Lepenik},\ and\ \citenamefont {Stahlhofen}}]{Hoang:2019fze}%
  \BibitemOpen
  \bibfield  {author} {\bibinfo {author} {\bibfnamefont {A.~H.}\ \bibnamefont
  {Hoang}}, \bibinfo {author} {\bibfnamefont {C.}~\bibnamefont {Lepenik}},\
  and\ \bibinfo {author} {\bibfnamefont {M.}~\bibnamefont {Stahlhofen}},\
  }\bibfield  {title} {\bibinfo {title} {{Two-Loop Massive Quark Jet Functions
  in SCET}},\ }\href {https://doi.org/10.1007/JHEP08(2019)112} {\bibfield
  {journal} {\bibinfo  {journal} {JHEP}\ }\textbf {\bibinfo {volume} {08}},\
  \bibinfo {pages} {112}},\ \Eprint {https://arxiv.org/abs/1904.12839}
  {arXiv:1904.12839 [hep-ph]} \BibitemShut {NoStop}%
\bibitem [{\citenamefont {Sakurai}\ and\ \citenamefont
  {Napolitano}(2020)}]{Sakurai:2011zz}%
  \BibitemOpen
  \bibfield  {author} {\bibinfo {author} {\bibfnamefont {J.~J.}\ \bibnamefont
  {Sakurai}}\ and\ \bibinfo {author} {\bibfnamefont {J.}~\bibnamefont
  {Napolitano}},\ }\href {https://doi.org/10.1017/9781108587280} {\emph
  {\bibinfo {title} {{Modern Quantum Mechanics}}}},\ \bibinfo {edition} {3rd}\
  ed.,\ Quantum physics, quantum information and quantum computation\ (\bibinfo
   {publisher} {Cambridge University Press},\ \bibinfo {year}
  {2020})\BibitemShut {NoStop}%
\bibitem [{\citenamefont {Toth}\ and\ \citenamefont
  {Turyshev}(2022)}]{Toth:2021cpx}%
  \BibitemOpen
  \bibfield  {author} {\bibinfo {author} {\bibfnamefont {V.~T.}\ \bibnamefont
  {Toth}}\ and\ \bibinfo {author} {\bibfnamefont {S.~G.}\ \bibnamefont
  {Turyshev}},\ }\bibfield  {title} {\bibinfo {title} {{Efficient trace-free
  decomposition of symmetric tensors of arbitrary rank}},\ }\href
  {https://doi.org/10.1142/S0219887822502012} {\bibfield  {journal} {\bibinfo
  {journal} {Int. J. Geom. Meth. Mod. Phys.}\ }\textbf {\bibinfo {volume}
  {19}},\ \bibinfo {pages} {2250201} (\bibinfo {year} {2022})},\ \Eprint
  {https://arxiv.org/abs/2109.11743} {arXiv:2109.11743 [gr-qc]} \BibitemShut
  {NoStop}%
\bibitem [{\citenamefont {Cho}(1992)}]{Cho:1992gg}%
  \BibitemOpen
  \bibfield  {author} {\bibinfo {author} {\bibfnamefont {P.~L.}\ \bibnamefont
  {Cho}},\ }\bibfield  {title} {\bibinfo {title} {{Chiral perturbation theory
  for hadrons containing a heavy quark: The Sequel}},\ }\href
  {https://doi.org/10.1016/0370-2693(92)91314-Y} {\bibfield  {journal}
  {\bibinfo  {journal} {Phys. Lett. B}\ }\textbf {\bibinfo {volume} {285}},\
  \bibinfo {pages} {145} (\bibinfo {year} {1992})},\ \Eprint
  {https://arxiv.org/abs/hep-ph/9203225} {arXiv:hep-ph/9203225} \BibitemShut
  {NoStop}%
\bibitem [{\citenamefont {Yan}\ \emph {et~al.}(1992)\citenamefont {Yan},
  \citenamefont {Cheng}, \citenamefont {Cheung}, \citenamefont {Lin},
  \citenamefont {Lin},\ and\ \citenamefont {Yu}}]{Yan:1992gz}%
  \BibitemOpen
  \bibfield  {author} {\bibinfo {author} {\bibfnamefont {T.-M.}\ \bibnamefont
  {Yan}}, \bibinfo {author} {\bibfnamefont {H.-Y.}\ \bibnamefont {Cheng}},
  \bibinfo {author} {\bibfnamefont {C.-Y.}\ \bibnamefont {Cheung}}, \bibinfo
  {author} {\bibfnamefont {G.-L.}\ \bibnamefont {Lin}}, \bibinfo {author}
  {\bibfnamefont {Y.~C.}\ \bibnamefont {Lin}},\ and\ \bibinfo {author}
  {\bibfnamefont {H.-L.}\ \bibnamefont {Yu}},\ }\bibfield  {title} {\bibinfo
  {title} {{Heavy quark symmetry and chiral dynamics}},\ }\href
  {https://doi.org/10.1103/PhysRevD.46.1148} {\bibfield  {journal} {\bibinfo
  {journal} {Phys. Rev. D}\ }\textbf {\bibinfo {volume} {46}},\ \bibinfo
  {pages} {1148} (\bibinfo {year} {1992})},\ \bibinfo {note} {[Erratum:
  Phys.Rev.D 55, 5851 (1997)]}\BibitemShut {NoStop}%
\bibitem [{\citenamefont {Lee}\ and\ \citenamefont
  {Moult}(2026)}]{Lee:2026bcq}%
  \BibitemOpen
  \bibfield  {author} {\bibinfo {author} {\bibfnamefont {K.}~\bibnamefont
  {Lee}}\ and\ \bibinfo {author} {\bibfnamefont {I.}~\bibnamefont {Moult}},\
  }\bibfield  {title} {\bibinfo {title} {{Flavor Flux Correlators,
  Forward-Backward Asymmetries, and Anomalies}},\ }\href@noop {} {\  (\bibinfo
  {year} {2026})},\ \Eprint {https://arxiv.org/abs/2609.13366}
  {arXiv:2609.13366 [hep-ph]} \BibitemShut {NoStop}%
\bibitem [{\citenamefont {Ko}\ \emph {et~al.}(2026)\citenamefont {Ko},
  \citenamefont {Lee}, \citenamefont {Liko}, \citenamefont {Moult},
  \citenamefont {Qian},\ and\ \citenamefont {Zhang}}]{Ko:2026ojy}%
  \BibitemOpen
  \bibfield  {author} {\bibinfo {author} {\bibfnamefont {S.}~\bibnamefont
  {Ko}}, \bibinfo {author} {\bibfnamefont {K.}~\bibnamefont {Lee}}, \bibinfo
  {author} {\bibfnamefont {D.}~\bibnamefont {Liko}}, \bibinfo {author}
  {\bibfnamefont {I.}~\bibnamefont {Moult}}, \bibinfo {author} {\bibfnamefont
  {S.}~\bibnamefont {Qian}},\ and\ \bibinfo {author} {\bibfnamefont
  {J.}~\bibnamefont {Zhang}},\ }\bibfield  {title} {\bibinfo {title}
  {{Forward-backward asymmetries from bottom flux correlators with DELPHI open
  data}},\ }\href@noop {} {\  (\bibinfo {year} {2026})},\ \Eprint
  {https://arxiv.org/abs/2611.35105} {arXiv:2611.35105 [hep-ph]} \BibitemShut
  {NoStop}%
\bibitem [{\citenamefont {Korchemsky}\ and\ \citenamefont
  {Radyushkin}(1992)}]{Korchemsky:1991zp}%
  \BibitemOpen
  \bibfield  {author} {\bibinfo {author} {\bibfnamefont {G.~P.}\ \bibnamefont
  {Korchemsky}}\ and\ \bibinfo {author} {\bibfnamefont {A.~V.}\ \bibnamefont
  {Radyushkin}},\ }\bibfield  {title} {\bibinfo {title} {{Infrared
  factorization, Wilson lines and the heavy quark limit}},\ }\href
  {https://doi.org/10.1016/0370-2693(92)90405-S} {\bibfield  {journal}
  {\bibinfo  {journal} {Phys. Lett. B}\ }\textbf {\bibinfo {volume} {279}},\
  \bibinfo {pages} {359} (\bibinfo {year} {1992})},\ \Eprint
  {https://arxiv.org/abs/hep-ph/9203222} {arXiv:hep-ph/9203222} \BibitemShut
  {NoStop}%
\bibitem [{\citenamefont {Bauer}\ \emph {et~al.}(2002)\citenamefont {Bauer},
  \citenamefont {Pirjol},\ and\ \citenamefont {Stewart}}]{Bauer:2001yt}%
  \BibitemOpen
  \bibfield  {author} {\bibinfo {author} {\bibfnamefont {C.~W.}\ \bibnamefont
  {Bauer}}, \bibinfo {author} {\bibfnamefont {D.}~\bibnamefont {Pirjol}},\ and\
  \bibinfo {author} {\bibfnamefont {I.~W.}\ \bibnamefont {Stewart}},\
  }\bibfield  {title} {\bibinfo {title} {{Soft collinear factorization in
  effective field theory}},\ }\href
  {https://doi.org/10.1103/PhysRevD.65.054022} {\bibfield  {journal} {\bibinfo
  {journal} {Phys. Rev. D}\ }\textbf {\bibinfo {volume} {65}},\ \bibinfo
  {pages} {054022} (\bibinfo {year} {2002})},\ \Eprint
  {https://arxiv.org/abs/hep-ph/0109045} {arXiv:hep-ph/0109045} \BibitemShut
  {NoStop}%
\bibitem [{\citenamefont {Ebert}\ \emph {et~al.}(2022)\citenamefont {Ebert},
  \citenamefont {Gao},\ and\ \citenamefont {Stewart}}]{Ebert:2021jhy}%
  \BibitemOpen
  \bibfield  {author} {\bibinfo {author} {\bibfnamefont {M.~A.}\ \bibnamefont
  {Ebert}}, \bibinfo {author} {\bibfnamefont {A.}~\bibnamefont {Gao}},\ and\
  \bibinfo {author} {\bibfnamefont {I.~W.}\ \bibnamefont {Stewart}},\
  }\bibfield  {title} {\bibinfo {title} {{Factorization for azimuthal
  asymmetries in SIDIS at next-to-leading power}},\ }\href
  {https://doi.org/10.1007/JHEP06(2022)007} {\bibfield  {journal} {\bibinfo
  {journal} {JHEP}\ }\textbf {\bibinfo {volume} {06}},\ \bibinfo {pages}
  {007}},\ \bibinfo {note} {[Erratum: JHEP 07, 096 (2023)]},\ \Eprint
  {https://arxiv.org/abs/2112.07680} {arXiv:2112.07680 [hep-ph]} \BibitemShut
  {NoStop}%
\bibitem [{\citenamefont {Bacchetta}\ \emph {et~al.}(2007)\citenamefont
  {Bacchetta}, \citenamefont {Diehl}, \citenamefont {Goeke}, \citenamefont
  {Metz}, \citenamefont {Mulders},\ and\ \citenamefont
  {Schlegel}}]{Bacchetta:2006tn}%
  \BibitemOpen
  \bibfield  {author} {\bibinfo {author} {\bibfnamefont {A.}~\bibnamefont
  {Bacchetta}}, \bibinfo {author} {\bibfnamefont {M.}~\bibnamefont {Diehl}},
  \bibinfo {author} {\bibfnamefont {K.}~\bibnamefont {Goeke}}, \bibinfo
  {author} {\bibfnamefont {A.}~\bibnamefont {Metz}}, \bibinfo {author}
  {\bibfnamefont {P.~J.}\ \bibnamefont {Mulders}},\ and\ \bibinfo {author}
  {\bibfnamefont {M.}~\bibnamefont {Schlegel}},\ }\bibfield  {title} {\bibinfo
  {title} {{Semi-inclusive deep inelastic scattering at small transverse
  momentum}},\ }\href {https://doi.org/10.1088/1126-6708/2007/02/093}
  {\bibfield  {journal} {\bibinfo  {journal} {JHEP}\ }\textbf {\bibinfo
  {volume} {02}},\ \bibinfo {pages} {093}},\ \Eprint
  {https://arxiv.org/abs/hep-ph/0611265} {arXiv:hep-ph/0611265} \BibitemShut
  {NoStop}%
\end{thebibliography}%

\end{document}